\documentclass[journal]{IEEEtran}

\usepackage{amsmath,amsfonts,amssymb,mathtools,amsthm}
\usepackage{graphicx}
\usepackage[caption=false,font=footnotesize,labelfont=sf,textfont=sf]{subfig}
\usepackage{booktabs}             
\usepackage{multirow}
\usepackage{algorithm}
\usepackage{algorithmic}          
\usepackage{array}
\usepackage{cite}                 
\usepackage{url}
\usepackage{hyperref}
\usepackage[capitalize,noabbrev]{cleveref}
\usepackage{xcolor}               
\usepackage{orcidlink}

\theoremstyle{plain}

\theoremstyle{definition}

\theoremstyle{remark}

\begin{document}

\title{DASM: Domain-Aware Sharpness Minimization\\for Multi-Domain Voice Stream Steganalysis}

\author{
        Pengcheng~Zhou\orcidlink{0009-0001-2659-1166}\textsuperscript{\dag},
        Pianran~Guo\orcidlink{0009-0008-8916-7937}\textsuperscript{\dag},
        Shuhua~Chen\orcidlink{0009-0007-3729-6334},
        Mengqin~Zhao\orcidlink{0009-0009-5036-2257},
        Zhongliang~Yang\orcidlink{0009-0003-8708-5202},
        and~Linna~Zhou%
\thanks{
( \textsuperscript{\dag} Pengcheng Zhou and  Pianran Guo contributed equally to this work.) (Corresponding authors: Zhongliang Yang; Linna Zhou.)}%
\thanks{Pengcheng Zhou is with the Department of Electrical and Computer Engineering, National University of Singapore, Singapore 117583, Singapore.}%
\thanks{Pianran Guo, Shuhua Chen, Zhongliang Yang and Linna Zhou are with the School of Cyberspace Security, Beijing University of Posts and Telecommunications, Beijing 100876, China (e-mail: yangzl@bupt.edu.cn; zhoulinna@bupt.edu.cn).}%
\thanks{Mengqin Zhao is with the College of Communication Engineering, Jilin University, Changchun 130012, China.}%
}

\markboth{IEEE Transactions on Information Forensics and Security}%
{Lastname1 \MakeLowercase{\textit{et al.}}: DASM for Multi-Domain Voice Stream Steganalysis}

\maketitle

\begin{abstract}
The growing use of information hiding in network streaming media for covert
communication poses a significant security threat, necessitating the development
of robust detection technologies. However, existing steganalysis methods for
network voice streams mostly rely on data distributions in specific scenarios,
making it difficult to adapt to the practical detection needs of non-homologous
data distributions. Through Hessian analysis, we find that the loss landscapes
of mainstream models are dominated by numerous saddle points and sharp local
minima, rendering them highly sensitive to data distribution shifts and
fundamentally limiting generalization. Therefore, we propose a new optimizer,
Domain-Aware Sharpness Minimization (DASM). The core mechanisms of DASM
consist of two aspects: first, it integrates domain-supervised contrastive
learning with sharpness-aware optimization, explicitly preserving inter-domain
feature separation while seeking flat minima; second, we design an adaptive
domain gap modulation strategy that dynamically calibrates the optimization loss
weights by sensing the real-time feature separability of different domains.
Extensive experimental results demonstrate that our method outperforms the
state-of-the-art methods by a large margin and achieves excellent generalization
and robustness. Our codes are available at \href{https://github.com/CamelliaLilium/DASM/}{DASM}.
\end{abstract}


\section{Introduction}
\label{sec:intro}

\begin{figure}[t]
  \centering
  \includegraphics[width=\columnwidth]{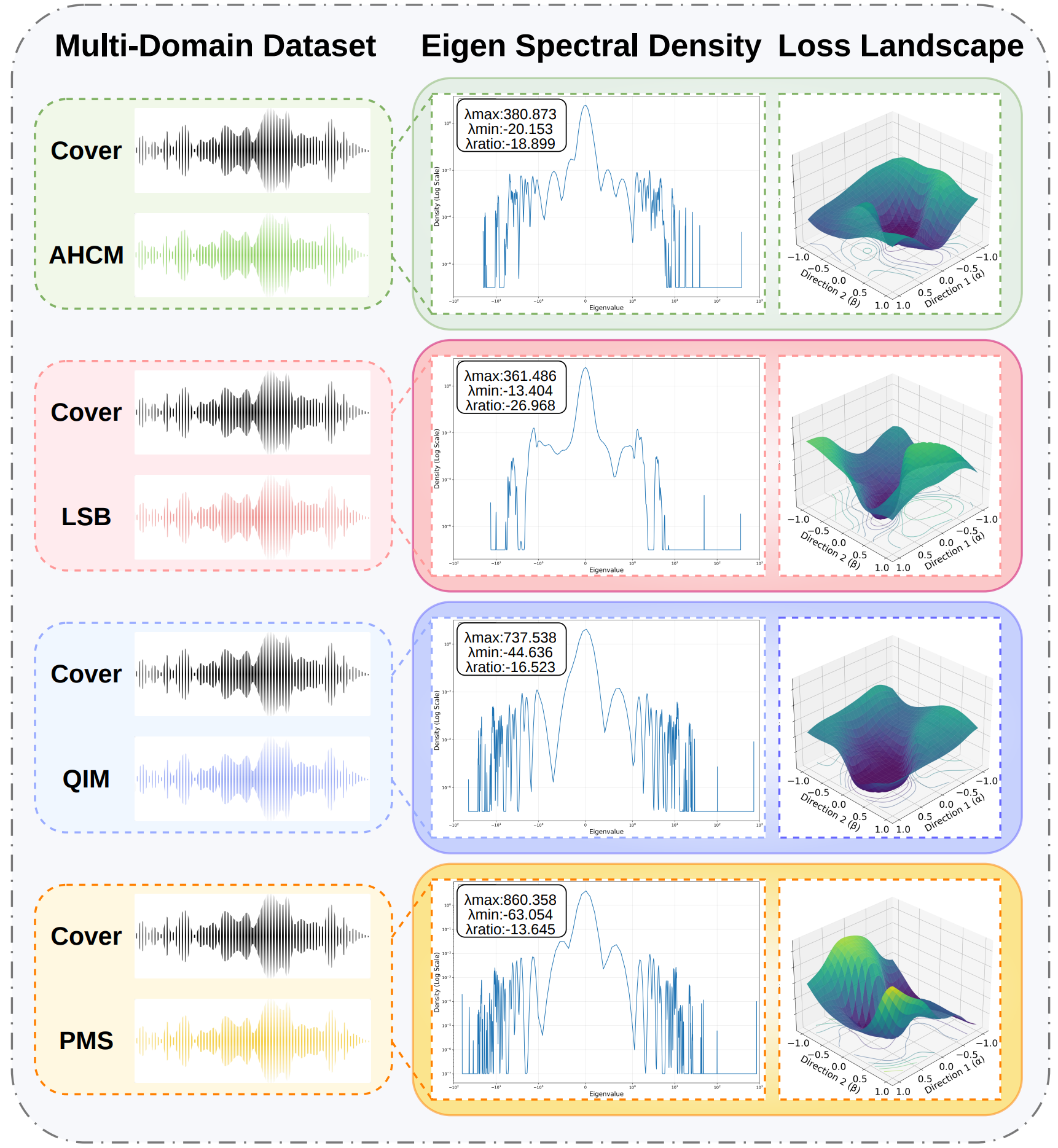}
  \caption{Multi-domain Hessian analysis. Eigenvalue spectral density and loss
  landscape visualizations reveal that AHCM and LSB converge to flat minima while
  QIM and PMS are dominated by saddle points, highlighting optimization challenges
  in heterogeneous distributions.}
  \label{fig:hessian}
  \vspace{-1em}
\end{figure}

\IEEEPARstart{V}{oice} over Internet Protocol (VoIP) streams have become a
ubiquitous carrier for covert communication due to their high throughput and
ephemeral nature, posing potential security risks. To counteract these threats,
steganalysis techniques have evolved from traditional correlation-based
methods~\cite{CCN, SS-QCCN} to advanced deep learning
frameworks~\cite{KFEF, SFFN, FS-MDP, LStegT, DAEF-VS, DVSF}, achieving
remarkable success in controlled environments. However, these methods encounter a
significant challenge where they often suffer from performance degradation when
facing non-homologous data distributions. This phenomenon arises from the
distinct statistical footprints introduced by different steganographic algorithms,
specifically Quantization Index Modulation (QIM)~\cite{CNV_QIM}, Pitch Modulation
Steganography (PMS)~\cite{PMS}, Least Significant Bit (LSB), and Adaptive Huffman
Code Mapping (AHCM)~\cite{AHCM}. Through rigorous Hessian spectral analysis
(Fig.~\ref{fig:hessian}), we reveal a critical finding that the loss landscapes of
mainstream models are dominated by numerous saddle points and sharp local minima,
particularly in domains with complex embedding mechanisms. This geometric
characteristic renders models highly sensitive to distribution shifts, restricting
their generalization capability.

To mitigate the issue of saddle points and enhance generalization under
distributional shifts, optimization algorithms that seek flat minima, most notably
Sharpness-Aware Minimization (SAM) and its variants, have become established
solutions in the general machine learning
community~\cite{foret2021sharpness, du2022efficient, wang2023sharpness, zhang2024friendly}.
Specifically, recent works have explored leveraging sharpness-aware minimization
to escape saddle points and handle domain
shifts~\cite{SAM_SaddlePointt, DISAM, DGSAM}. However, directly transferring
generic SAM algorithms to multi-domain voice stream steganalysis proves suboptimal
due to the unique nature of steganographic data. Unlike computer vision tasks
characterized by explicit semantic domain shifts, the domain discrepancies in VoIP
steganalysis are both minute and imbalanced. First, the domain gap is extremely
minute because the primary goal of steganography is imperceptibility, resulting in
critically weak distinguishing features. Second, the domain gaps are imbalanced
because different algorithms exhibit varying levels of detection difficulty.

To address these challenges, we propose a novel optimizer, Domain-Aware Sharpness
Minimization (DASM), designed to navigate the complex optimization landscape of
multi-algorithm steganalysis. First, to address the minute domain gaps caused by
high imperceptibility, we introduce Domain-Supervised Contrastive Learning (DSCL),
which explicitly widens the separation between different steganographic domains
during the perturbation step to enhance feature discriminability. Second, to handle
the imbalanced domain gaps caused by varying detection difficulties, we design
Adaptive Domain Gap Modulation (ADGM), which dynamically calibrates the
optimization weights by sensing the real-time feature separability of different
domains, ensuring the optimizer focuses on the most challenging directions to
effectively escape saddle points. Extensive experiments on datasets containing
QIM, PMS, LSB, and AHCM algorithms demonstrate that DASM significantly outperforms
state-of-the-art methods, achieving superior detection accuracy and robustness
against distribution shifts.

Our main contributions are summarized as follows:
\begin{itemize}
    \item We perform the first Hessian analysis in voice stream steganalysis,
    identifying that poor generalization on non-homologous data stems from
    convergence to saddle points and sharp minima in the loss landscape.
    \item We propose DASM, which integrates DSCL to amplify minute feature
    differences and employs ADGM to balance uneven detection difficulties,
    effectively enabling the model to escape saddle points.
    \item Extensive experiments on datasets containing QIM, PMS, LSB, and AHCM
    algorithms demonstrate that DASM significantly outperforms state-of-the-art
    methods, establishing new benchmarks for generalization and robustness.
\end{itemize}

\section{Related Work}

\subsection{VoIP Steganalysis Techniques}

Voice over IP (VoIP) has emerged as a primary carrier for covert communication,
allowing secret data to be imperceptibly embedded into real-time speech streams.
The evolution of steganography has progressed from rudimentary Least Significant
Bit (LSB) substitution~\cite{LSB2000,LSB2002,LSB2014,LSB2023} to highly
sophisticated mechanisms deeply integrated into speech encoding standards.
Prominent techniques include Quantization Index Modulation
(QIM)~\cite{QIM, CNV, LPC1, LPC2, SPM}, which alters quantization indices during
encoding; Pitch Modulation Steganography
(PMS)~\cite{PMS,nishimura2009data,huang2012steganography,janicki2016pitch,wu2023method},
which manipulates adaptive codebook parameters; and Adaptive Huffman Code Mapping
(AHCM)~\cite{ahcm2019}, which modifies entropy coding. However, the distinct
statistical artifacts introduced by these diverse embedding algorithms create a
highly heterogeneous feature space with minute, imperceptible modifications. This
poses a significant challenge for detecting low-rate steganography across varying
domains, necessitating a geometry-aware optimization approach to capture these
subtle signal variations robustly.

\subsection{VoIP Steganalysis Methods}

Steganalysis has evolved from handcrafted statistical
features~\cite{qiao2019mdii, liu2020cooccurrence} to advanced deep learning
architectures. Significant progress includes mechanism-aware detectors like
CCN~\cite{CCN} and SS-QCCN~\cite{SS-QCCN}; models leveraging temporal-spectral
representations such as SFFN~\cite{SFFN} and KFEF~\cite{KFEF}; detectors
capturing frame-level correlations like FS-MDP~\cite{FS-MDP} and
LStegT~\cite{LStegT}; and streaming architectures including DVSF~\cite{DVSF} and
DAEF-VS~\cite{DAEF-VS}. Despite these advancements, most approaches rely on
Empirical Risk Minimization, which excels on i.i.d.\ data but falters under
distribution shifts caused by varying codecs and network conditions. However,
standard domain adaptation methods fail to address the granular and imbalanced
domain gaps inherent to low-rate embedding, often causing models to converge to
sharp minima or saddle points. This highlights the critical need to explicitly
model the geometric structure of the multi-domain loss landscape for improved
generalization.

\subsection{Sharpness-Aware Optimization}

Generalization capability is closely linked to the geometry of the loss landscape.
Theoretical and empirical studies indicate that models converging to flat minima
exhibit better generalization~\cite{keskar2017large}. Sharpness-Aware Minimization
(SAM) was proposed to guide optimization toward flat regions by minimizing the
maximum loss within a neighborhood~\cite{foret2021sharpness}. Subsequent works
have improved the efficiency and effectiveness of SAM: ESAM enhances computational
efficiency~\cite{du2022efficient}, FSAM removes harmful gradient components to
improve generalization~\cite{zhang2024friendly}, and SAGM uses gradient matching
for domain generalization~\cite{wang2023sharpness}. Recent domain-inspired
variants like DISAM~\cite{DISAM} and DGSAM~\cite{DGSAM} further explore
sharpness-aware training under distribution shifts. However, a fundamental
limitation persists: these methods primarily choose a uniform or isotropic
smoothness constraint over the parameter neighborhood, which is misaligned with
problems where discriminative signals are microscopic and heterogeneously
distributed across domains, as in low-rate steganalysis. An isotropic perturbation
risks obliterating the very faint, domain-specific features that are crucial for
detection.

\section{Problem Formulation}
\label{sec:problem}

We define multi-domain audio steganalysis as a binary classification task over a
heterogeneous dataset $\mathcal{D} = \{\mathcal{D}_1, \dots, \mathcal{D}_S\}$
consisting of $S$ distinct source domains. Each domain $\mathcal{D}_s$ corresponds
to a specific steganographic embedding algorithm (e.g., QIM, PMS, LSB, AHCM) with
a unique data distribution $P_s(X, Y)$. The dataset for the $s$-th domain is
defined as $\mathcal{D}_s = \{(x_i^s, y_i^s)\}_{i=1}^{N_s}$, where
$x_i^s \in \mathbb{R}^d$ represents the input audio feature vector (e.g., deep
representations of cover or stego audio), and $y_i^s \in \{0, 1\}$ is the
corresponding binary label ($0$ for cover, $1$ for stego). Given the low embedding
rates, the feature discrepancy between cover and stego samples is microscopic,
while the distribution shift between domains $P_i \neq P_j$ is significant. We
define the training set as the union of all source domains
$\mathcal{S}_{train} = \bigcup_{s=1}^S \mathcal{D}_s$. The goal is to learn a
robust mapping function $f_\theta: \mathbb{R}^d \rightarrow \{0, 1\}$,
parameterized by $\theta$, using $\mathcal{S}_{train}$. The objective is to find
a parameter configuration $\theta^*$ that correctly assigns the binary label
$y \in \{0, 1\}$ to an input query from the target domain $\mathcal{D}_{target}$,
by learning discriminative features robust to the minute and imbalanced domain
gaps across heterogeneous steganographic algorithms.

\section{Methodology}
\label{sec:method}

\begin{figure*}[t]
    \centering
    \includegraphics[width=\linewidth]{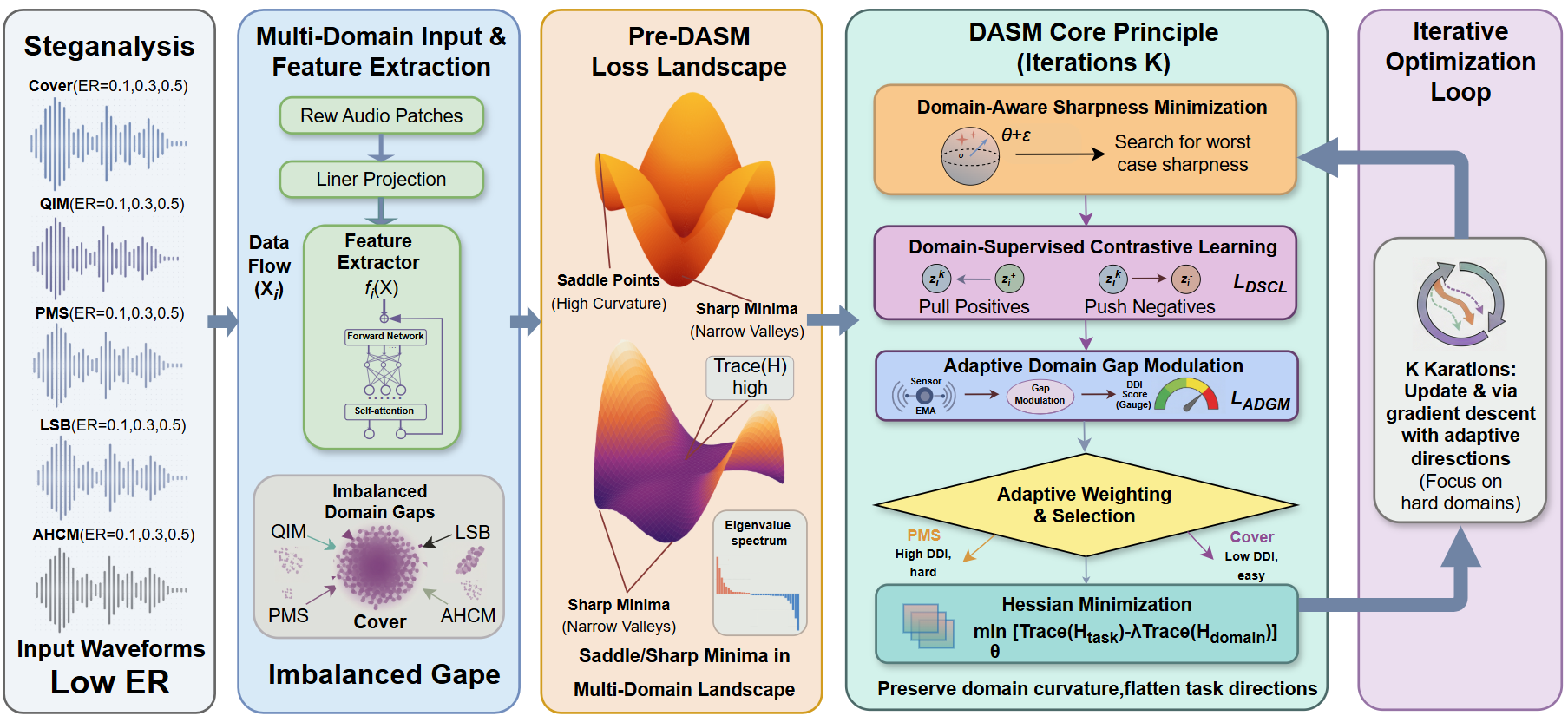}
    \caption{Overview of DASM for multi-domain voice steganalysis. (1)~Input:
    Cover and steganographic audio patches with multiple embedding rates.
    (2)~Feature Extraction: Raw audio projected into a shared representation
    space, revealing imbalanced domain gaps. (3)~Pre-DASM Loss Landscape:
    Dominated by saddle points and sharp minima. (4)~DASM Core Principle: Combines
    domain-supervised contrastive learning $\mathcal{L}_{\text{DSCL}}$ and
    adaptive domain gap modulation $\mathcal{L}_{\text{ADGM}}$ with
    sharpness-aware perturbation to reshape the landscape. (5)~Iterative
    Optimization Loop: Solves
    $\min_\theta \max_{\|\epsilon\| \le \rho} \mathcal{L}_{\text{total}}(\theta + \epsilon)$
    with adaptive weights focusing on hard domains, enabling robust detection
    across distribution shifts.}
    \label{fig:Method_fig}
\end{figure*}

Building upon the problem formulation, we propose Domain-Aware Sharpness
Minimization (DASM). Fig.~\ref{fig:Method_fig} provides an overview of the DASM
framework. To robustly solve the multi-domain detection task, we model the training
process as a minimax optimization problem, aiming to find parameters that minimize
the worst-case loss within a perturbation neighborhood.

\subsection{Overview of DASM}
\label{subsec:overview}

Standard training minimizes the empirical risk, which often leads to sharp minima.
In contrast, our method seeks a flat minimum by solving the following
sharpness-aware objective:
\begin{equation}
\label{eq:objective}
\min_{\theta} \mathbb{E}_{(\mathbf{x}, y) \sim \mathcal{S}_{train}}
\left[ \max_{\|\epsilon\|_2 \le \rho} \mathcal{L}_{\text{total}}(\theta + \epsilon) \right],
\end{equation}
where $\rho$ represents the perturbation radius, and $\epsilon$ is the adversarial
weight perturbation.

Standard SAM approximates the inner maximization of \eqref{eq:objective} by solely
maximizing the cross-entropy loss. However, in the context of multi-domain
steganalysis, a perturbation that simply maximizes classification error does not
necessarily reflect the geometric collapse of minute steganographic features.
Therefore, we redesign the optimization objective to enforce that the model
maintains both feature separation and domain balance even under the worst-case
perturbation.

Specifically, at each training step $t$, let
$\mathcal{B} = \{(x_i, y_i, d_i)\}_{i=1}^B$ be a mini-batch, where $d_i$ denotes
the domain index. We define a composite domain-aware loss $\mathcal{L}_{\text{total}}$
as the sum of three components:
\begin{equation}
\label{eq:total_loss}
\mathcal{L}_{\text{total}}(\theta) =
  \mathcal{L}_{\text{CE}}(\theta) +
  \mathcal{L}_{\text{DSCL}}(\theta) +
  \mathcal{L}_{\text{ADGM}}(\theta),
\end{equation}
where $\mathcal{L}_{\text{CE}}$ is the standard cross-entropy loss for
classification, $\mathcal{L}_{\text{DSCL}}$ is the Domain-Supervised Contrastive
Learning loss that preserves inter-domain feature separability, and
$\mathcal{L}_{\text{ADGM}}$ is the Adaptive Domain Gap Modulation loss that
balances optimization across domains. The three loss terms are designed to be
self-normalized to comparable scales, eliminating the need for manual balancing
coefficients.

The proposed method performs a two-step optimization. First, it computes an
adversarial perturbation $\hat{\epsilon}$ that maximizes $\mathcal{L}_{\text{total}}$
within the local neighborhood defined by $\rho$ via a first-order Taylor expansion:
\begin{equation}
\label{eq:perturbation}
\hat{\epsilon} = \rho \frac{\nabla_\theta \mathcal{L}_{\text{total}}(\theta_t)}
                           {\|\nabla_\theta \mathcal{L}_{\text{total}}(\theta_t)\|_2}.
\end{equation}
Second, the model parameters are updated using the gradient evaluated at the
perturbed state $\theta_t + \hat{\epsilon}$:
\begin{equation}
\label{eq:update}
\theta_{t+1} = \theta_t - \eta \nabla_\theta \mathcal{L}_{\text{total}}(\theta_t + \hat{\epsilon}),
\end{equation}
where $\eta$ is the learning rate. Note that $\mathcal{L}_{\text{total}}$ in
\eqref{eq:total_loss} incorporates adaptive weights $w_k$ through
$\mathcal{L}_{\text{ADGM}}(w_k)$, where $w_k$ are computed adaptively based on
domain gaps as described in Section~\ref{subsec:adgm}. By incorporating
$\mathcal{L}_{\text{DSCL}}$ and $\mathcal{L}_{\text{ADGM}}$ into the perturbation
generation process, the optimizer is guided to find a flat region where the loss
landscape is robust not only in terms of classification accuracy but also in terms
of domain separability and balance. The complete training procedure is summarized
in Algorithm~\ref{alg:dasm}.

\begin{algorithm}[t]
\caption{Domain-Aware Sharpness Minimization (DASM)}
\label{alg:dasm}
\begin{algorithmic}[1]
\REQUIRE Training dataset $\mathcal{D}$, batch size $B$, learning rate $\eta$,
         perturbation radius $\rho$, contrastive temperature $\tau$, EMA
         momentum $\mu$
\ENSURE  Trained parameters $\theta$
\STATE Initialize model parameters $\theta_0$, domain centers
       $\mathbf{C} = \{\mathbf{c}_k\}_{k=0}^K$, step $t \leftarrow 0$
\WHILE{not converged}
    \STATE Sample a minibatch $\mathcal{B} = \{(x_i, y_i, d_i)\}_{i=1}^B$
           from $\mathcal{D}$
    \STATE Extract features and update domain centers $\mathbf{c}_k$ via EMA
    \STATE Compute domain gaps $g_k = \|\mathbf{c}_k - \mathbf{c}_{\text{cover}}\|_2$
    \STATE Compute adaptive temperature $\tau_g = \text{std}(\{g_k\}) + \xi$
    \STATE Compute adaptive weights $w_k = \text{softmax}(-g_k / \tau_g)$
    \STATE Compute composite loss
           $\mathcal{L}_{\text{total}} = \mathcal{L}_{\text{CE}} +
           \mathcal{L}_{\text{DSCL}} + \mathcal{L}_{\text{ADGM}}$
    \STATE Compute gradient $g = \nabla_{\theta} \mathcal{L}_{\text{total}}(\theta_t)$
    \STATE Compute adversarial perturbation $\hat{\epsilon} = \rho \cdot g / \|g\|_2$
    \STATE Compute perturbed gradient
           $g_{\text{adv}} = \nabla_{\theta} \mathcal{L}_{\text{total}}(\theta_t + \hat{\epsilon})$
    \STATE Update parameters $\theta_{t+1} \leftarrow \theta_t - \eta\, g_{\text{adv}}$
    \STATE $t \leftarrow t + 1$
\ENDWHILE
\STATE \textbf{return} $\theta_t$
\end{algorithmic}
\end{algorithm}

\vspace{-1em}
\subsection{Domain-Supervised Contrastive Learning}
\label{subsec:dscl}

The domain discrepancies between cover and steganographic streams are inherently
minute. Standard optimization often leads to feature representations that are
inseparable near the decision boundary, making the model vulnerable to slight
distribution shifts. To counteract this, we introduce Domain-Supervised
Contrastive Learning (DSCL). The core insight is to enforce feature separability
under worst-case perturbation by pulling together samples from the same domain
while pushing apart samples from different domains.

Let $\mathbf{z}_i = f_\theta(x_i) / \|f_\theta(x_i)\|_2$ be the L2-normalized
feature representation of sample $x_i$. For each anchor sample $i$ in the batch,
we define the set of positive samples $P(i) = \{j : d_j = d_i,\, j \neq i\}$ as
samples sharing the same domain label, and the set of negative samples
$N(i) = \{j : d_j \neq d_i\}$ as samples from different domains. The DSCL loss
aggregates positive and negative similarities:
\begin{equation}
\label{eq:sim}
S_i^+ = \sum_{p \in P(i)} \exp(\mathbf{z}_i^\top \mathbf{z}_p / \tau), \quad
S_i^- = \sum_{n \in N(i)} \exp(\mathbf{z}_i^\top \mathbf{z}_n / \tau),
\end{equation}
where $\tau$ is a temperature hyperparameter. The DSCL loss follows an InfoNCE
formulation:
\begin{equation}
\label{eq:dscl}
\mathcal{L}_{\text{DSCL}} =
  -\frac{1}{|\mathcal{B}|} \sum_{i \in \mathcal{B}}
  \log \frac{S_i^+}{S_i^+ + S_i^-}.
\end{equation}
A smaller $\tau$ produces sharper similarity distributions that enforce stronger
separation between domains. By minimizing \eqref{eq:dscl} within the
sharpness-aware optimization loop, the method prevents the feature boundaries of
different steganographic algorithms from collapsing into the cover domain, even
when the model is subjected to adversarial weight perturbations.

\subsection{Adaptive Domain Gap Modulation}
\label{subsec:adgm}

In multi-domain steganalysis, different algorithms exhibit varying detection
difficulties, leading to imbalanced domain gaps. Standard optimizers naturally
prioritize easier domains with larger gradients, causing the model to converge to
saddle points with respect to harder domains. To resolve this, we propose Adaptive
Domain Gap Modulation (ADGM), which dynamically adjusts the optimization focus
based on real-time domain separability.

Drawing inspiration from center-based feature learning~\cite{wen2016discriminative}
and online prototype clustering~\cite{caron2020unsupervised}, we maintain a running
centroid $\mathbf{c}_k$ for each domain $k$ using Exponential Moving Average (EMA):
\begin{equation}
\label{eq:ema}
\mathbf{c}_k^{(t)} = \mu \cdot \mathbf{c}_k^{(t-1)} + (1 - \mu) \cdot \bar{\mathbf{z}}_k^{(t)},
\end{equation}
where $\bar{\mathbf{z}}_k^{(t)}$ is the mean feature of domain $k$ in the current
batch and $\mu$ is the momentum coefficient. We separately track the cover domain
center $\mathbf{c}_{\text{cover}}$ and each steganographic domain center
$\mathbf{c}_k$ for $k \in \{1, \dots, S\}$.

Based on the tracked centers, we define the domain gap
$g_k = \|\mathbf{c}_k - \mathbf{c}_{\text{cover}}\|_2$ as the Euclidean distance
between the $k$-th steganographic domain center and the cover center. A smaller
$g_k$ implies a harder detection task. To automatically focus on hard domains, we
compute adaptive weights via softmax over the negated gaps:
\begin{equation}
\label{eq:adaptive_weight}
w_k = \frac{\exp(-g_k / \tau_g)}{\sum_{j=1}^{S} \exp(-g_j / \tau_g)},
\end{equation}
where $\tau_g = \text{std}(\{g_1, \dots, g_S\}) + \xi$ is an adaptive temperature
computed from the current gap distribution. This formulation assigns larger weights
to domains with smaller gaps, enabling the optimizer to automatically discover and
prioritize hard-to-separate domains.

Using these adaptive weights, we define the ADGM loss as:
\begin{equation}
\label{eq:adgm}
\mathcal{L}_{\text{ADGM}} = 1 - \frac{\sum_{k=1}^{S} w_k \cdot g_k}{\max_{k}(g_k) + \xi},
\end{equation}
where $\xi$ is a small constant for numerical stability. The numerator computes a
weighted average of domain gaps emphasizing harder domains, while the denominator
normalizes by the maximum gap, making the loss self-normalized to $[0, 1)$. By
minimizing \eqref{eq:adgm}, the optimizer dynamically focuses on expanding the
feature distance of hard domains, acting as a geometric regularizer that prevents
the optimization trajectory from stagnating in saddle points.

\section{Experiments}
\label{sec:experiments}

\subsection{Experimental Settings}
\label{sec:exp_settings}

\textbf{Datasets and Evaluation.}
We construct a binary classification dataset comprising natural audio samples and
steganographic audio samples. The steganographic samples are generated using four
distinct algorithms: QIM, PMS, LSB, and AHCM, forming four separate domains with
equal sample sizes. The audio sources exhibit high diversity, covering different
genders, languages, and various content types. The datasets encompass five
embedding rates: 0.1, 0.2, 0.3, 0.4 and 0.5. Each embedding rate maintains
balanced positive and negative samples across all four domains. Training employs
mixed data from all four domains simultaneously, while evaluation is conducted
domain-specifically to assess cross-algorithm generalization capabilities.

\textbf{Baselines.}
To comprehensively evaluate the proposed DASM framework, we compare it with two
categories of state-of-the-art methods. First, we consider advanced deep learning
models for VoIP steganalysis: CCN, SS-QCCN, SFFN, KFEF, FS-MDP, LStegT, DVSF,
and DAEF-VS. Second, we include prominent sharpness-aware optimizers: standard SAM
and ERM; and domain-inspired variants DISAM, DGSAM, FSAM, and SAGM. These
baselines are implemented with identical network architectures and training
configurations as DASM to ensure a fair comparison.

\textbf{Implementation Details.}
All experiments are conducted on NVIDIA vGPU-32GB. Training
configuration includes batch size 128, learning rate 0.001, and Adam optimizer for
baseline models with a Transformer backbone for SAM variants. For DASM, we set
the perturbation radius $\rho = 0.03$, contrastive temperature $\tau = 0.1$, and
EMA momentum $\mu = 0.9$ for domain center tracking. The three loss components in
\eqref{eq:total_loss} are self-normalized and directly summed without additional
balancing coefficients. All models train for 100 epochs with early stopping based
on validation loss, using detection accuracy and AUC as evaluation metrics.

\subsection{Main Results}
\label{subsec:main_results}

We evaluate DASM against the comprehensive baseline suite on balanced test sets
across all four domains. Table~\ref{tab:main_results} summarizes the detection
accuracy at embedding rate 0.5, with all results averaged over three independent
runs.

\begin{table*}[t]
\caption{Detection Accuracy (\%) at ER=0.5. DASM Consistently Outperforms
Specialized Steganalysis Methods and Advanced Optimization Strategies,
Particularly in PMS and QIM. Best Results in Bold.}
\label{tab:main_results}
\centering
\begin{tabular}{l|cccc|c}
\toprule
Algorithm (Reference) & QIM & PMS & LSB & AHCM & Average \\
\midrule
CCN~\cite{CCN}
  & 90.79\,{\scriptsize\textcolor{green!60!black}{$_{+8.47}$}}
  & 50.36\,{\scriptsize\textcolor{red}{$_{-22.58}$}}
  & 50.01\,{\scriptsize\textcolor{red}{$_{-33.04}$}}
  & 50.57\,{\scriptsize\textcolor{red}{$_{-39.32}$}}
  & 60.43\,{\scriptsize\textcolor{red}{$_{-21.62}$}} \\
SS-QCCN~\cite{SS-QCCN}
  & 91.66\,{\scriptsize\textcolor{green!60!black}{$_{+9.34}$}}
  & 51.06\,{\scriptsize\textcolor{red}{$_{-21.88}$}}
  & 47.96\,{\scriptsize\textcolor{red}{$_{-35.09}$}}
  & 50.22\,{\scriptsize\textcolor{red}{$_{-39.67}$}}
  & 60.23\,{\scriptsize\textcolor{red}{$_{-21.82}$}} \\
LStegT~\cite{LStegT}
  & 82.17\,{\scriptsize\textcolor{red}{$_{-0.15}$}}
  & 63.11\,{\scriptsize\textcolor{red}{$_{-9.83}$}}
  & 75.56\,{\scriptsize\textcolor{red}{$_{-7.49}$}}
  & 76.29\,{\scriptsize\textcolor{red}{$_{-13.60}$}}
  & 74.28\,{\scriptsize\textcolor{red}{$_{-7.77}$}} \\
KFEF~\cite{KFEF}
  & 90.97\,{\scriptsize\textcolor{green!60!black}{$_{+8.65}$}}
  & 71.89\,{\scriptsize\textcolor{red}{$_{-1.05}$}}
  & 85.93\,{\scriptsize\textcolor{green!60!black}{$_{+2.88}$}}
  & 80.17\,{\scriptsize\textcolor{red}{$_{-9.72}$}}
  & 82.24\,{\scriptsize\textcolor{green!60!black}{$_{+0.19}$}} \\
SFFN~\cite{SFFN}
  & 69.20\,{\scriptsize\textcolor{red}{$_{-13.12}$}}
  & 62.93\,{\scriptsize\textcolor{red}{$_{-10.01}$}}
  & 77.93\,{\scriptsize\textcolor{red}{$_{-5.12}$}}
  & 91.48\,{\scriptsize\textcolor{green!60!black}{$_{+1.59}$}}
  & 75.39\,{\scriptsize\textcolor{red}{$_{-6.66}$}} \\
FS-MDP~\cite{FS-MDP}
  & 87.41\,{\scriptsize\textcolor{green!60!black}{$_{+5.09}$}}
  & 52.44\,{\scriptsize\textcolor{red}{$_{-20.50}$}}
  & 88.38\,{\scriptsize\textcolor{green!60!black}{$_{+5.33}$}}
  & 80.47\,{\scriptsize\textcolor{red}{$_{-9.42}$}}
  & 77.18\,{\scriptsize\textcolor{red}{$_{-4.87}$}} \\
DAEF-VS~\cite{DAEF-VS}
  & 89.91\,{\scriptsize\textcolor{green!60!black}{$_{+7.59}$}}
  & 73.31\,{\scriptsize\textcolor{green!60!black}{$_{+0.37}$}}
  & 89.68\,{\scriptsize\textcolor{green!60!black}{$_{+6.63}$}}
  & 89.24\,{\scriptsize\textcolor{red}{$_{-0.65}$}}
  & 85.54\,{\scriptsize\textcolor{green!60!black}{$_{+3.49}$}} \\
DVSF~\cite{DVSF}
  & 89.37\,{\scriptsize\textcolor{green!60!black}{$_{+7.05}$}}
  & 51.03\,{\scriptsize\textcolor{red}{$_{-21.91}$}}
  & 76.62\,{\scriptsize\textcolor{red}{$_{-6.43}$}}
  & 52.20\,{\scriptsize\textcolor{red}{$_{-37.69}$}}
  & 67.31\,{\scriptsize\textcolor{red}{$_{-14.74}$}} \\
\midrule
Transformer
  & 82.32 & 72.94 & 83.05 & 89.89 & 82.05 \\
+\,ERM
  & 88.18\,{\scriptsize\textcolor{green!60!black}{$_{+5.86}$}}
  & 70.14\,{\scriptsize\textcolor{red}{$_{-2.80}$}}
  & 92.49\,{\scriptsize\textcolor{green!60!black}{$_{+9.44}$}}
  & 93.72\,{\scriptsize\textcolor{green!60!black}{$_{+3.83}$}}
  & 86.13\,{\scriptsize\textcolor{green!60!black}{$_{+4.08}$}} \\
+\,SAM~\cite{foret2021sharpness}
  & 92.09\,{\scriptsize\textcolor{green!60!black}{$_{+9.77}$}}
  & 71.76\,{\scriptsize\textcolor{red}{$_{-1.18}$}}
  & 94.76\,{\scriptsize\textcolor{green!60!black}{$_{+11.71}$}}
  & 93.24\,{\scriptsize\textcolor{green!60!black}{$_{+3.35}$}}
  & 87.96\,{\scriptsize\textcolor{green!60!black}{$_{+5.91}$}} \\
+\,DISAM~\cite{DISAM}
  & 85.11\,{\scriptsize\textcolor{green!60!black}{$_{+2.79}$}}
  & 70.91\,{\scriptsize\textcolor{red}{$_{-2.03}$}}
  & 87.63\,{\scriptsize\textcolor{green!60!black}{$_{+4.58}$}}
  & 92.44\,{\scriptsize\textcolor{green!60!black}{$_{+2.55}$}}
  & 84.02\,{\scriptsize\textcolor{green!60!black}{$_{+1.97}$}} \\
+\,FSAM~\cite{zhang2024friendly}
  & 86.26\,{\scriptsize\textcolor{green!60!black}{$_{+3.94}$}}
  & 76.32\,{\scriptsize\textcolor{green!60!black}{$_{+3.38}$}}
  & 90.87\,{\scriptsize\textcolor{green!60!black}{$_{+7.82}$}}
  & 94.46\,{\scriptsize\textcolor{green!60!black}{$_{+4.57}$}}
  & 86.98\,{\scriptsize\textcolor{green!60!black}{$_{+4.93}$}} \\
+\,DGSAM~\cite{DGSAM}
  & 86.72\,{\scriptsize\textcolor{green!60!black}{$_{+4.40}$}}
  & 68.56\,{\scriptsize\textcolor{red}{$_{-4.38}$}}
  & 93.51\,{\scriptsize\textcolor{green!60!black}{$_{+10.46}$}}
  & 91.15\,{\scriptsize\textcolor{green!60!black}{$_{+1.26}$}}
  & 84.99\,{\scriptsize\textcolor{green!60!black}{$_{+2.94}$}} \\
+\,SAGM~\cite{wang2023sharpness}
  & 89.13\,{\scriptsize\textcolor{green!60!black}{$_{+6.81}$}}
  & 71.29\,{\scriptsize\textcolor{red}{$_{-1.65}$}}
  & 95.27\,{\scriptsize\textcolor{green!60!black}{$_{+12.22}$}}
  & 94.66\,{\scriptsize\textcolor{green!60!black}{$_{+4.77}$}}
  & 87.59\,{\scriptsize\textcolor{green!60!black}{$_{+5.54}$}} \\
\textbf{+\,DASM (Ours)}
  & \textbf{93.72}\,{\scriptsize\textcolor{green!60!black}{$_\textbf{+11.40}$}}
  & \textbf{82.38}\,{\scriptsize\textcolor{green!60!black}{$_\textbf{+9.44}$}}
  & \textbf{96.68}\,{\scriptsize\textcolor{green!60!black}{$_\textbf{+13.63}$}}
  & \textbf{99.44}\,{\scriptsize\textcolor{green!60!black}{$_\textbf{+9.55}$}}
  & \textbf{93.06}\,{\scriptsize\textcolor{green!60!black}{$_\textbf{+11.01}$}} \\
\bottomrule
\end{tabular}
\end{table*}

The experimental results reveal several key findings.

\textbf{Superiority over State-of-the-Art.} DASM achieves the highest average
accuracy of 93.06\%, surpassing the best domain-specific method DAEF-VS by 7.52\%.
The improvement is most pronounced in the challenging PMS domain, where DASM
attains 82.38\% compared to 73.31\% for DAEF-VS. This 9.07\% gain demonstrates
that our domain-aware optimization effectively captures subtle steganographic
artifacts that specialized architectures fail to detect.

\textbf{Effectiveness against Saddle Points.} On the Transformer backbone,
standard SAM improves over ERM from 86.13\% to 87.96\%, validating the benefit of
flat minima. However, domain-aware variants DISAM and DGSAM fail to consistently
outperform SAM because their isotropic perturbations cannot address the polarized
domain gaps revealed by our PAD analysis in Appendix~\ref{app:domain_gap_analysis}.
DASM overcomes this limitation through adaptive gap modulation, achieving a 5.10\%
improvement over SAM. The loss landscape visualizations in Appendix~\ref{app:loss_landscape} further corroborate that DASM transforms
the rugged optimization terrain into smooth, flat basins across all domains.

\begin{table}[t]
\caption{Performance Comparison Across Embedding Rates (\%). Best Results in Bold.}
\label{tab:er_robustness_transposed}
\centering
\begin{tabular}{c|c|ccccc}
\toprule
Opt. & ER & QIM & PMS & LSB & AHCM & Avg. \\
\midrule
\multirow{5}{*}{Adam}
& 0.1 & 52.79 & 51.73 & 90.88 & 95.88 & 72.82 \\
& 0.2 & 66.52 & 58.34 & 85.67 & 92.30 & 75.71 \\
& 0.3 & 72.52 & 63.15 & 82.65 & 90.84 & 77.29 \\
& 0.4 & 78.65 & 68.32 & 83.44 & 89.26 & 79.92 \\
& 0.5 & 82.32 & 72.94 & 83.05 & 89.89 & 82.05 \\
\midrule
\multirow{5}{*}{SAM}
& 0.1 & 60.64 & 52.31 & 93.63 & 99.29 & 76.47 \\
& 0.2 & 72.97 & 58.31 & 92.60 & 95.52 & 79.85 \\
& 0.3 & 80.57 & 65.02 & 92.20 & 98.36 & 84.04 \\
& 0.4 & 85.45 & 72.91 & 91.53 & 96.25 & 86.53 \\
& 0.5 & 92.09 & 71.76 & 94.76 & 93.24 & 87.96 \\
\midrule
\multirow{5}{*}{\textbf{DASM}}
& 0.1 & \textbf{63.21} & \textbf{52.08} & \textbf{97.08} & \textbf{99.84} & \textbf{78.05} \\
& 0.2 & \textbf{75.68} & \textbf{57.12} & \textbf{96.29} & \textbf{98.30} & \textbf{81.85} \\
& 0.3 & \textbf{84.18} & \textbf{65.73} & \textbf{95.72} & \textbf{99.48} & \textbf{86.28} \\
& 0.4 & \textbf{89.40} & \textbf{74.25} & \textbf{95.88} & \textbf{99.15} & \textbf{89.67} \\
& 0.5 & \textbf{93.72} & \textbf{82.38} & \textbf{96.68} & \textbf{99.44} & \textbf{93.06} \\
\bottomrule
\end{tabular}
\end{table}

\textbf{Robustness across Embedding Rates.}
Table~\ref{tab:er_robustness_transposed} presents performance across embedding
rates from 0.1 to 0.5, with
trends visualized in Appendix~\ref{app:er_dynamics}. As
ER decreases, steganographic signals weaken critically. Adam degrades to
near-random accuracy of 54.63\% at ER=0.1, while SAM improves only marginally
to 63.80\%. DASM maintains 78.05\% average accuracy at ER=0.1, representing a
14.25\% gain over SAM. The advantage is most significant in the PMS domain, where
DASM outperforms SAM by 11.48\% at ER=0.3. This robustness stems from the
adaptive weighting mechanism that dynamically prioritizes hard domains even when
their feature gaps collapse toward zero, as quantified by the PAD analysis in Appendix~\ref{app:domain_gap_analysis} showing PMS gap compression from 1.064 to
0.328 across ERs.

\begin{figure}[t]
    \centering
    \subfloat[Adam]{\label{fig:tsne_adam}%
      \includegraphics[width=0.46\columnwidth]{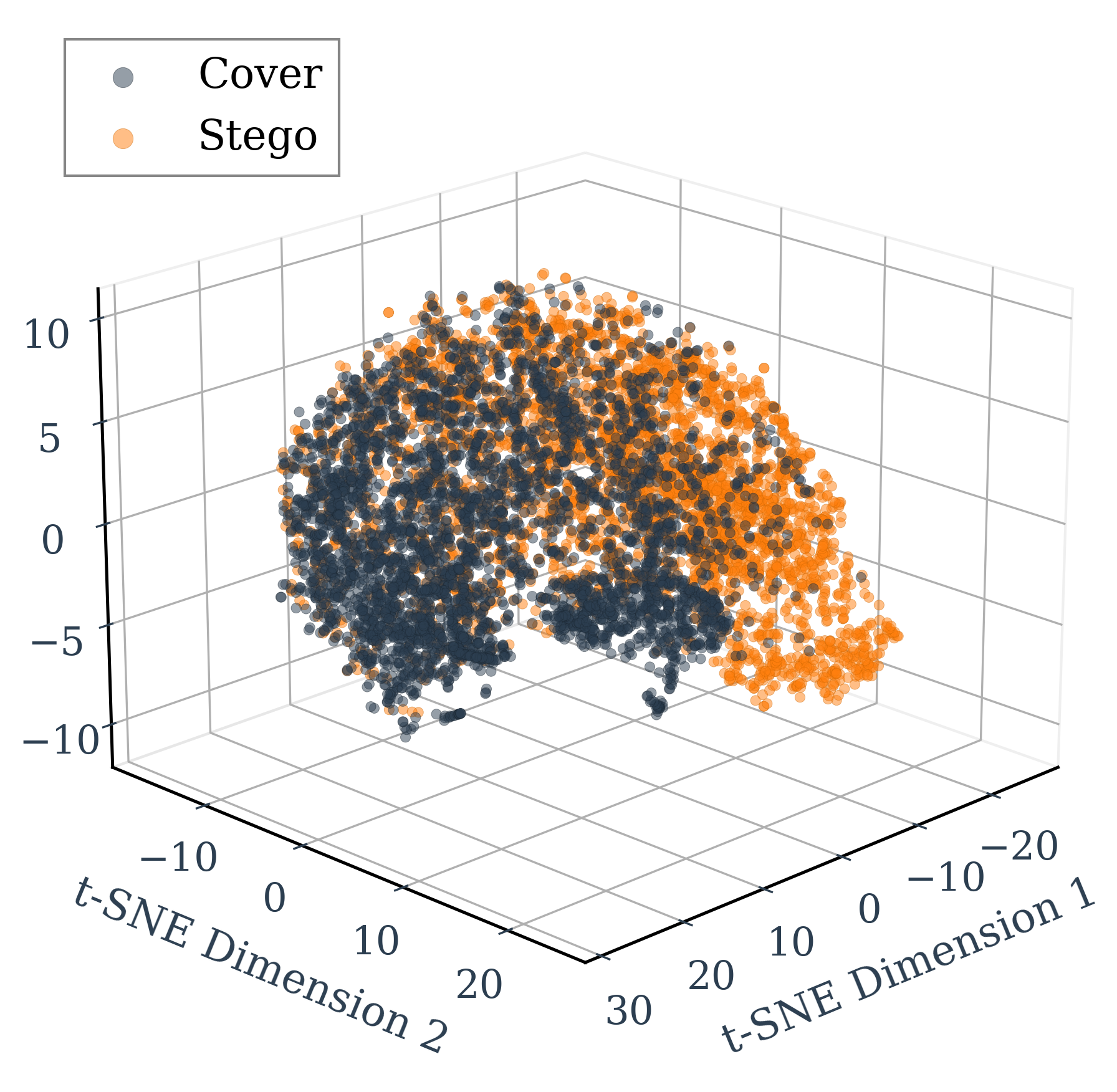}}%
    \hfill
    \subfloat[SAM]{\label{fig:tsne_sam}%
      \includegraphics[width=0.46\columnwidth]{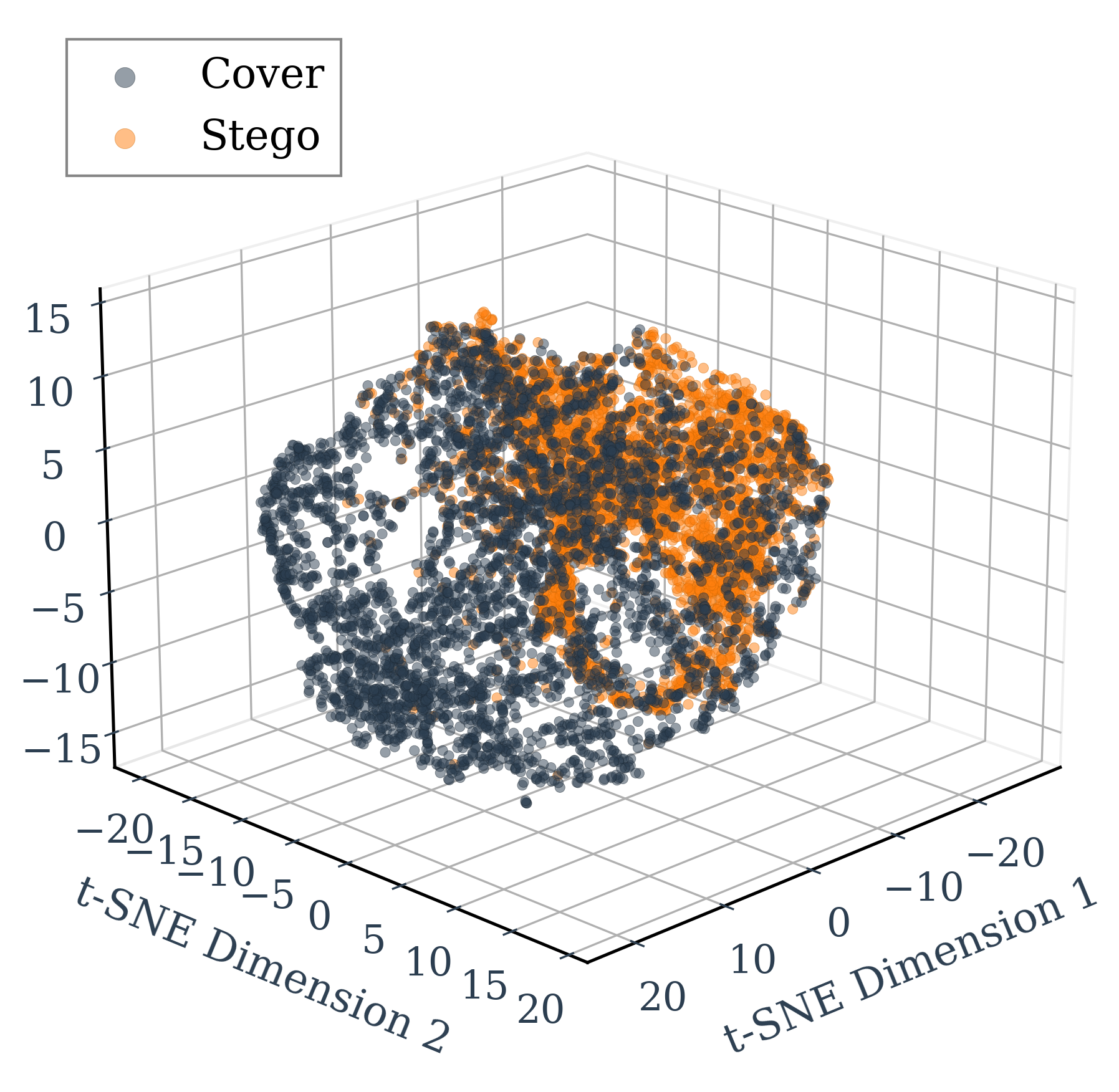}}%

    \vskip 0.08in

    \subfloat[DAEF-VS]{\label{fig:tsne_daefvs}%
      \includegraphics[width=0.46\columnwidth]{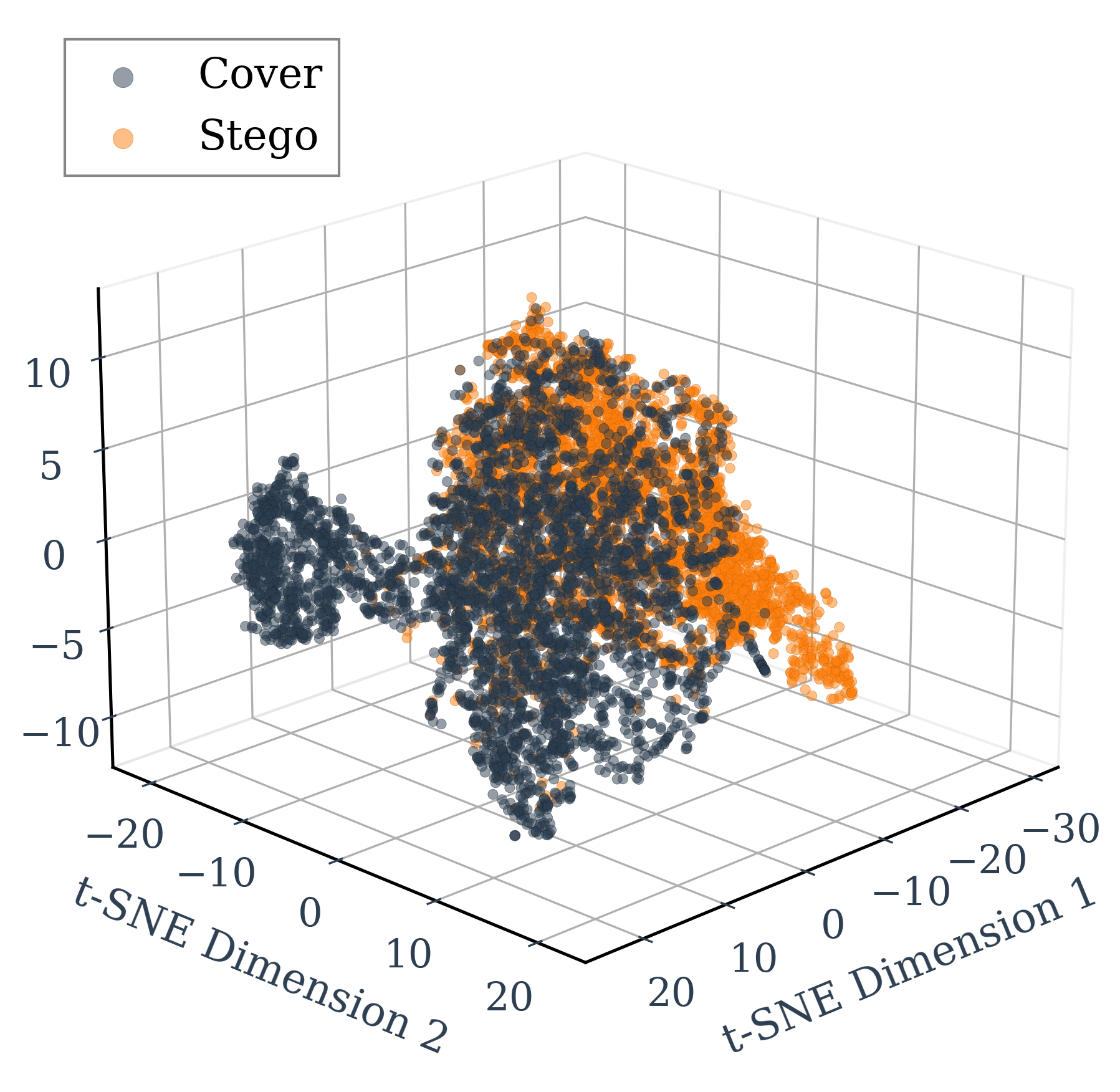}}%
    \hfill
    \subfloat[DASM]{\label{fig:tsne_dasm}%
      \includegraphics[width=0.46\columnwidth]{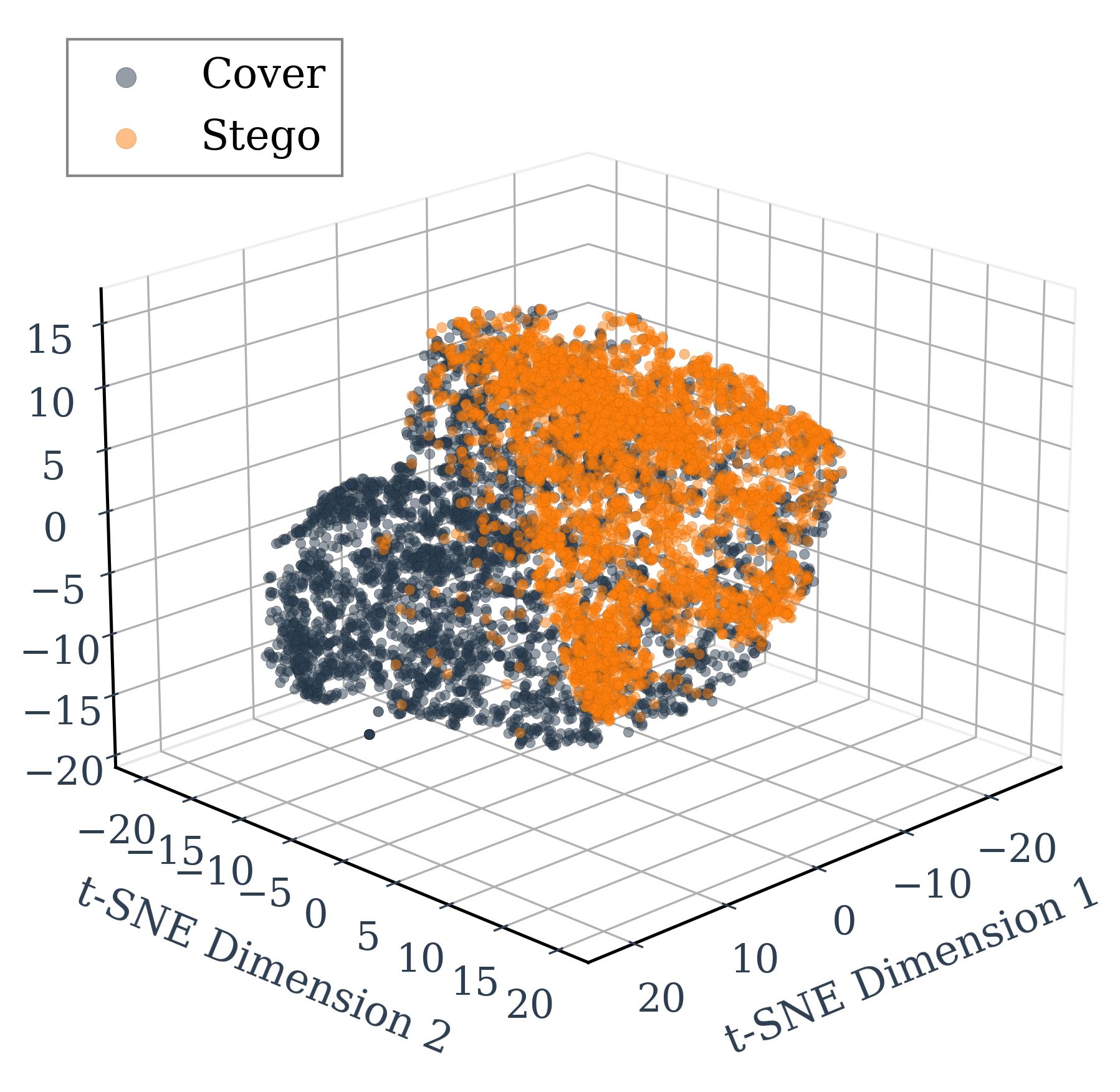}}%

    \caption{3-D t-SNE visualization of feature distributions. DASM achieves the
    clearest separation between Cover and Stego samples, while baseline methods
    exhibit significant overlap. Per-domain visualizations are provided in
    Appendix~\ref{app:tsne_detailed}.}
    \label{fig:tsne_main}
    \vspace{-1.5em}
\end{figure}

\textbf{Feature Space Geometry.}
Fig.~\ref{fig:tsne_main} provides qualitative validation through t-SNE
visualization. Adam produces severely entangled Cover-Stego features, explaining
its poor generalization. SAM provides marginal improvement but maintains
substantial overlap in the PMS and QIM regions. DAEF-VS, despite its specialized
architecture, fails to achieve complete separation. In contrast, DASM produces
well-structured feature clusters with clear Cover-Stego boundaries across all
domains. This geometric organization directly correlates with the detection
accuracy in Table~\ref{tab:main_results} and validates that the domain-supervised
contrastive loss effectively prevents feature collapse. Per-domain visualizations
in Appendix~\ref{app:tsne_detailed} confirm consistent separation even for the
challenging PMS domain.

\subsection{Ablation Study}
\label{subsec:ablation}

We investigate the individual contributions of the core components in DASM:
Domain-Supervised Contrastive Learning ($\mathcal{L}_{\text{DSCL}}$) and Adaptive
Domain Gap Modulation ($\mathcal{L}_{\text{ADGM}}$). Taking the standard
Transformer-based classifier as the \textit{Baseline}, we evaluate three variants:
(1)~\textbf{DSCL Only}, which incorporates contrastive loss to enforce feature
separability; (2)~\textbf{ADGM Only}, which employs adaptive weighting to balance
optimization; and (3)~\textbf{DASM (Full)}, the complete framework.

As summarized in Table~\ref{tab:ablation}, each component provides a significant performance boost. A visual comparison of these improvements is provided in Appendix~\ref{app:visual_ablation}.

\begin{table}[t]
\caption{Ablation Study of DASM Components at ER=0.5 (\%). Best Results in Bold.}
\label{tab:ablation}
\centering
\begin{tabular}{l|cccc|c}
\toprule
Variant & QIM & PMS & LSB & AHCM & Avg. \\
\midrule
Adam (Baseline) & 82.32 & 72.94 & 83.05 & 89.89 & 82.05 \\
DSCL Only       & 90.63 & 77.57 & 90.81 & 97.49 & 89.13 \\
ADGM Only       & 91.53 & 80.62 & 91.75 & 98.83 & 90.68 \\
\midrule
\textbf{DASM (Full)} & \textbf{93.72} & \textbf{82.38} & \textbf{96.68} & \textbf{99.44} & \textbf{93.06} \\
\bottomrule
\end{tabular}
\end{table}

\textbf{Effectiveness of DSCL.} Incorporating $\mathcal{L}_{\text{DSCL}}$
improves average accuracy to 89.13\%. By maximizing inter-domain distance in the
feature space, DSCL prevents different steganographic algorithms from collapsing
into the cover domain, which is crucial for distinguishing algorithms with minute
embedding traces like QIM.

\textbf{Effectiveness of ADGM.} The ADGM module achieves 90.68\% average
accuracy. Notably, it provides a substantial gain in the PMS domain (from 72.94\%
to 80.62\%). This confirms that by dynamically up-weighting domains with smaller
gaps, the optimizer successfully escapes saddle points associated with hard-to-detect
domains.

\textbf{Synergy in DASM.} The full framework achieves the highest accuracy of
93.06\%. The synergy between DSCL and ADGM ensures that the model not only finds
a flat region in the loss landscape but also maintains structured and discriminative
feature boundaries across all domains.

\subsection{Hyperparameter Sensitivity Analysis}
\label{subsec:sensitivity}

We evaluate the sensitivity of DASM to two key parameters: the perturbation radius
$\rho$ and the contrastive temperature $\tau$. The corresponding line charts are illustrated in Appendix~\ref{app:sensitivity_trends}.

\textbf{Sensitivity to Perturbation Radius $\rho$.}
The radius $\rho$ controls the neighborhood size for sharpness estimation. We fix
$\tau=0.5$ and vary $\rho \in \{0.01, 0.03, 0.05, 0.08\}$. As shown in
Table~\ref{tab:rho_sensitivity}, performance is stable across a wide range.
$\rho=0.03$ yields the best average accuracy. A very small $\rho$ ($0.01$)
provides insufficient regularization to escape sharp minima, while a very large
$\rho$ ($0.08$) may over-smooth the landscape, potentially obscuring fine-grained
steganographic features.

\begin{table}[t]
\caption{Sensitivity Analysis of the Perturbation Radius $\rho$
(Fixed $\tau=0.5$, \%). }
\label{tab:rho_sensitivity}
\centering
\begin{tabular}{l|cccc|c}
\toprule
$\rho$ & QIM & PMS & LSB & AHCM & Avg. \\
\midrule
0.01 & 93.34 & 81.40 & 95.65 & 99.73 & 92.53 \\
\textbf{0.03} & 92.68 & \textbf{82.78} & 96.10 & \textbf{99.80} & \textbf{92.84} \\
0.05 & 93.16 & 82.14 & \textbf{96.25} & 99.65 & 92.80 \\
0.08 & \textbf{93.27} & 81.69 & 95.83 & 99.60 & 92.60 \\
\bottomrule
\end{tabular}
\end{table}

\textbf{Sensitivity to Contrastive Temperature $\tau$.}
Temperature $\tau$ regulates the similarity distribution in DSCL. We fix
$\rho=0.03$ and vary $\tau \in \{0.05, 0.1, 0.2, 0.5\}$. As shown in
Table~\ref{tab:tau_sensitivity}, $\tau=0.1$ is optimal. Lower values of $\tau$
enforce stricter domain separation, which benefits domains with high task
difficulty. However, an excessively small $\tau$ ($0.05$) can lead to optimization
instability by over-focusing on hard-negative samples.

\begin{table}[t]
\caption{Sensitivity Analysis of the Contrastive Temperature $\tau$
(Fixed $\rho=0.03$, \%).}
\label{tab:tau_sensitivity}
\centering
\begin{tabular}{l|cccc|c}
\toprule
$\tau$ & QIM & PMS & LSB & AHCM & Avg. \\
\midrule
0.05 & 93.14 & 81.95 & \textbf{96.86} & 99.59 & 92.89 \\
\textbf{0.10} & \textbf{93.72} & 82.38 & 96.68 & 99.44 & \textbf{93.06} \\
0.20 & 93.41 & 82.04 & 96.61 & 99.41 & 92.87 \\
0.50 & 92.68 & \textbf{82.78} & 96.10 & \textbf{99.80} & 92.84 \\
\bottomrule
\end{tabular}
\end{table}

\subsection{Sharpness Analysis}
\label{sec:sharpness_main}

To empirically corroborate the theoretical premise that DASM enhances
generalization by locating flatter minima, we conducted a quantitative analysis of
the loss landscape geometry using zeroth-order sharpness metrics with a
perturbation radius $\rho=0.05$. As detailed in Appendix~\ref{app:sharpness_details},
DASM achieves an exceptionally low mean sharpness of 0.252, representing a radical
reduction in curvature compared to both the Adam baseline of 2.334 and standard
SAM of 1.056. This finding provides a geometric justification for the superior
generalization performance reported in Table~\ref{tab:main_results}, confirming
that our method successfully converges to a wide, stable basin where the model is
insensitive to parameter perturbations.

Crucially, this flatness is not achieved by sacrificing performance on challenging
domains. In the PMS domain, where baseline methods often get trapped in sharp,
unstable minima as evidenced by Adam's sharpness of 2.272, DASM effectively
smooths the landscape to a sharpness of 0.371. This result directly validates the
efficacy of our adaptive domain gap modulation strategy. By dynamically
up-weighting harder domains during the perturbation step, DASM forces the optimizer
to escape the sharp valleys associated with subtle steganographic artifacts,
thereby resolving the high detection error rates typically observed in the PMS
domain.

Furthermore, DASM demonstrates superior optimization stability compared to prior
domain generalization approaches. While competitors often exhibit high variance in
sharpness across domains—such as DGSAM with a standard deviation of 1.296,
indicating an imbalance where easier domains are over-optimized at the expense of
harder ones—DASM maintains a negligible standard deviation of 0.080. This
uniformity confirms that our domain-aware mechanism ensures a consistently flat
landscape across all distributional shifts, preventing the formation of sharp
directions in any specific domain and guaranteeing robust test-time performance.

\subsection{Computational Overhead}
\label{subsec:efficiency}

Table~\ref{tab:efficiency} compares computational costs across optimization
strategies. SAM incurs approximately $2\times$ the training time of Adam due to
its two-step optimization. DASM introduces only 1.0\% overhead over SAM, arising
from the domain-supervised contrastive loss with $\mathcal{O}(B^2)$ pairwise
similarity calculations, maintaining $K$ domain feature centers via EMA, and the
adaptive gap modulation. The memory overhead is negligible as the contrastive loss
operates on already-extracted features. At 370.0\,ms per batch compared to SAM's
366.4\,ms, DASM demonstrates that domain-aware components introduce minimal
computational cost while providing substantial generalization gains.

\begin{table}[t]
\caption{Computational Complexity and Empirical Results. $P$: Model Parameters;
$K$: Domains; $D$: Feature Dimension; $B$: Batch Size. Empirical Measurements
on RTX~4090 with Batch Size~128.}
\label{tab:efficiency}
\centering
\renewcommand{\arraystretch}{1.1}
\begin{tabular}{@{}llll@{}}
\toprule
\multicolumn{4}{@{}l@{}}{\textit{(a) Theoretical Complexity}} \\
\midrule
Opt. & Fwd/Bwd & Time & Space \\
\midrule
Adam & 1/1 & $\mathcal{O}(P)$          & $\mathcal{O}(P)$ \\
SAM  & 2/2 & $\mathcal{O}(2P)$         & $\mathcal{O}(2P)$ \\
DASM & 2/2 & $\mathcal{O}(2P{+}B^{2})$ & $\mathcal{O}(2P{+}KD)$ \\
\midrule
\multicolumn{4}{@{}l@{}}{\textit{(b) Empirical Results}} \\
\midrule
Opt. & Mem.\ (GiB) & ms/batch & Rel.\ Time \\
\midrule
Adam & 11.3 & $182.0\pm0.1$ & $1.00\times$ \\
SAM  & 11.3 & $366.4\pm1.0$ & $2.01\times$ \\
DASM & 11.3 & $370.0\pm4.5$ & $2.03\times$ \\
\bottomrule
\end{tabular}
\end{table}

\section{Discussion and Future Work}
\label{sec:limitations}

While DASM establishes a new benchmark, two aspects merit future exploration.
First, the two-step optimization introduces computational overhead during training.
As quantified in Table~\ref{tab:efficiency}, DASM
incurs only 1.0\% overhead over SAM while providing substantial generalization
gains; developing efficient approximation techniques remains valuable. Second, the
current supervised setting requires domain labels; extending DASM to
semi-supervised or open-set scenarios would broaden its applicability. Future work
will investigate theoretical convergence guarantees and explore defending against
unknown steganographic algorithms via domain-invariant feature generalization.

\section{Conclusion}
\label{sec:conclusion}

In this paper, we identified that the generalization bottleneck in multi-domain
voice stream steganalysis stems from the model converging to saddle points, a
phenomenon induced by the minute and imbalanced gaps between steganographic
distributions. To address this, we proposed Domain-Aware Sharpness Minimization
(DASM), a novel optimization framework that synergizes domain-supervised
contrastive learning with adaptive gap modulation. By explicitly reshaping the loss
landscape to widen minute feature separations and inherently prioritizing
hard-to-separate domains, DASM effectively guides the optimization trajectory out
of saddle points toward robust flat minima. Extensive experiments on datasets
containing QIM, PMS, LSB, and AHCM algorithms demonstrate that DASM achieves
superior detection accuracy and robustness, offering a resilient solution for
securing VoIP communications.



\bibliography{main_TIFS}
\bibliographystyle{IEEEtran}

\newpage
\appendix

\subsection{Empirical Analysis of Domain Discrepancies via Proxy A-Distance}
\label{app:domain_gap_analysis}

To empirically validate the challenges of minute and imbalanced domain gaps
articulated in the Introduction, and to elucidate why generic Sharpness-Aware
Minimization methods fail in multi-domain VoIP steganalysis, we conducted a
rigorous quantitative analysis using the Proxy A-Distance (PAD)~\cite{PAD}. PAD is
a metric defined as $d_A = 2(1 - 2\epsilon)$, where $\epsilon$ represents the
generalization error of a domain classifier trained to distinguish between two
domains. A PAD value of 0 implies the domains are indistinguishable, while a value
of 2 indicates perfect separability.

\begin{figure*}[t]
    \centering
    \subfloat[ER = 0.1]{\label{fig:gap_er01}%
      \includegraphics[width=0.32\textwidth]{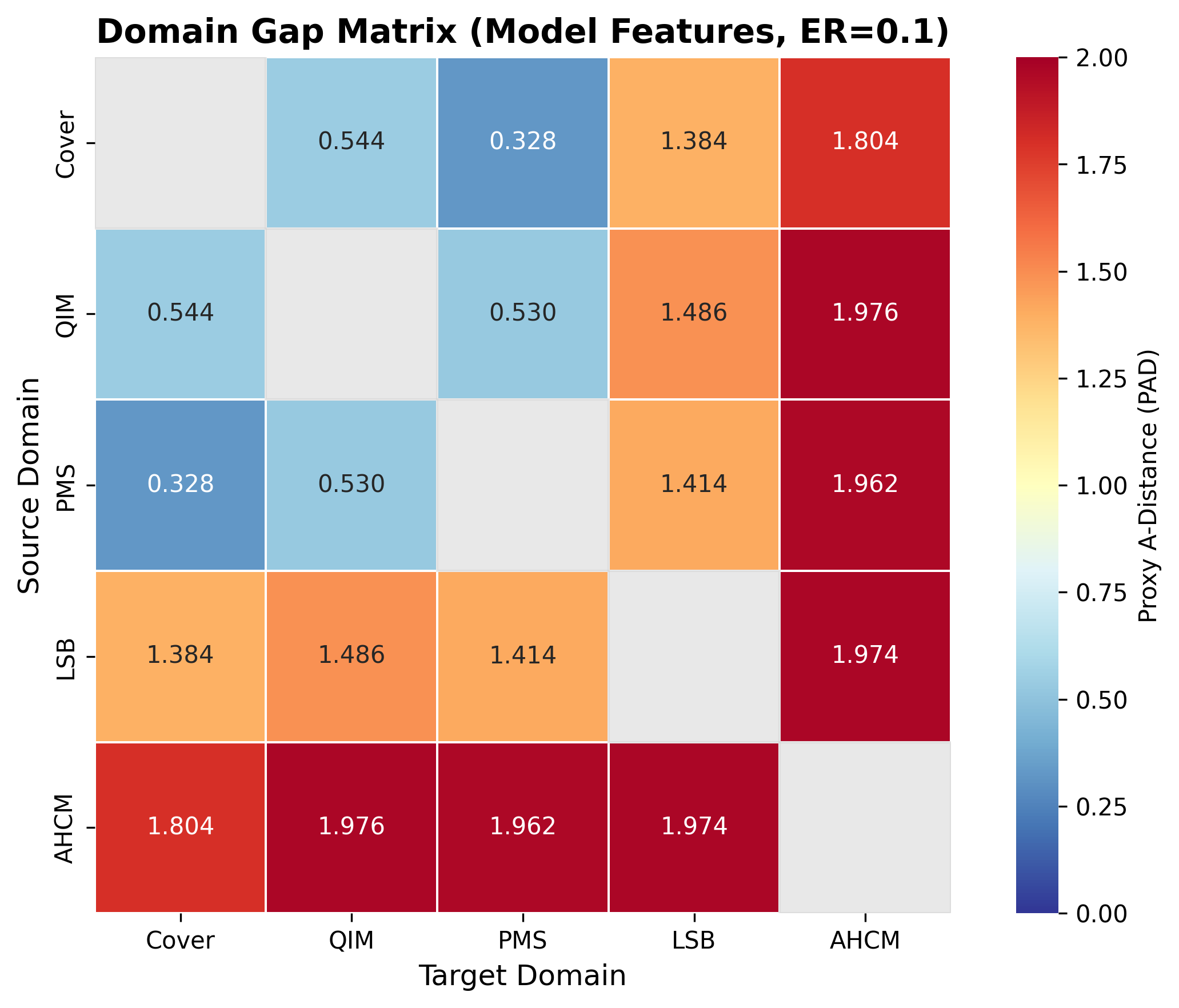}}%
    \hfill
    \subfloat[ER = 0.3]{\label{fig:gap_er03}%
      \includegraphics[width=0.32\textwidth]{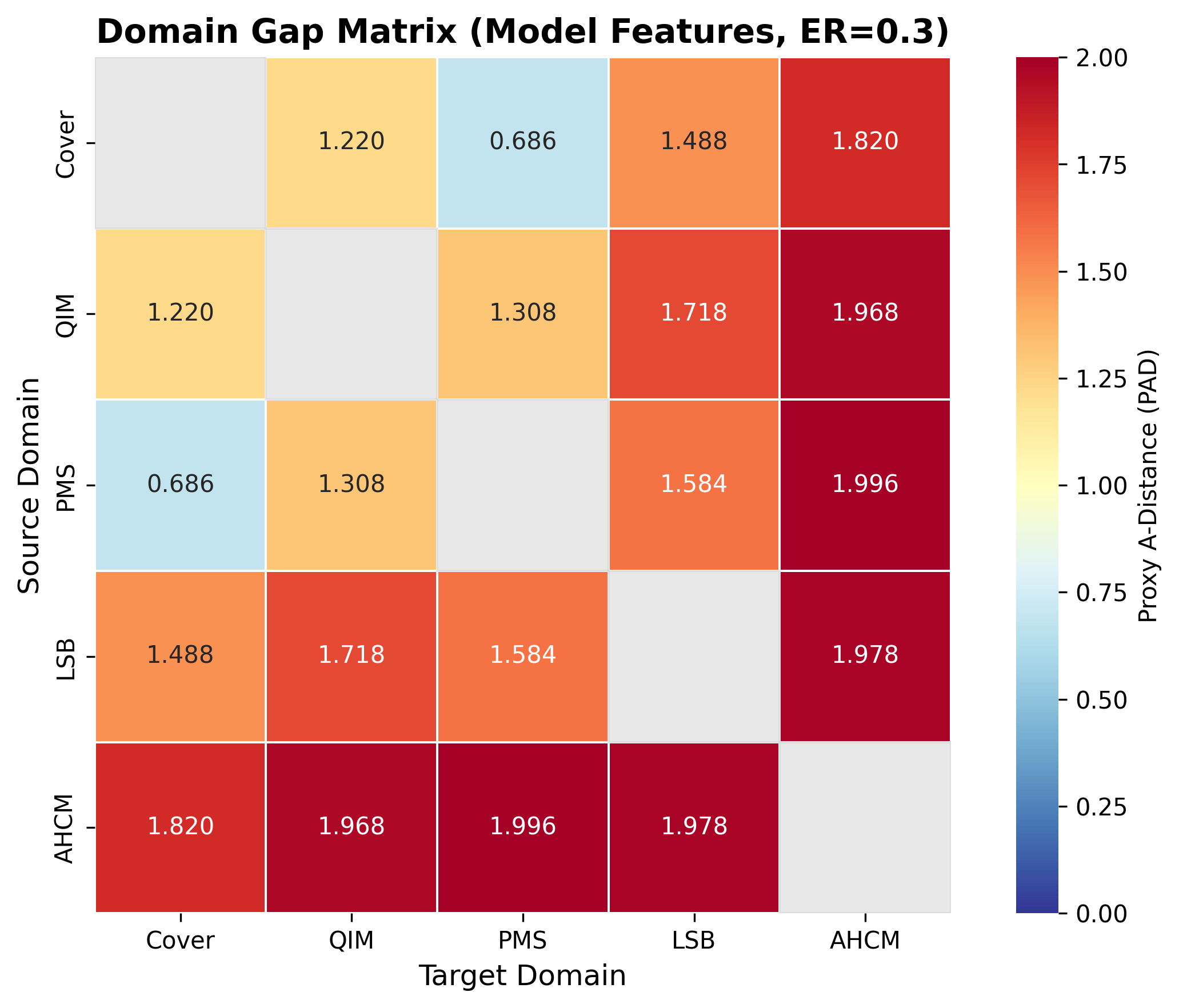}}%
    \hfill
    \subfloat[ER = 0.5]{\label{fig:gap_er05}%
      \includegraphics[width=0.32\textwidth]{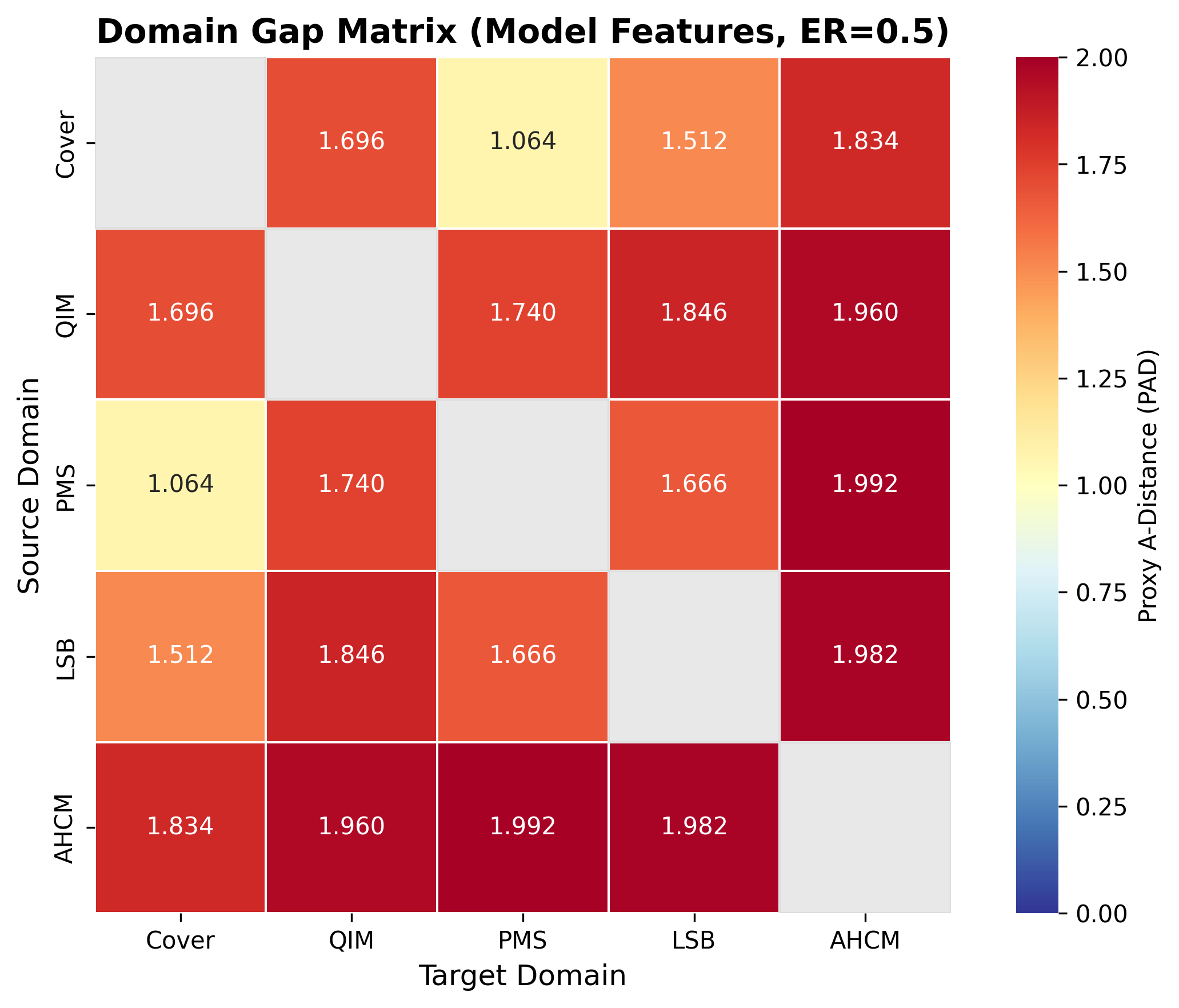}}%
    \caption{Pairwise PAD matrices across embedding rates. Lighter colors indicate
    smaller domain gaps, correlating with higher detection difficulty. PMS
    consistently exhibits the minimum gap, validating its status as the most
    challenging domain.}
    \label{fig:domain_gap_matrices}
\end{figure*}

Fig.~\ref{fig:domain_gap_matrices} visualizes the pairwise PAD matrices calculated
on model features under varying embedding rates. These heatmaps reveal three
critical findings regarding the optimization landscape.

First, the heatmaps demonstrate a strong correlation between small domain gaps and
high detection difficulty. Across all embedding rates, the PAD between PMS and
Cover is consistently the lowest, decreasing from 1.064 at ER=0.5 to 0.328 at
ER=0.1. This indicates that PMS features are statistically almost indistinguishable
from Cover, making it the most challenging steganographic algorithm. In optimization
terms, these minute gaps correspond to regions with vanishing gradients where
standard optimizers struggle to find descent directions, leading to the saddle
point convergence observed in our Hessian analysis.

Second, comparing Figs.~\ref{fig:gap_er01}--\ref{fig:gap_er05}, we observe
systemic compression of the feature space as embedding rate decreases. At ER=0.1,
the gaps for QIM and PMS collapse towards zero while AHCM remains relatively
distinct. This non-uniform compression creates a dynamic optimization landscape
where static perturbation strategies fail to adapt to varying signal-to-noise
ratios.

Third, the data reveals drastic polarization in domain discrepancies that renders
generic SAM suboptimal. At ER=0.1, the gap for AHCM is approximately 5.5 times
larger than that for PMS. Standard SAM applies isotropic perturbations based on
the average gradient, which is dominated by the easy AHCM domain with large
gradients. Consequently, the optimization trajectory is biased towards widening the
already sufficient margin of AHCM while neglecting the sharp minima associated with
PMS. This necessitates our Adaptive Domain Gap Modulation to dynamically up-weight
hard domains and Domain-Supervised Contrastive Learning to explicitly expand minute
gaps during perturbation.

\subsection{Loss Landscape Visualization}
\label{app:loss_landscape}

\begin{figure*}[t]
    \centering
    \subfloat[PMS (Adam)]{%
      \includegraphics[width=0.24\textwidth]{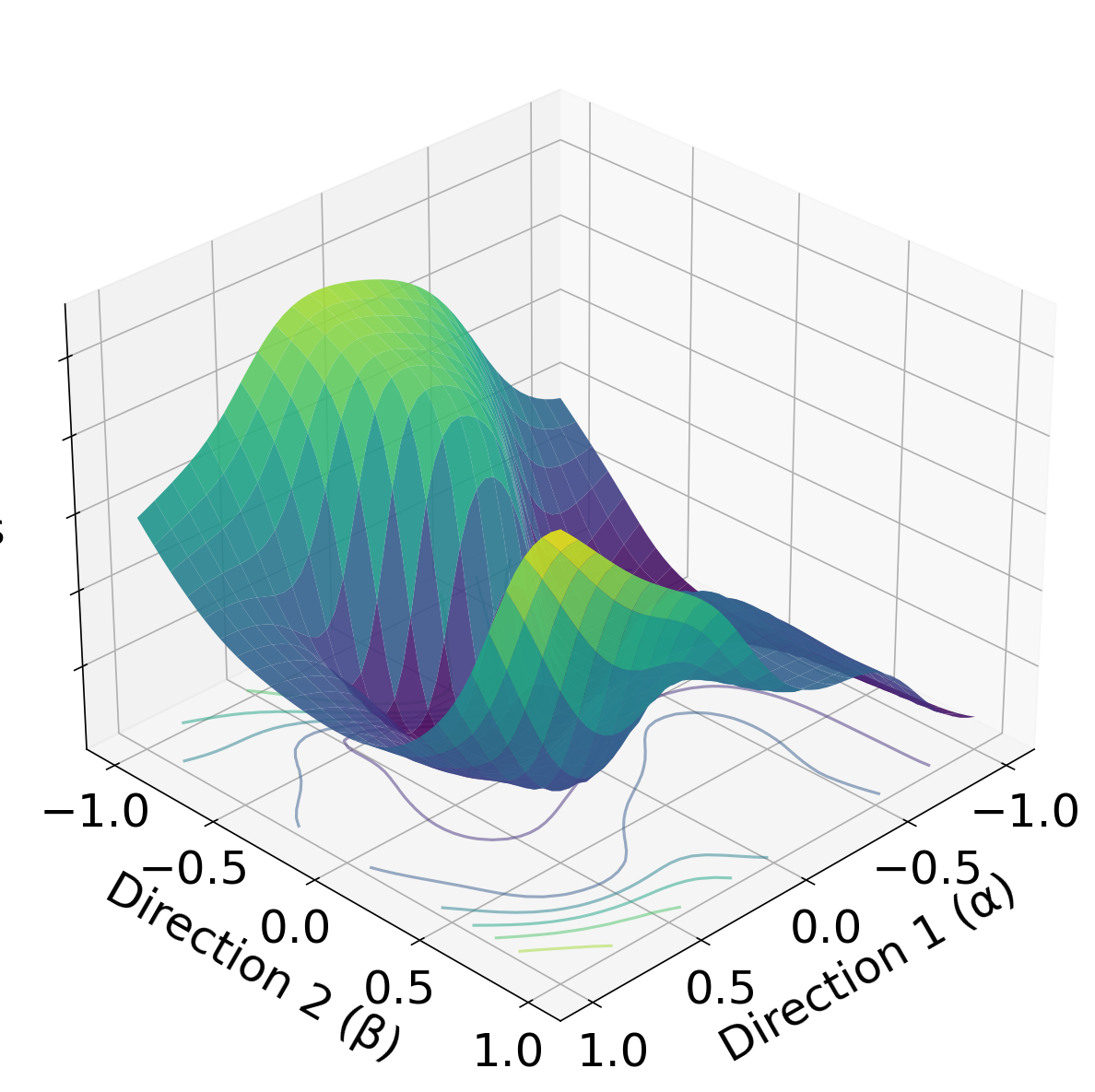}}%
    \hfill
    \subfloat[QIM (Adam)]{%
      \includegraphics[width=0.24\textwidth]{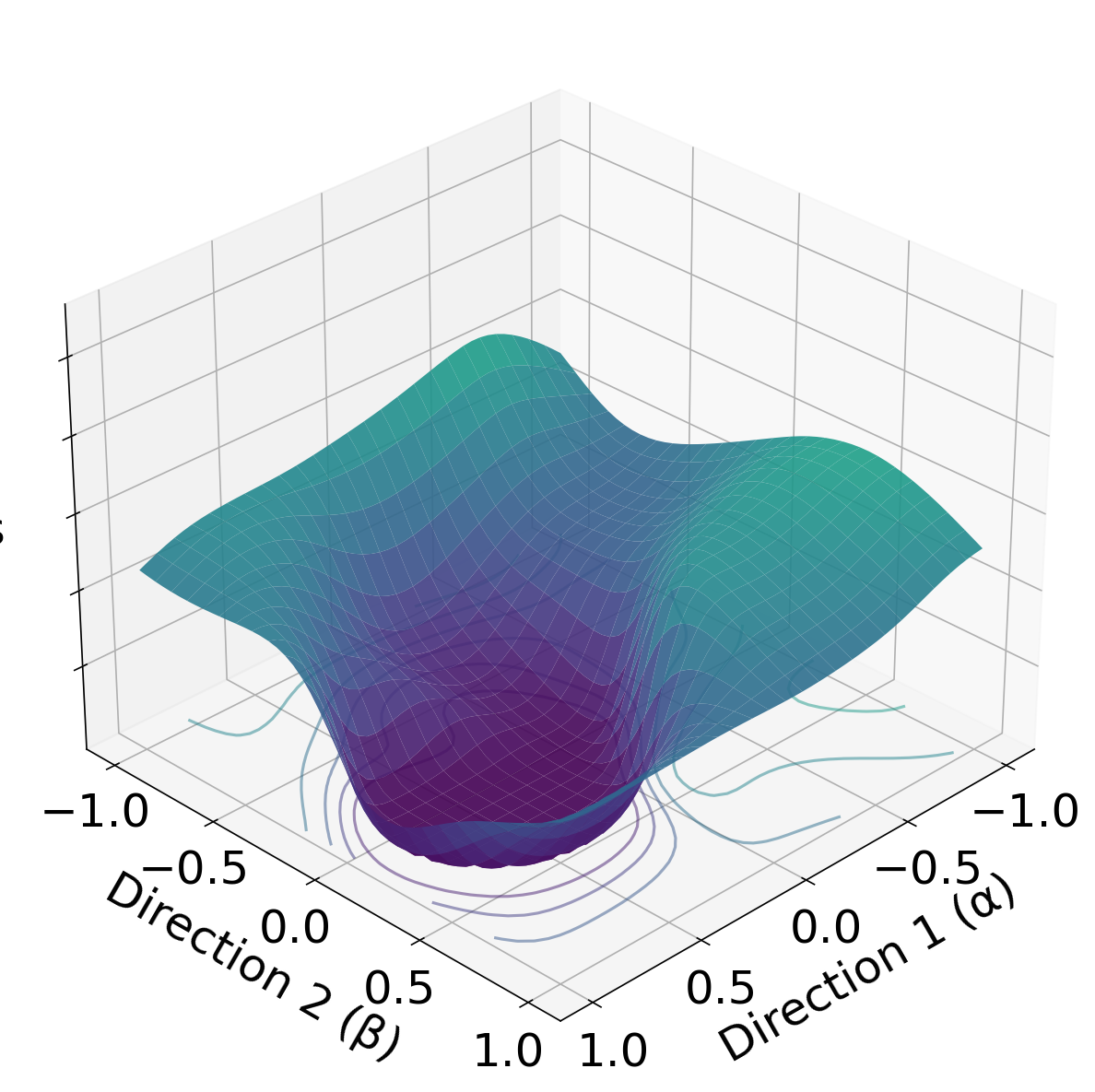}}%
    \hfill
    \subfloat[LSB (Adam)]{%
      \includegraphics[width=0.24\textwidth]{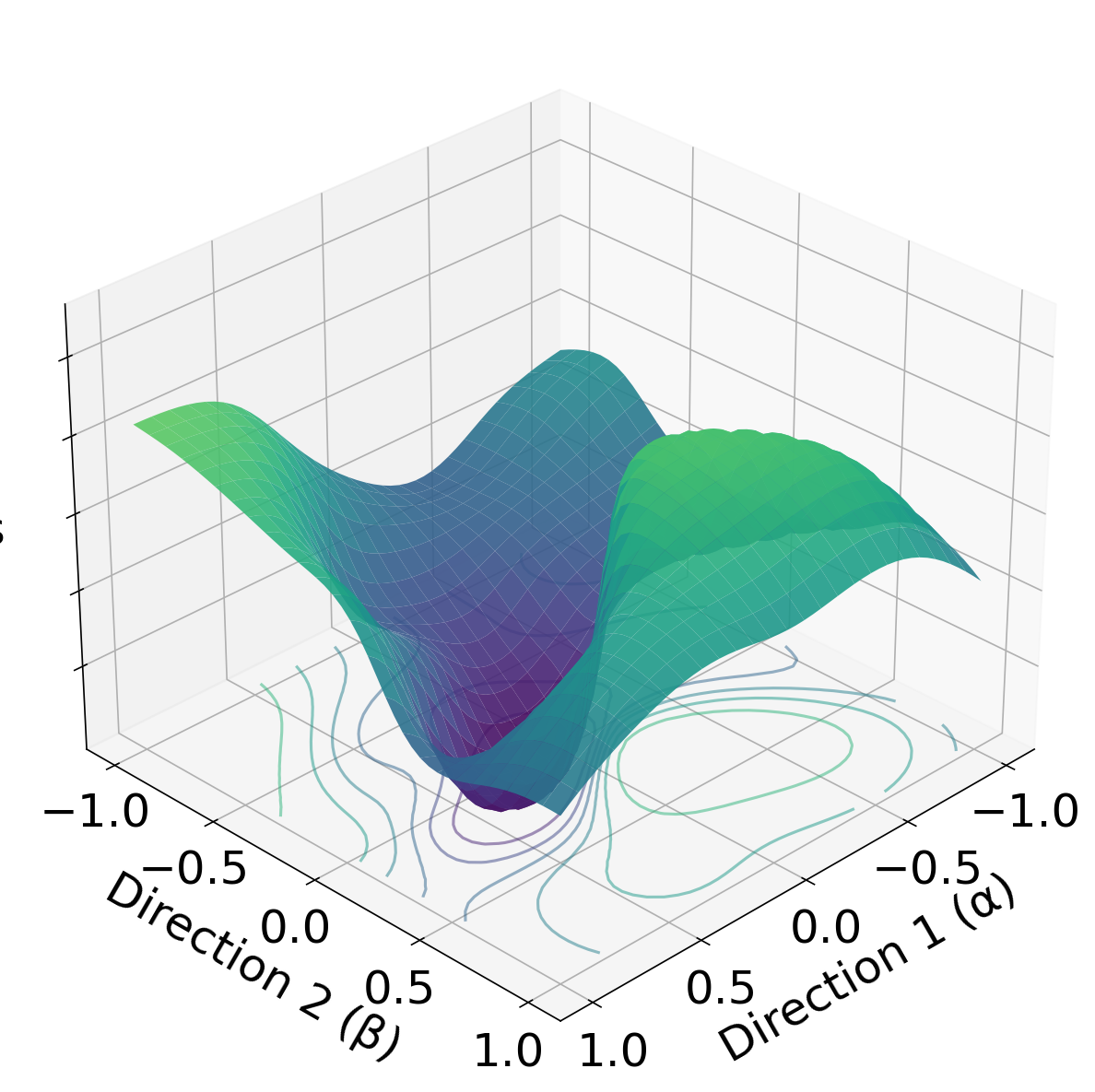}}%
    \hfill
    \subfloat[AHCM (Adam)]{%
      \includegraphics[width=0.24\textwidth]{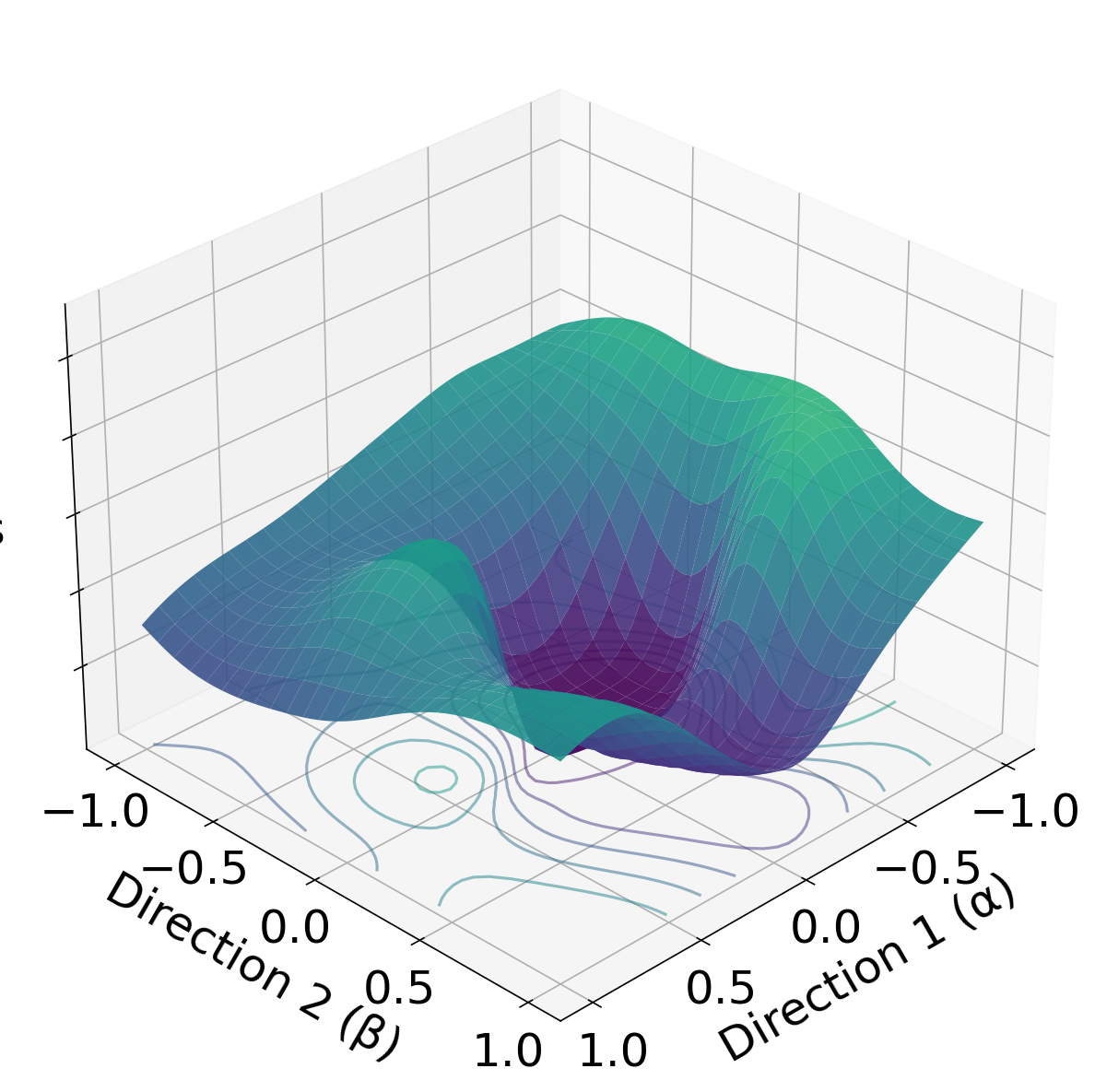}}%

    \vskip 0.06in

    \subfloat[PMS (DASM)]{%
      \includegraphics[width=0.24\textwidth]{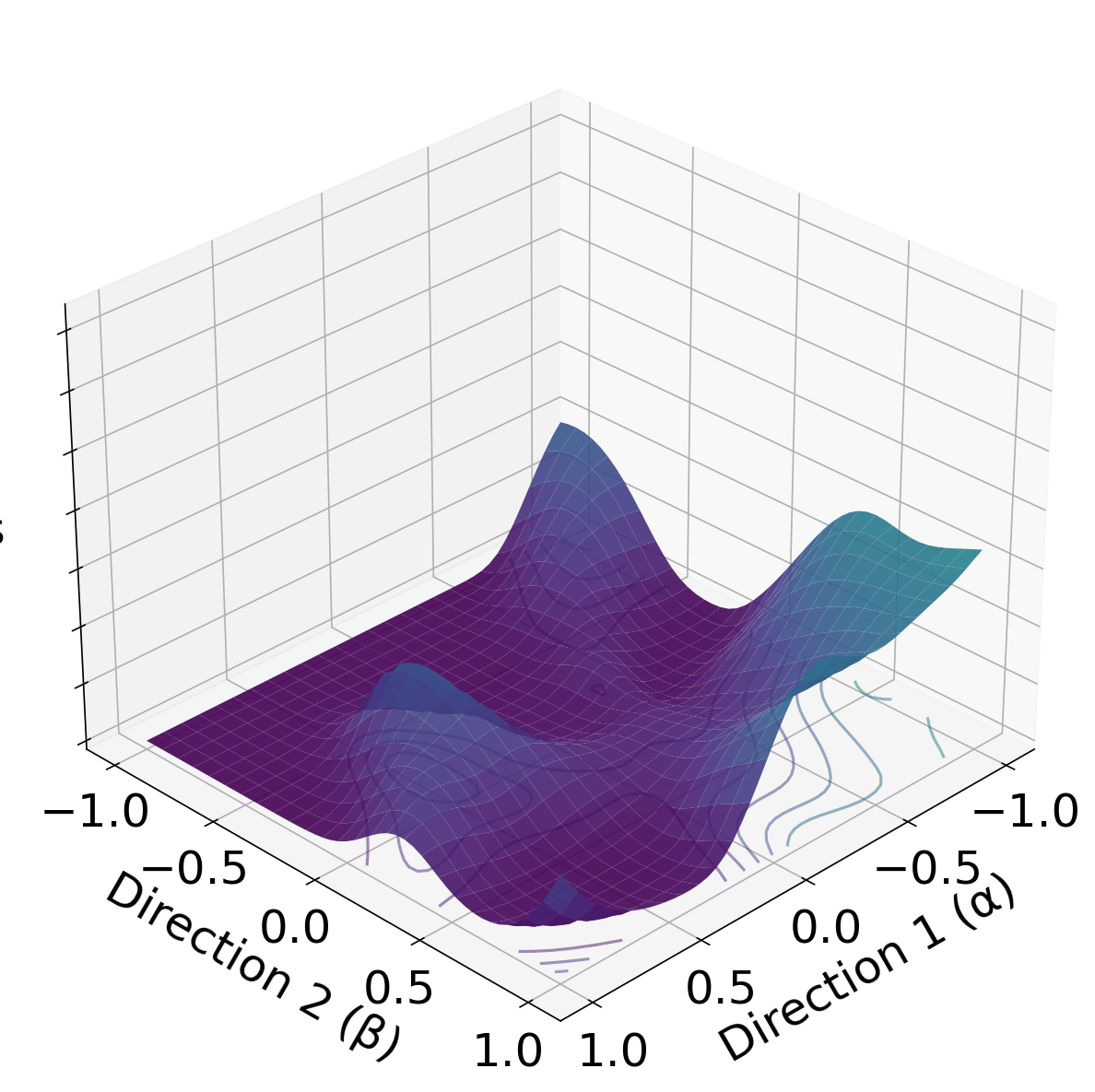}}%
    \hfill
    \subfloat[QIM (DASM)]{%
      \includegraphics[width=0.24\textwidth]{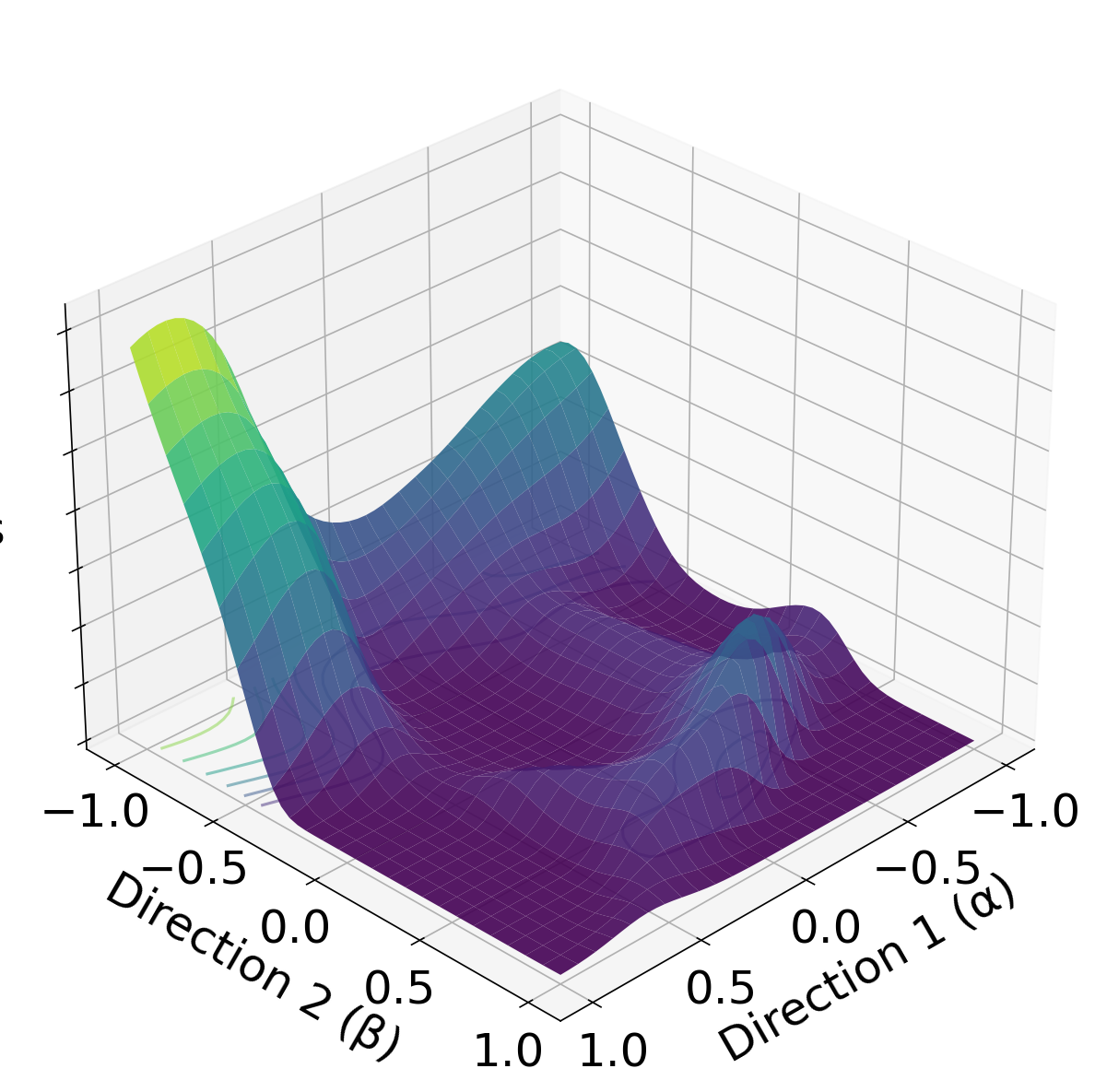}}%
    \hfill
    \subfloat[LSB (DASM)]{%
      \includegraphics[width=0.24\textwidth]{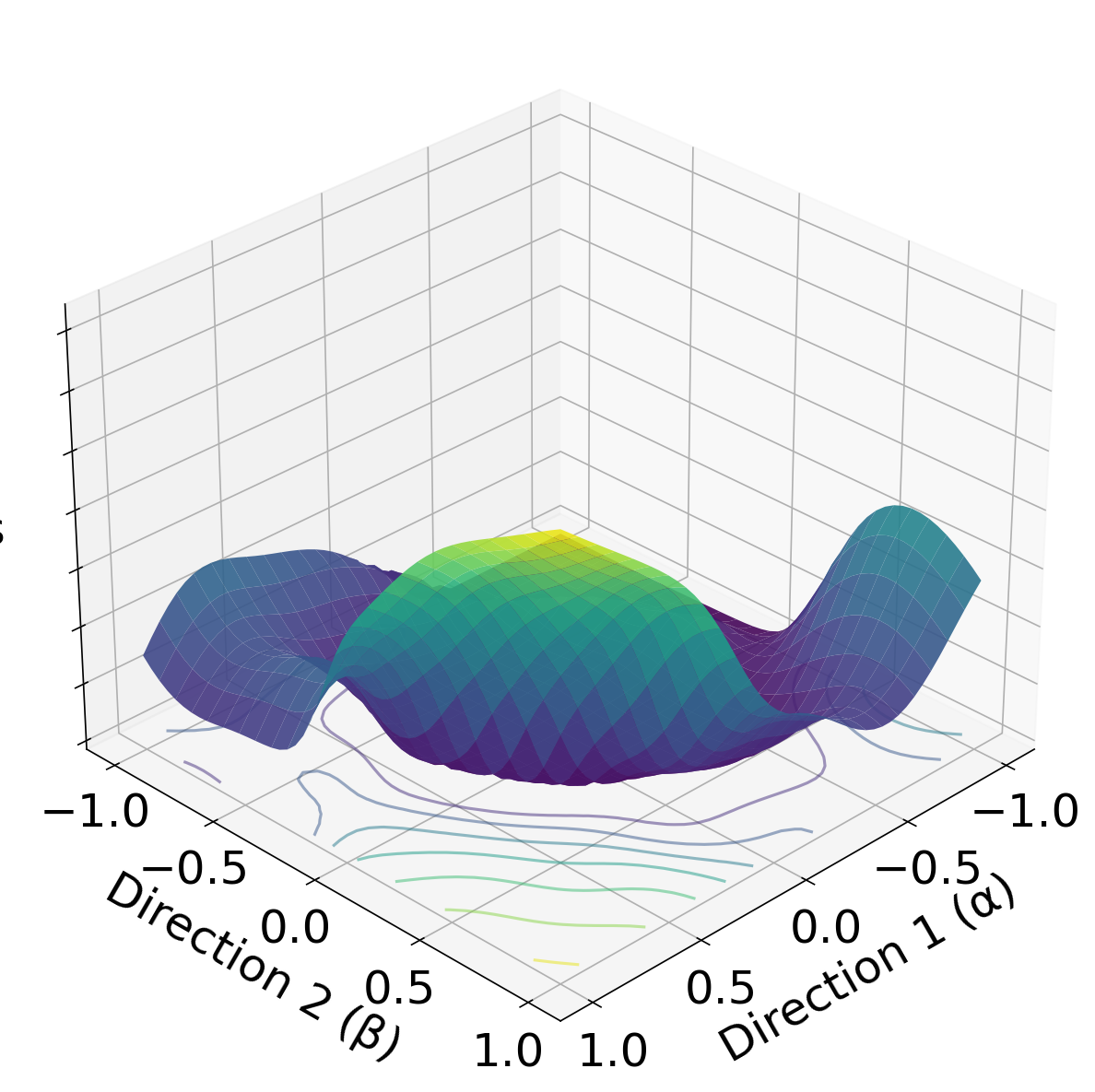}}%
    \hfill
    \subfloat[AHCM (DASM)]{%
      \includegraphics[width=0.24\textwidth]{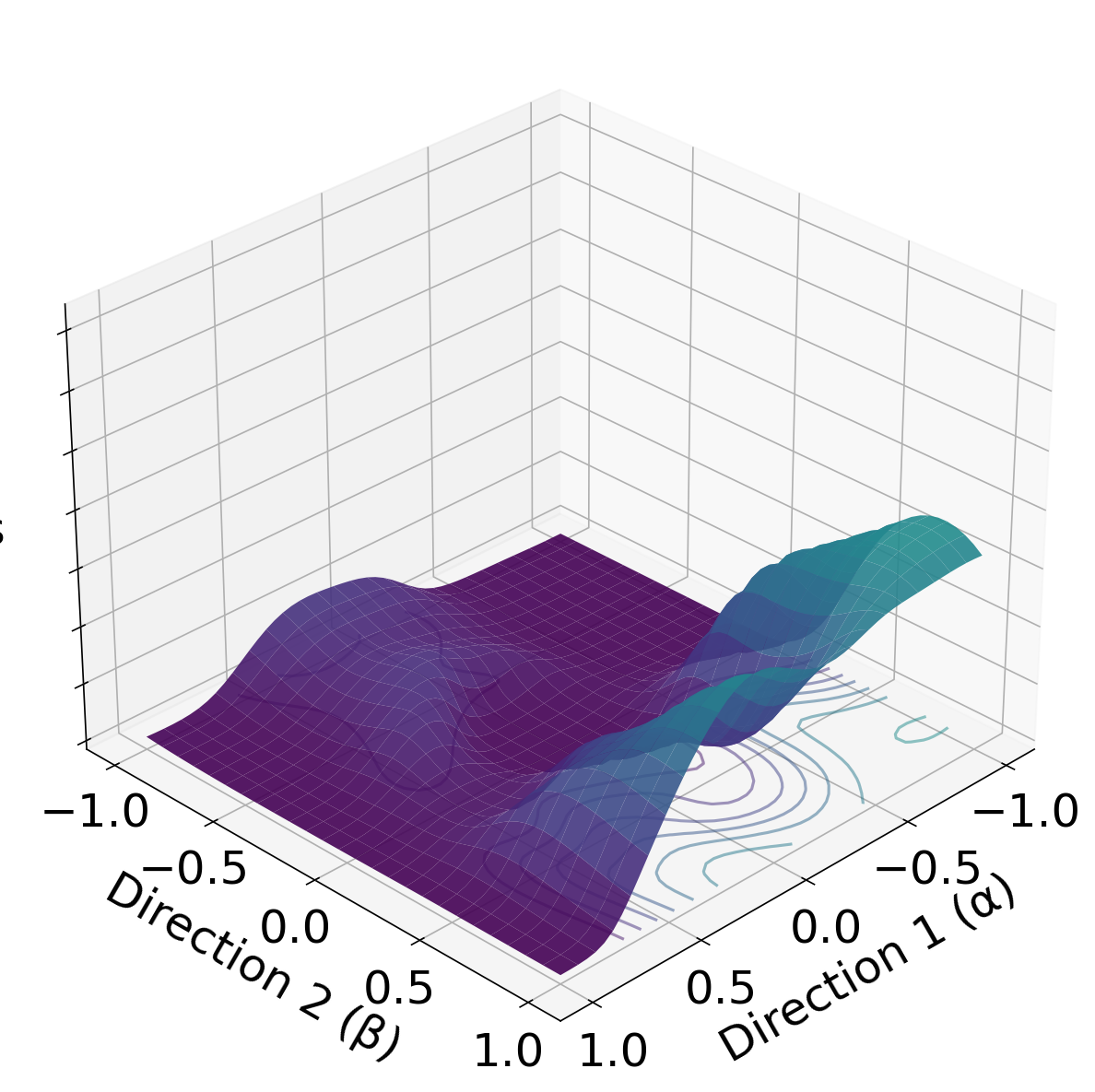}}%

    \caption{Loss landscape at ER=0.5. Top row: Adam converges to sharp minima
    with pronounced non-convexity in PMS and QIM. Bottom row: DASM smoothes the
    loss surfaces across all domains, widening basins and promoting convergence to
    flat minima.}
    \label{fig:loss_landscape_comparison}
\end{figure*}

We employ the visualization technique proposed by Li~\cite{li2018visualizing} to
plot the 3-D loss landscapes around converged model parameters at ER=0.5.
Fig.~\ref{fig:loss_landscape_comparison} presents the comparison between Adam and
DASM across all four steganographic domains.

For the baseline Adam optimizer, the loss surfaces exhibit pronounced non-convexity
characterized by numerous local maxima and saddle points. This geometric pathology
is particularly acute in PMS and QIM domains, correlating with their high detection
difficulty. Standard optimization struggles to navigate the intricate curvature
induced by minute steganographic perturbations, often converging to unstable sharp
minima.

The landscapes associated with DASM demonstrate significant smoothing across all
domains. By explicitly maximizing the loss within the perturbation neighborhood
while balancing domain gaps, DASM transforms the chaotic loss geometry into wide,
flat basins. Even for PMS, the rugged terrain is regularized into a smoother
surface. This geometric flatness facilitates escape from saddle points and underpins
the superior generalization observed in our main experiments.

\subsection{Performance Dynamics Across Embedding Rates}
\label{app:er_dynamics}

\begin{figure*}[t]
    \centering
    \subfloat[QIM]{\label{fig:er_qim}%
      \includegraphics[width=0.24\textwidth]{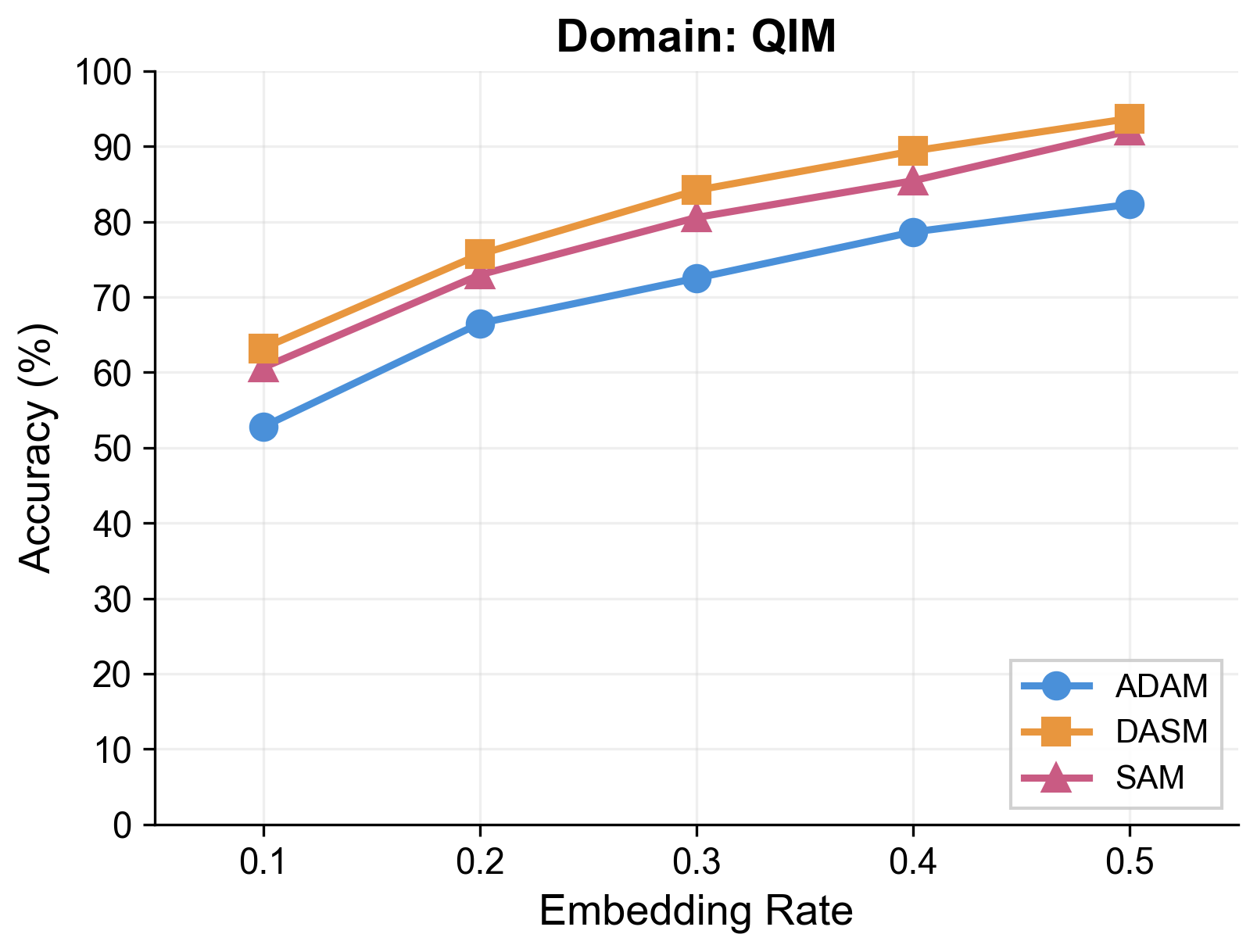}}%
    \hfill
    \subfloat[PMS]{\label{fig:er_pms}%
      \includegraphics[width=0.24\textwidth]{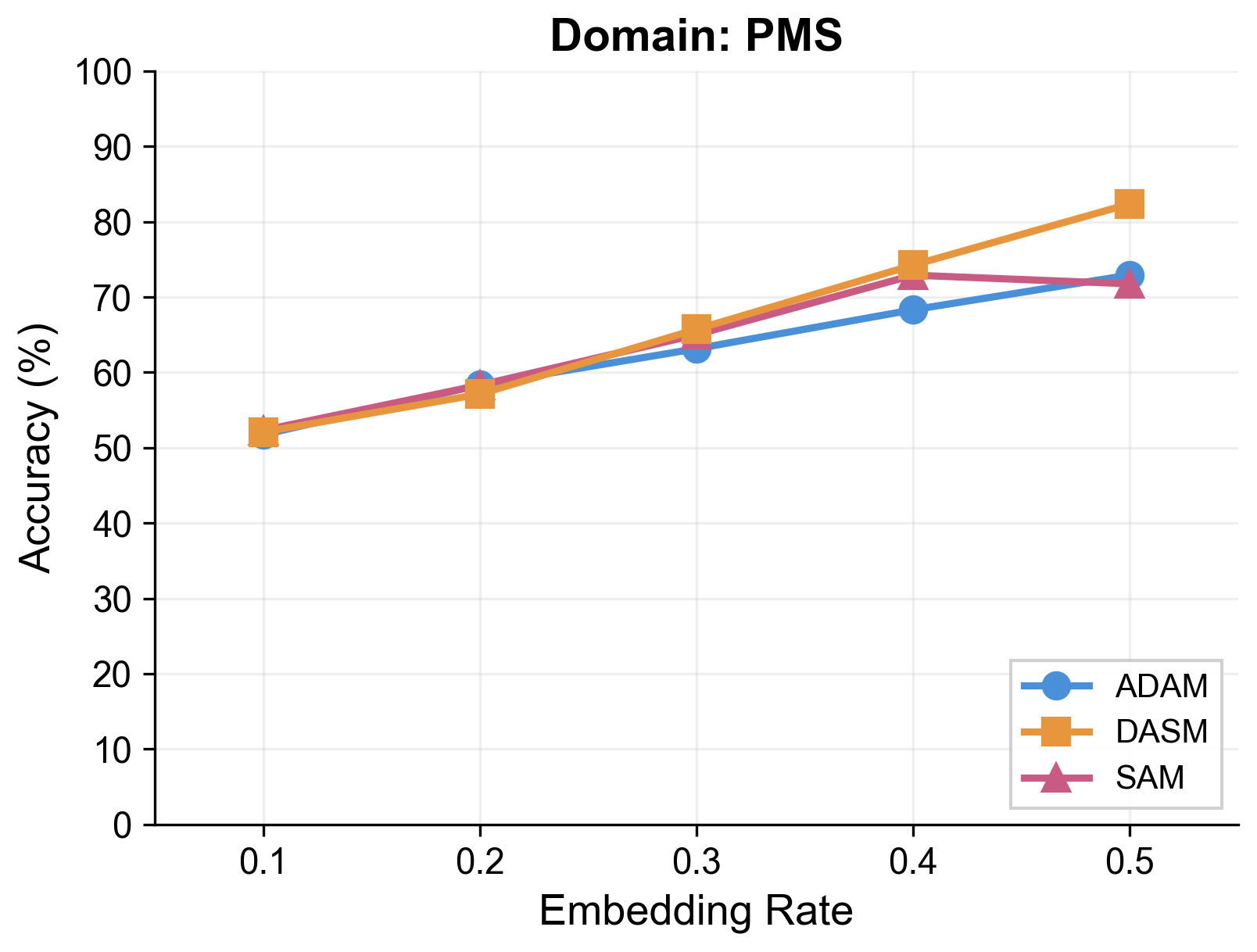}}%
    \hfill
    \subfloat[LSB]{\label{fig:er_lsb}%
      \includegraphics[width=0.24\textwidth]{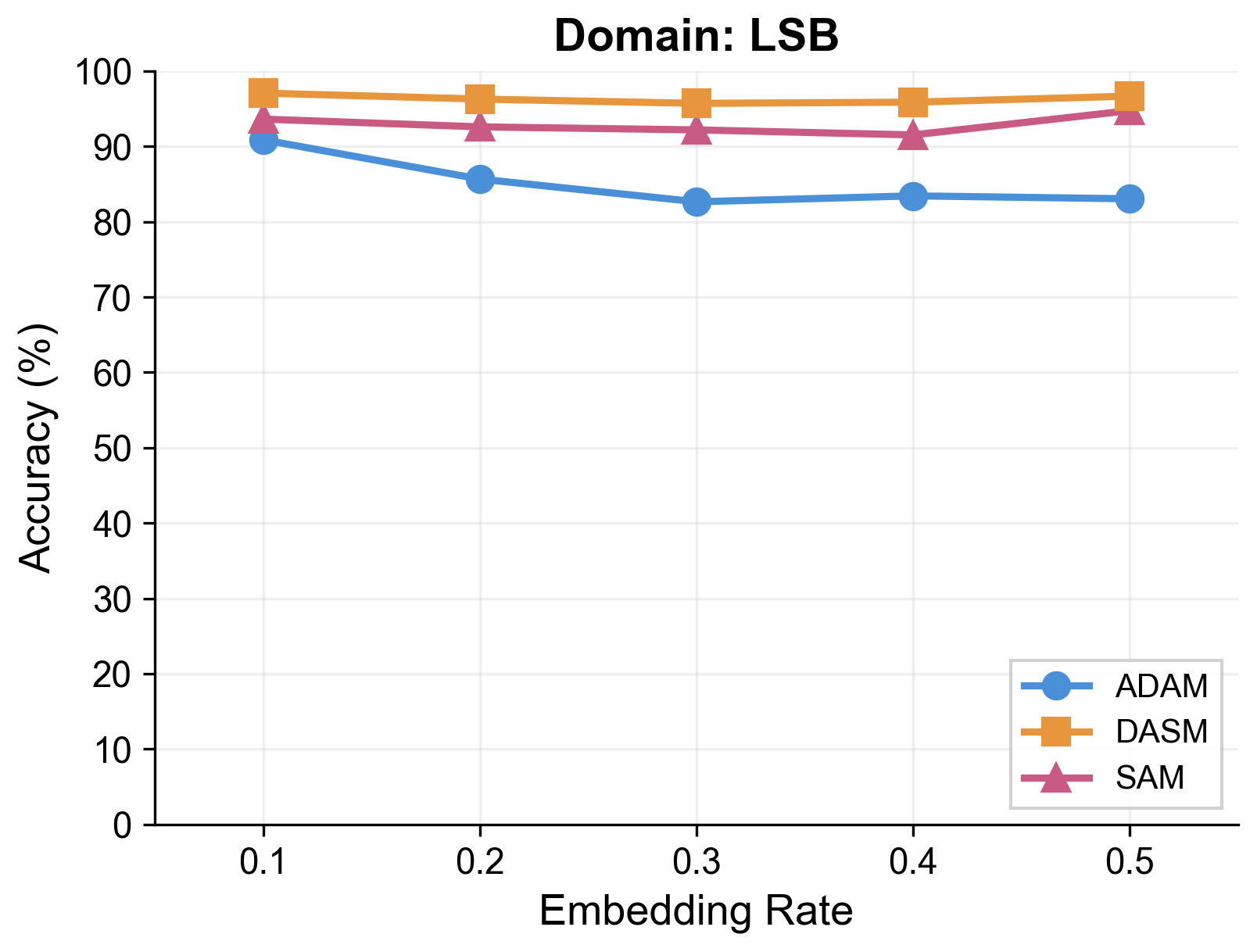}}%
    \hfill
    \subfloat[AHCM]{\label{fig:er_ahcm}%
      \includegraphics[width=0.24\textwidth]{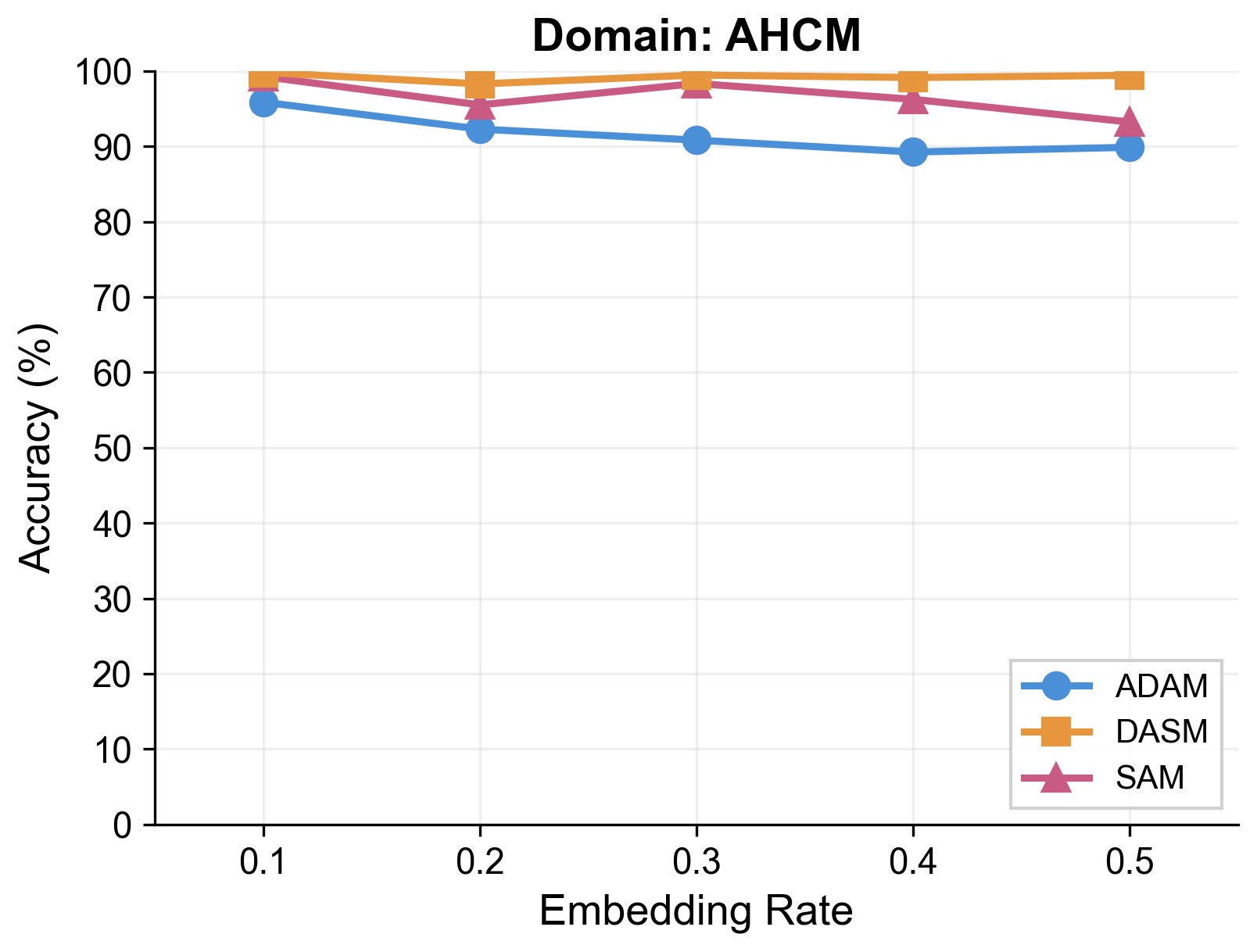}}%
    \caption{Detection accuracy across embedding rates from 0.1 to 0.5. DASM
    consistently outperforms Adam and SAM across all domains. The advantage is
    most pronounced in PMS and QIM at low embedding rates where steganographic
    signals are weakest.}
    \label{fig:er_dynamics}
\end{figure*}

Fig.~\ref{fig:er_dynamics} illustrates detection accuracy across five embedding
rates from 0.1 to 0.5 for Adam, SAM, and DASM.

\subsection{Detailed t-SNE Feature Space Visualization}
\label{app:tsne_detailed}

We present 3-D t-SNE visualizations of penultimate layer features at ER=0.5,
comparing Adam, SAM, DAEF-VS, and DASM. Gray points represent Cover samples while
colored points represent Stego samples.

\begin{figure*}[t]
    \centering
    \subfloat[AHCM]{%
      \includegraphics[width=0.24\textwidth]{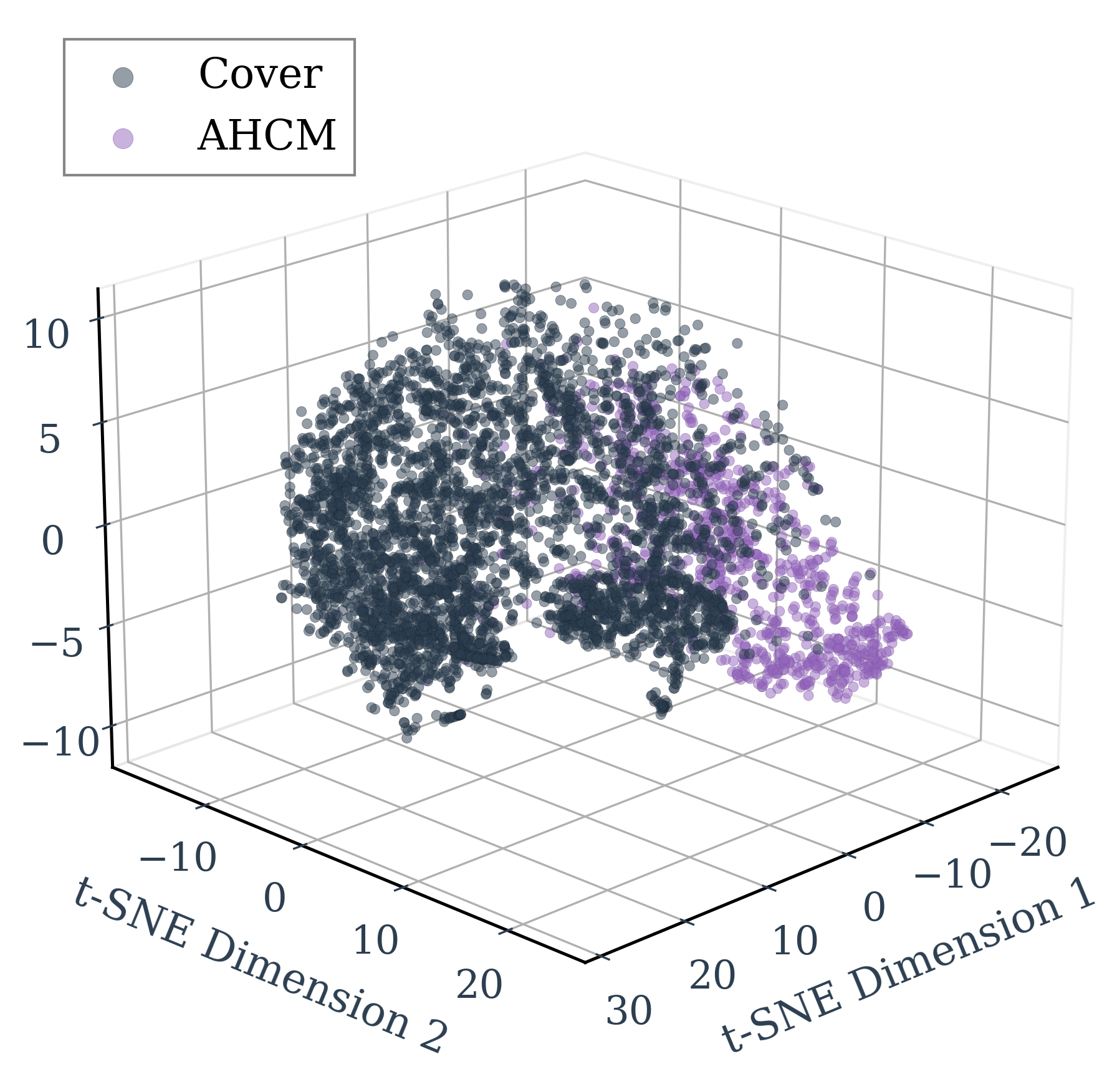}}%
    \hfill
    \subfloat[LSB]{%
      \includegraphics[width=0.24\textwidth]{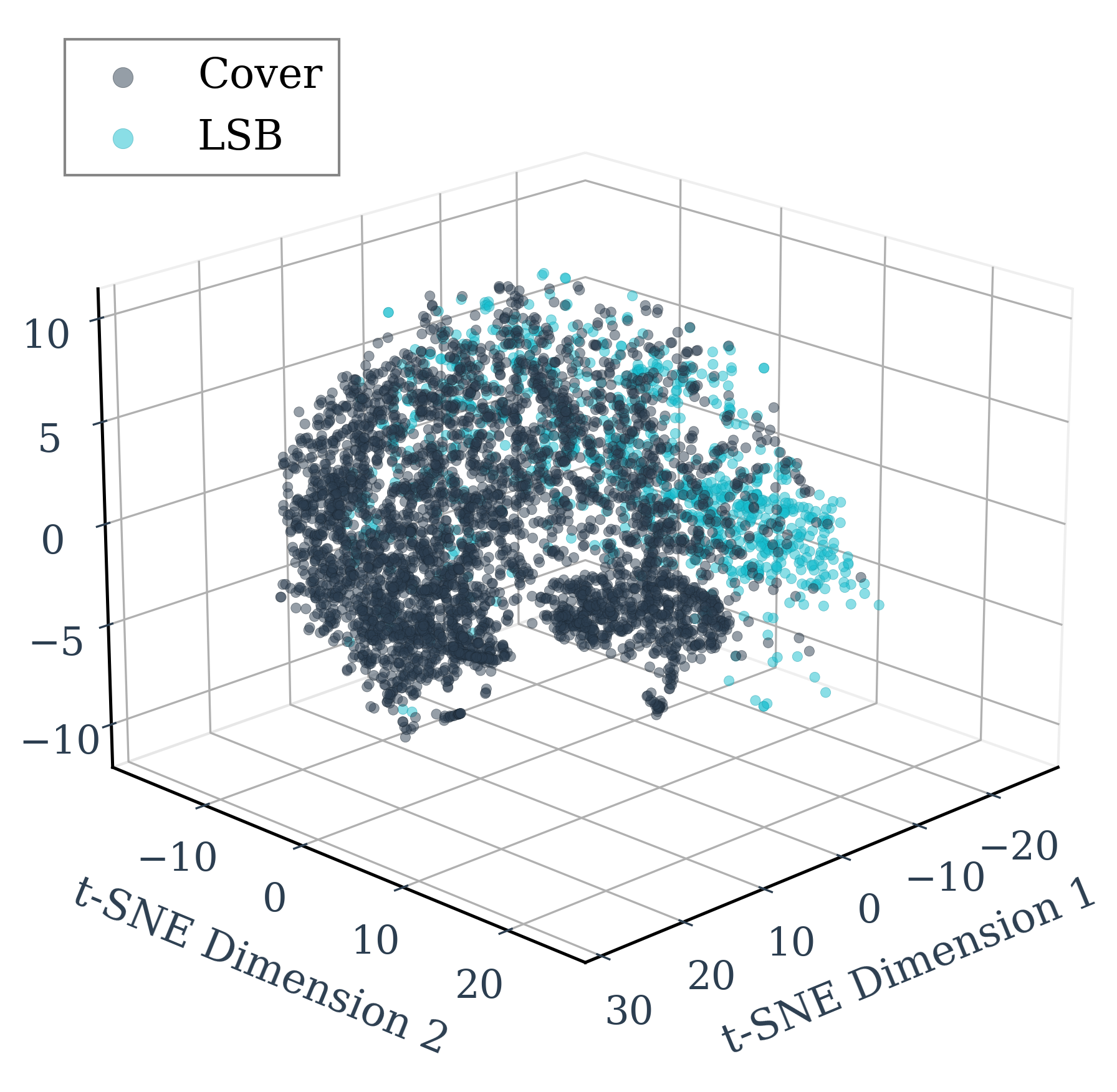}}%
    \hfill
    \subfloat[PMS]{%
      \includegraphics[width=0.24\textwidth]{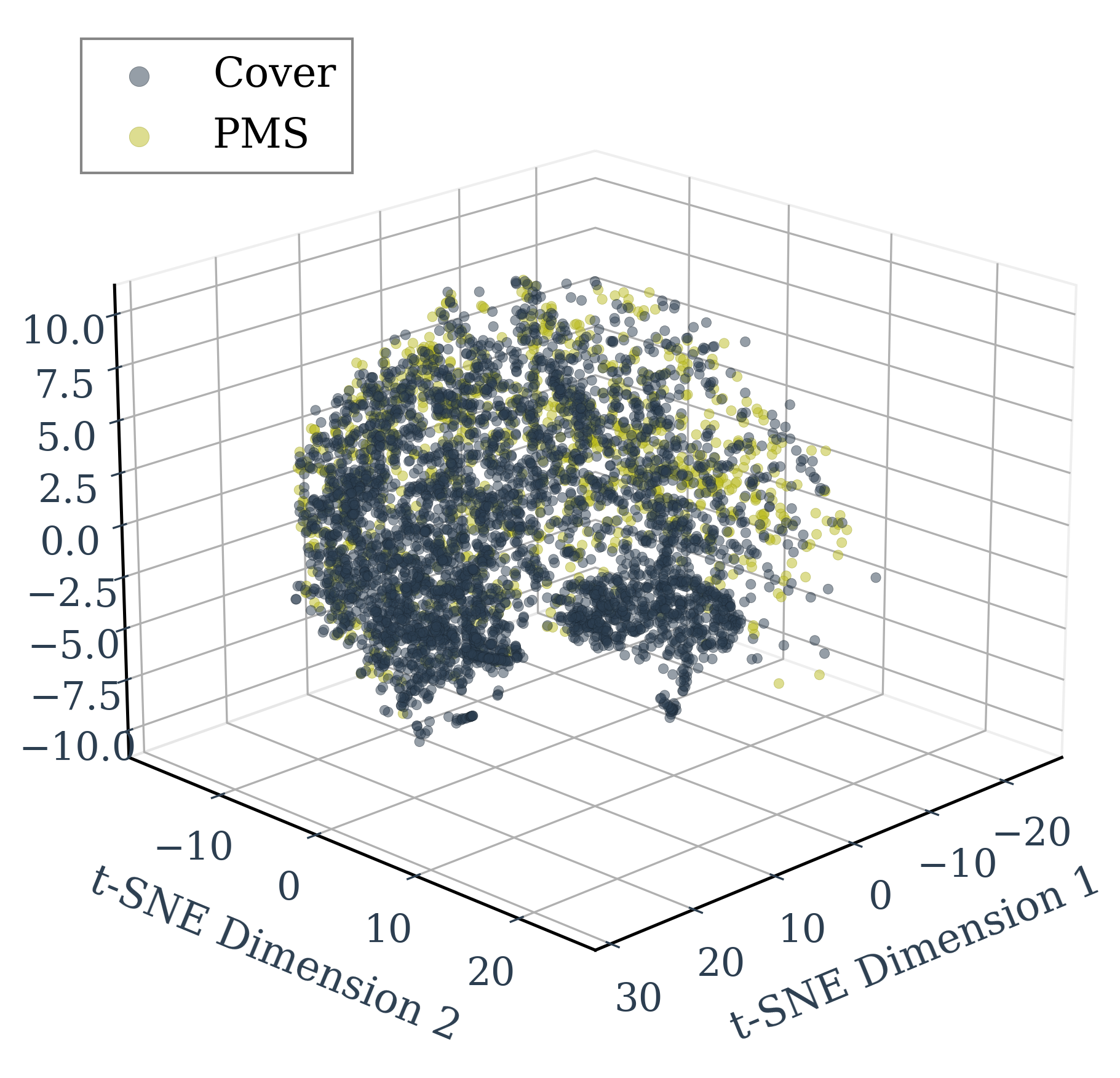}}%
    \hfill
    \subfloat[QIM]{%
      \includegraphics[width=0.24\textwidth]{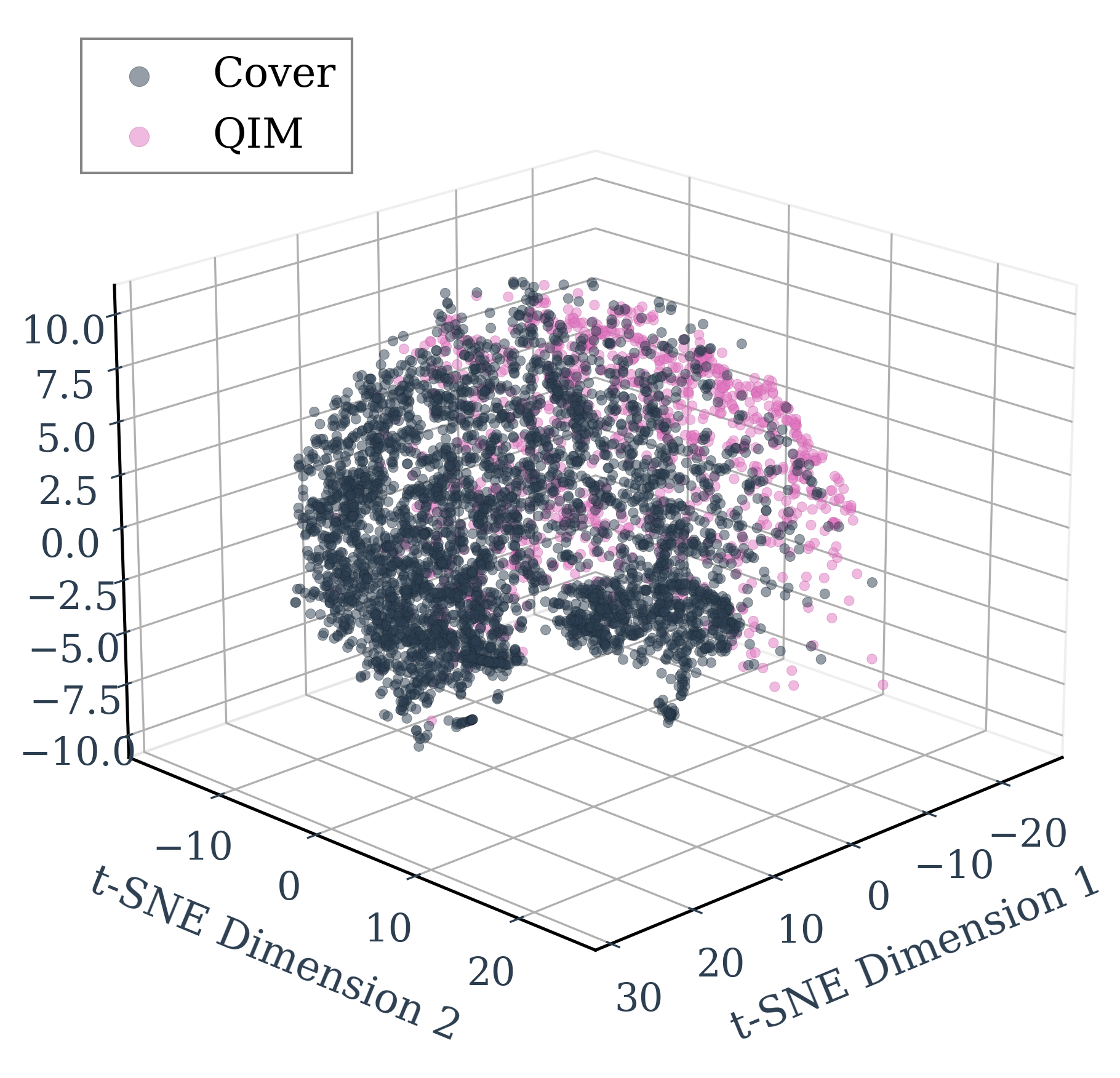}}%
    \caption{t-SNE visualization for Adam. Severe overlap between Cover and Stego
    features indicates convergence to suboptimal solutions.}
    \label{fig:tsne_adam_detailed}
\end{figure*}

\textbf{Adam Baseline.} Fig.~\ref{fig:tsne_adam_detailed} shows significant
overlap between Cover and Stego samples across all domains. This entanglement is
particularly severe for PMS and QIM, correlating with the near-random detection
accuracy observed in our main experiments.

\begin{figure*}[t]
    \centering
    \subfloat[AHCM]{%
      \includegraphics[width=0.24\textwidth]{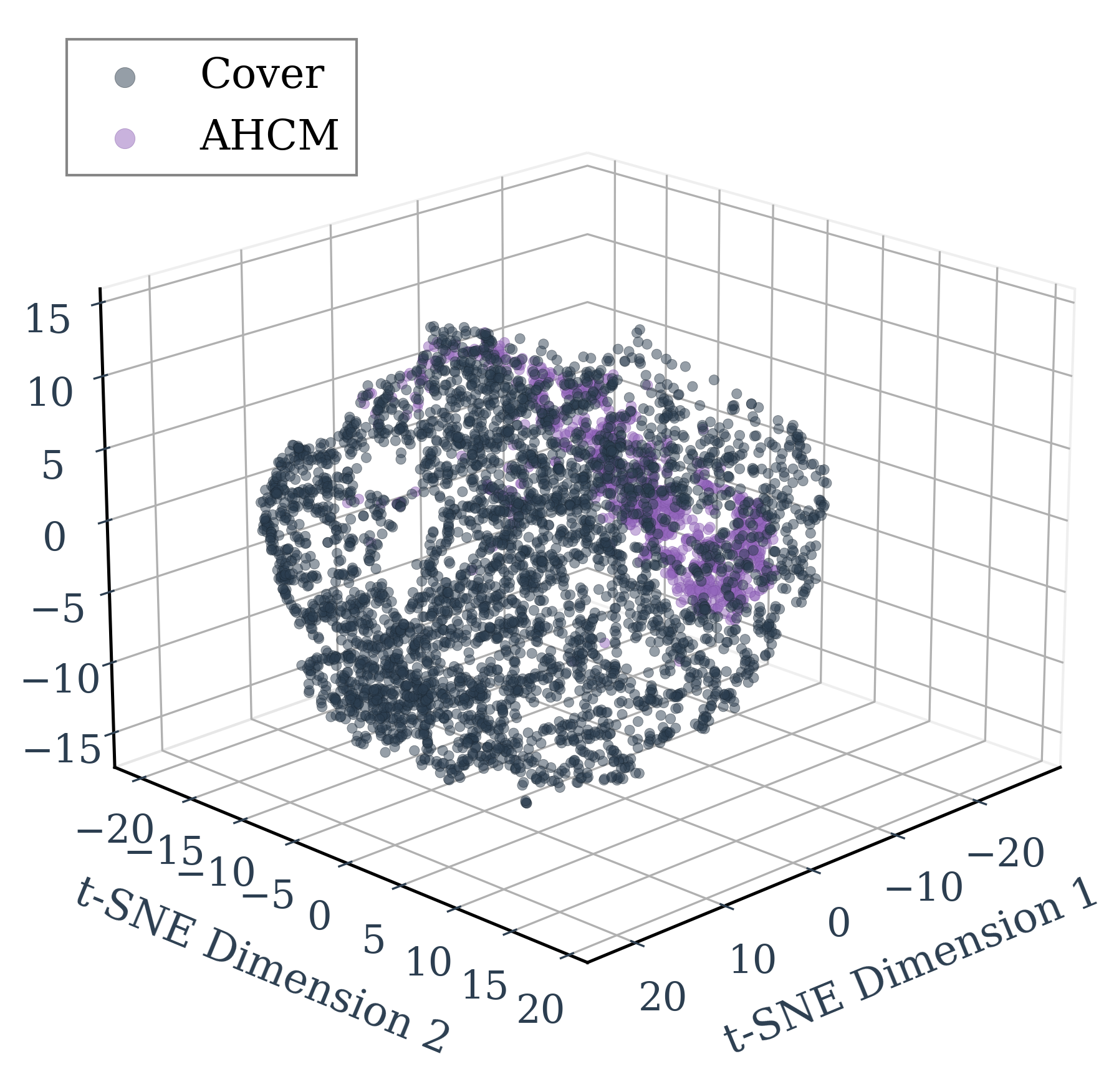}}%
    \hfill
    \subfloat[LSB]{%
      \includegraphics[width=0.24\textwidth]{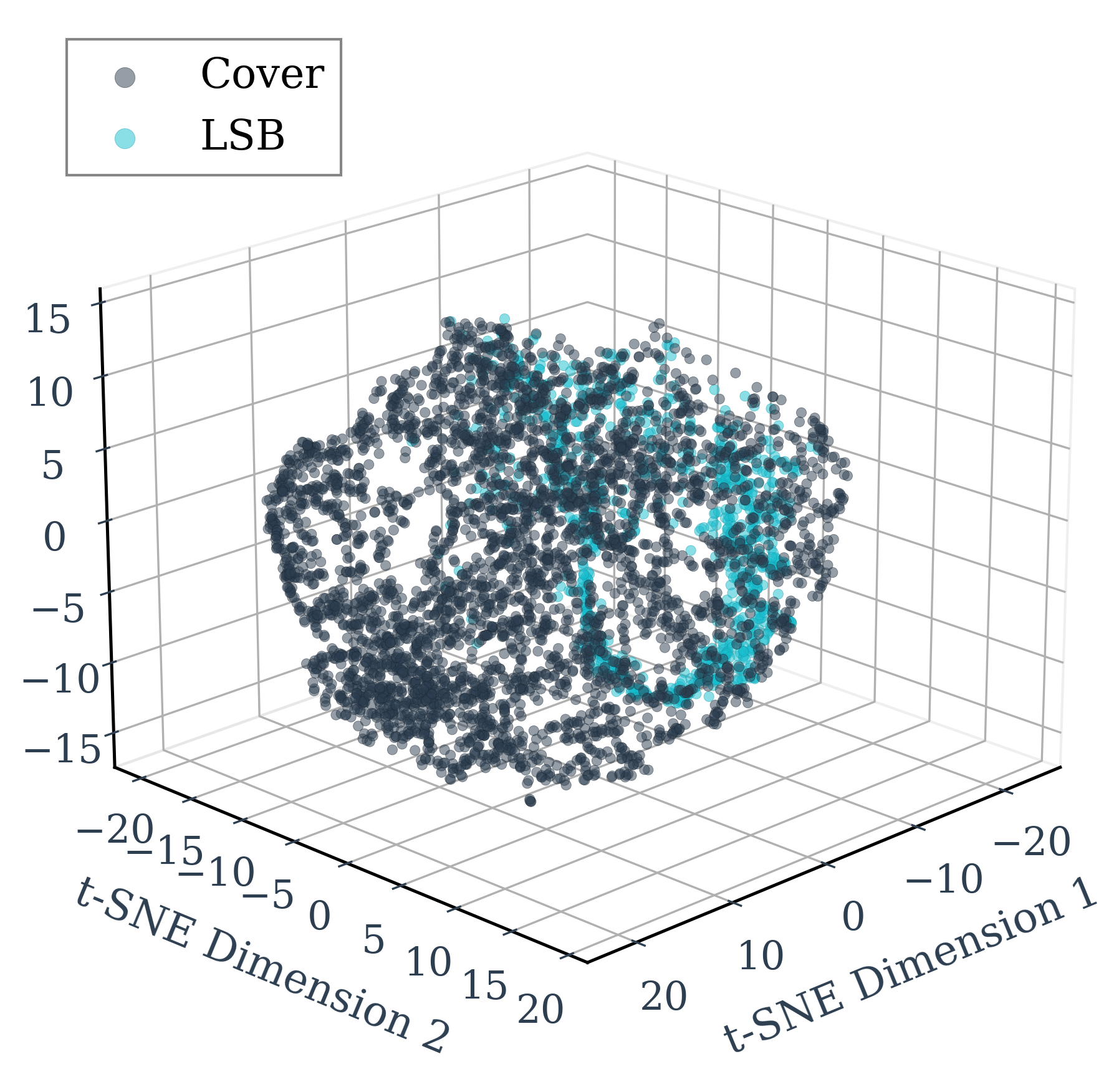}}%
    \hfill
    \subfloat[PMS]{%
      \includegraphics[width=0.24\textwidth]{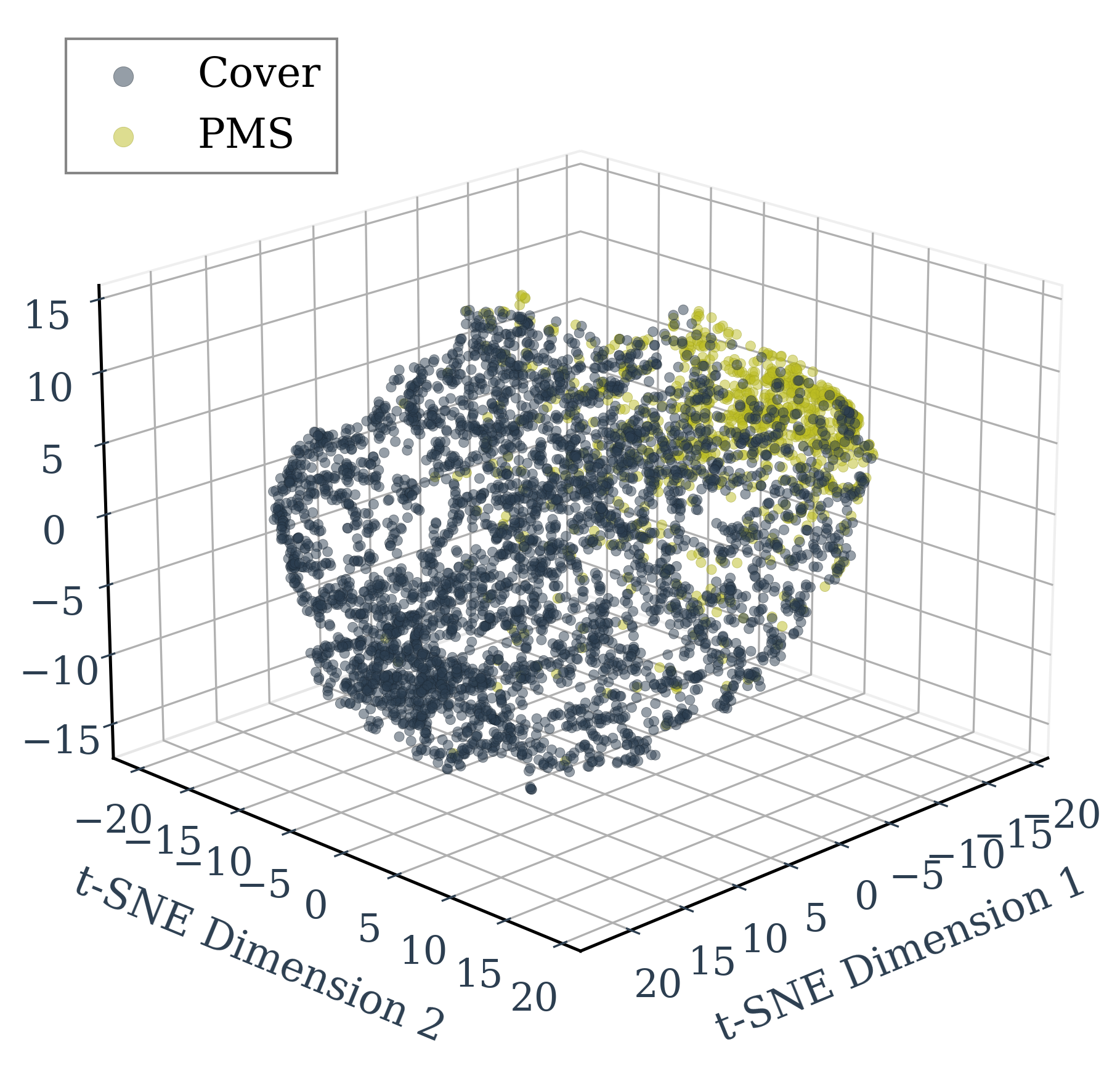}}%
    \hfill
    \subfloat[QIM]{%
      \includegraphics[width=0.24\textwidth]{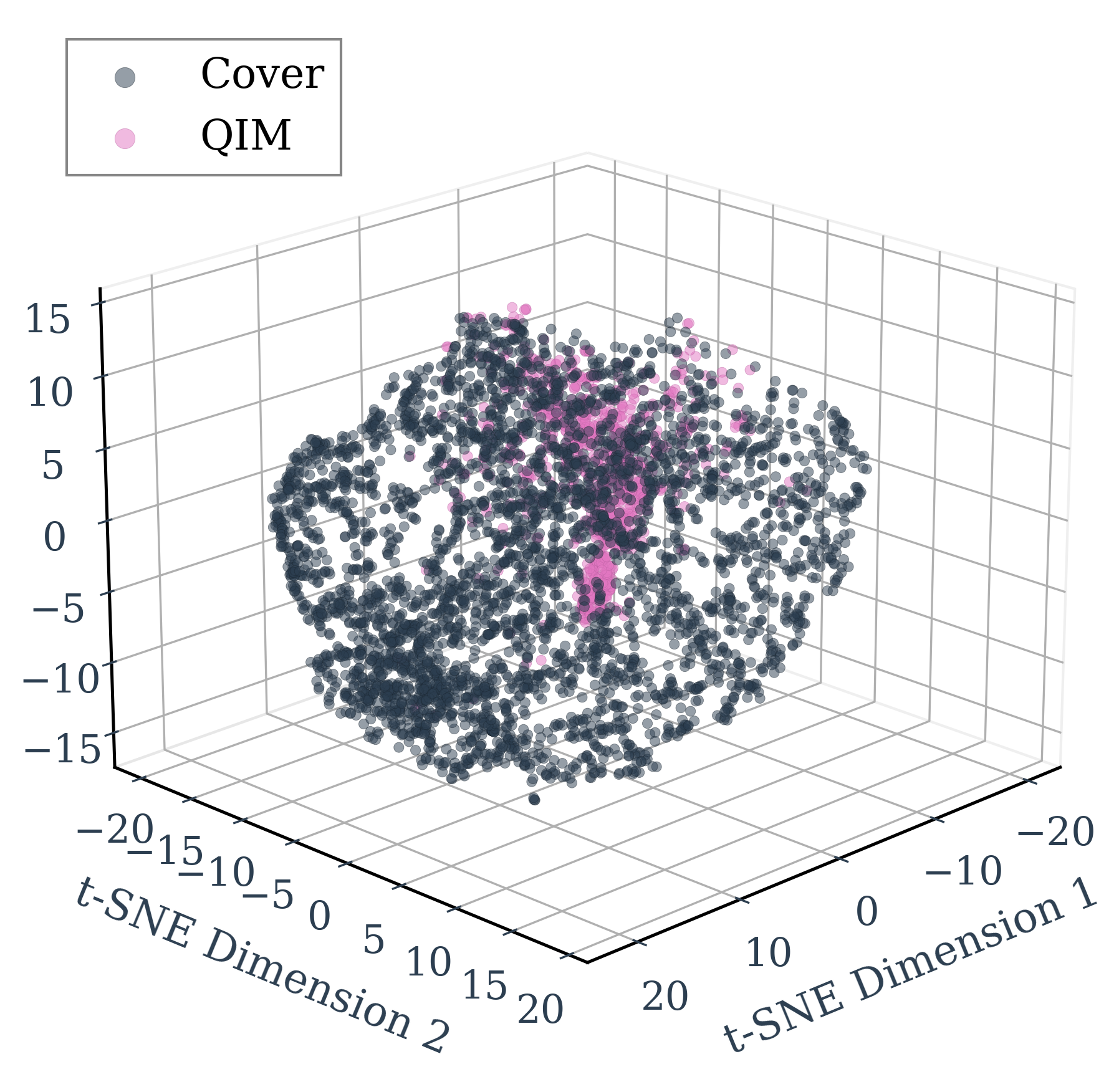}}%
    \caption{t-SNE visualization for SAM. Improved separation for AHCM and LSB,
    but PMS remains entangled due to limitations of isotropic perturbations.}
    \label{fig:tsne_sam_detailed}
\end{figure*}

\textbf{Standard SAM.} Fig.~\ref{fig:tsne_sam_detailed} shows improved separation
for AHCM and LSB compared to Adam, but PMS remains entangled. This validates that
isotropic perturbations fail to address imbalanced domain gaps.

\begin{figure*}[t]
    \centering
    \subfloat[AHCM]{%
      \includegraphics[width=0.24\textwidth]{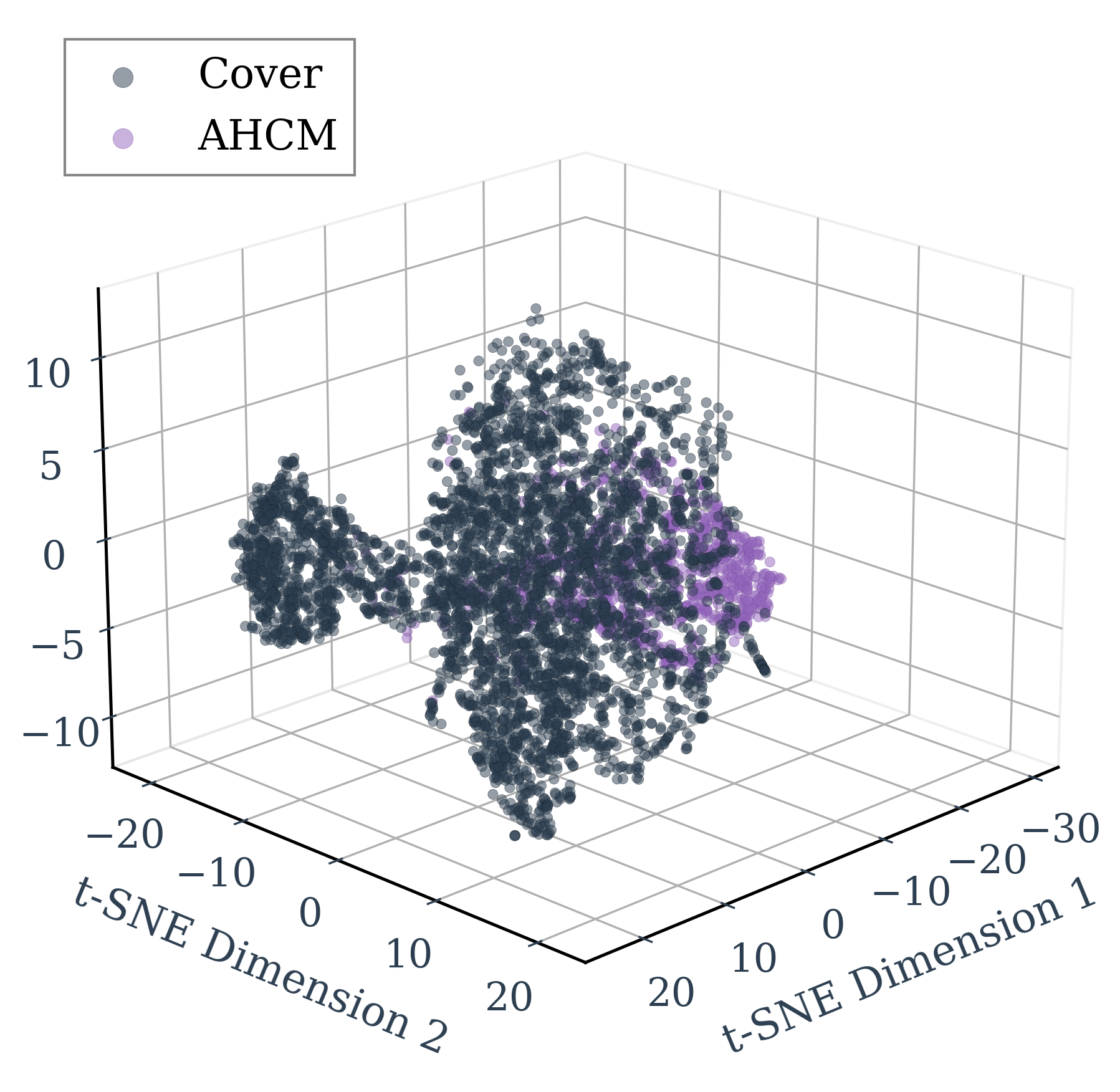}}%
    \hfill
    \subfloat[LSB]{%
      \includegraphics[width=0.24\textwidth]{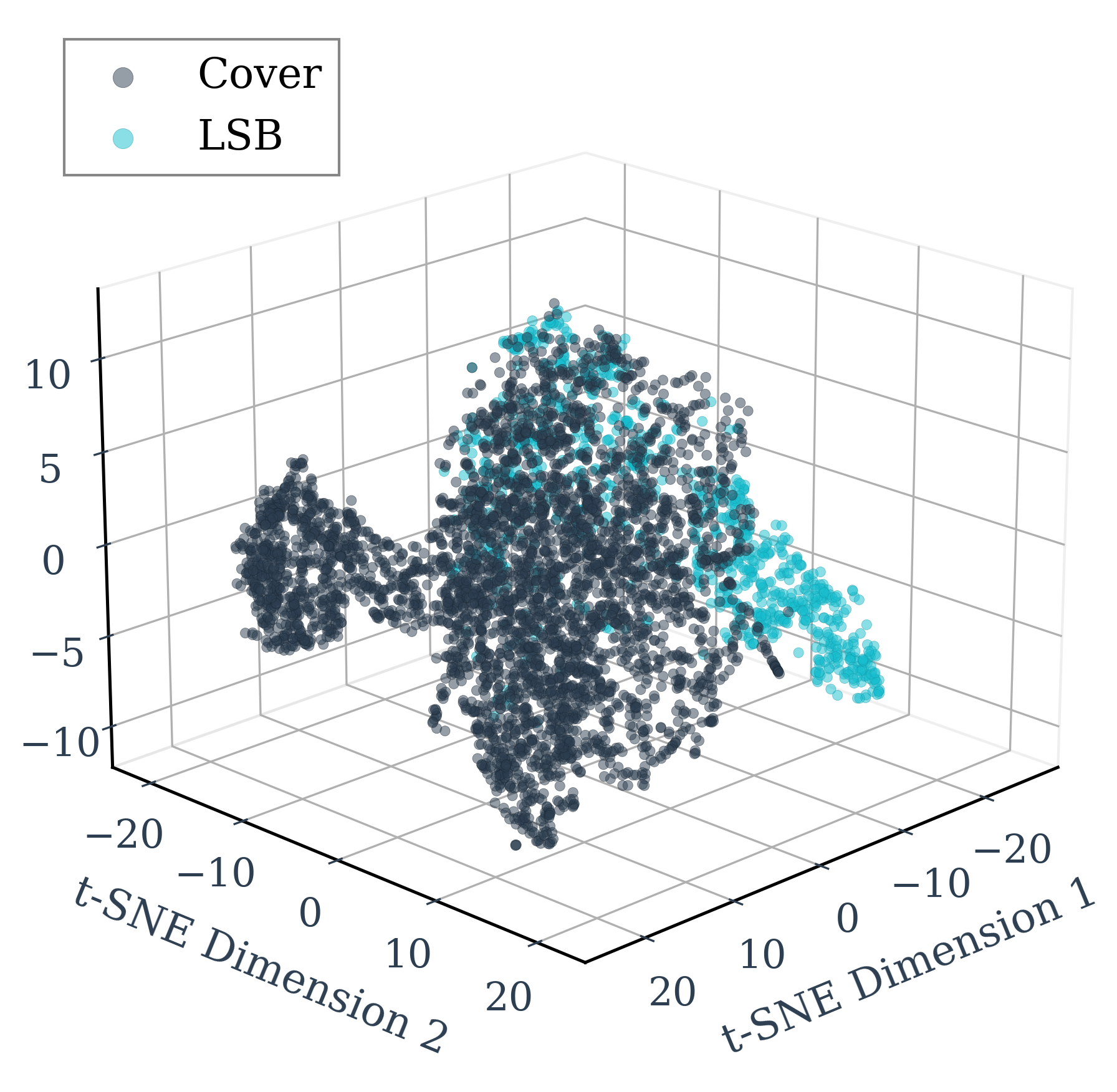}}%
    \hfill
    \subfloat[PMS]{%
      \includegraphics[width=0.24\textwidth]{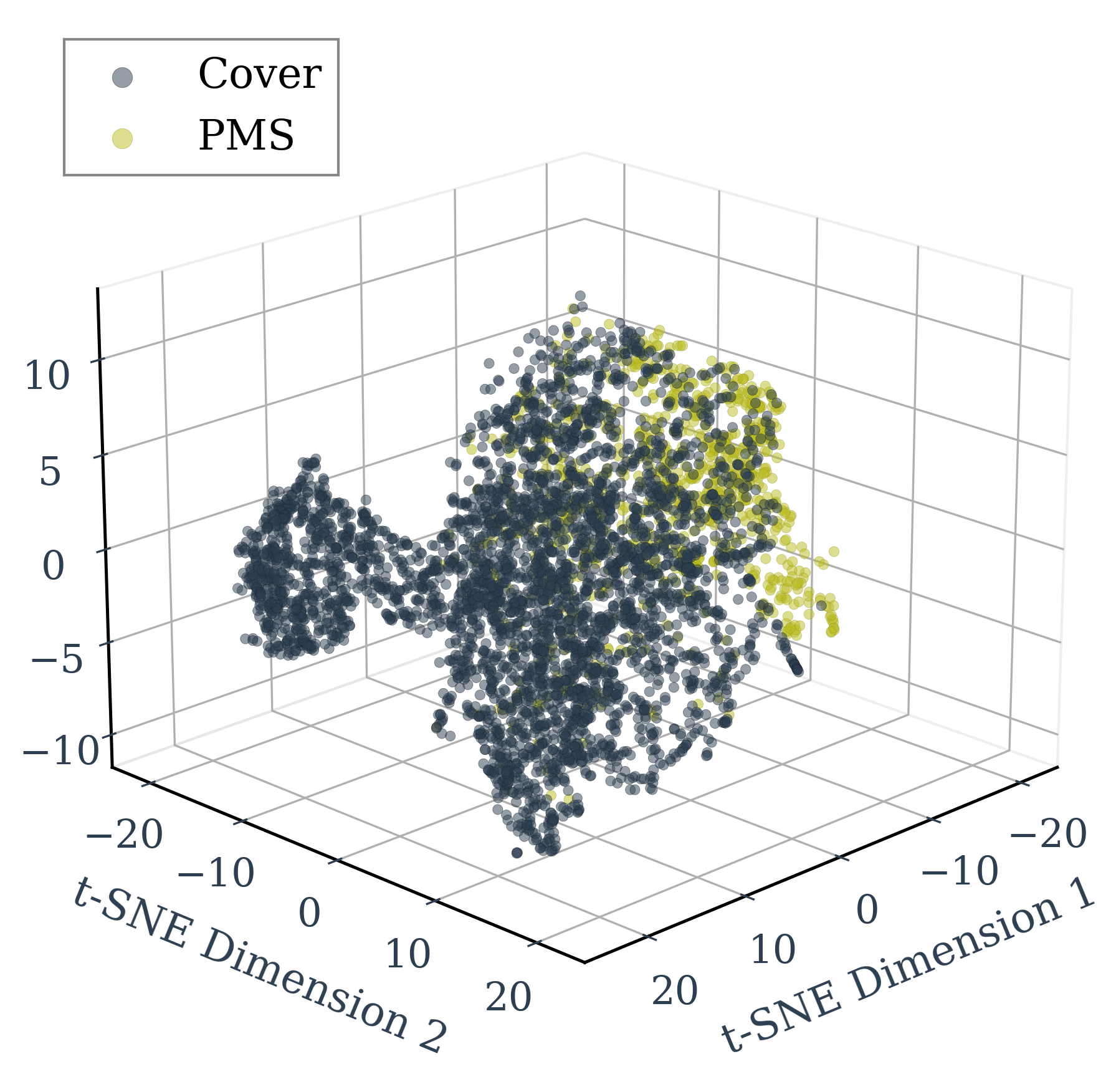}}%
    \hfill
    \subfloat[QIM]{%
      \includegraphics[width=0.24\textwidth]{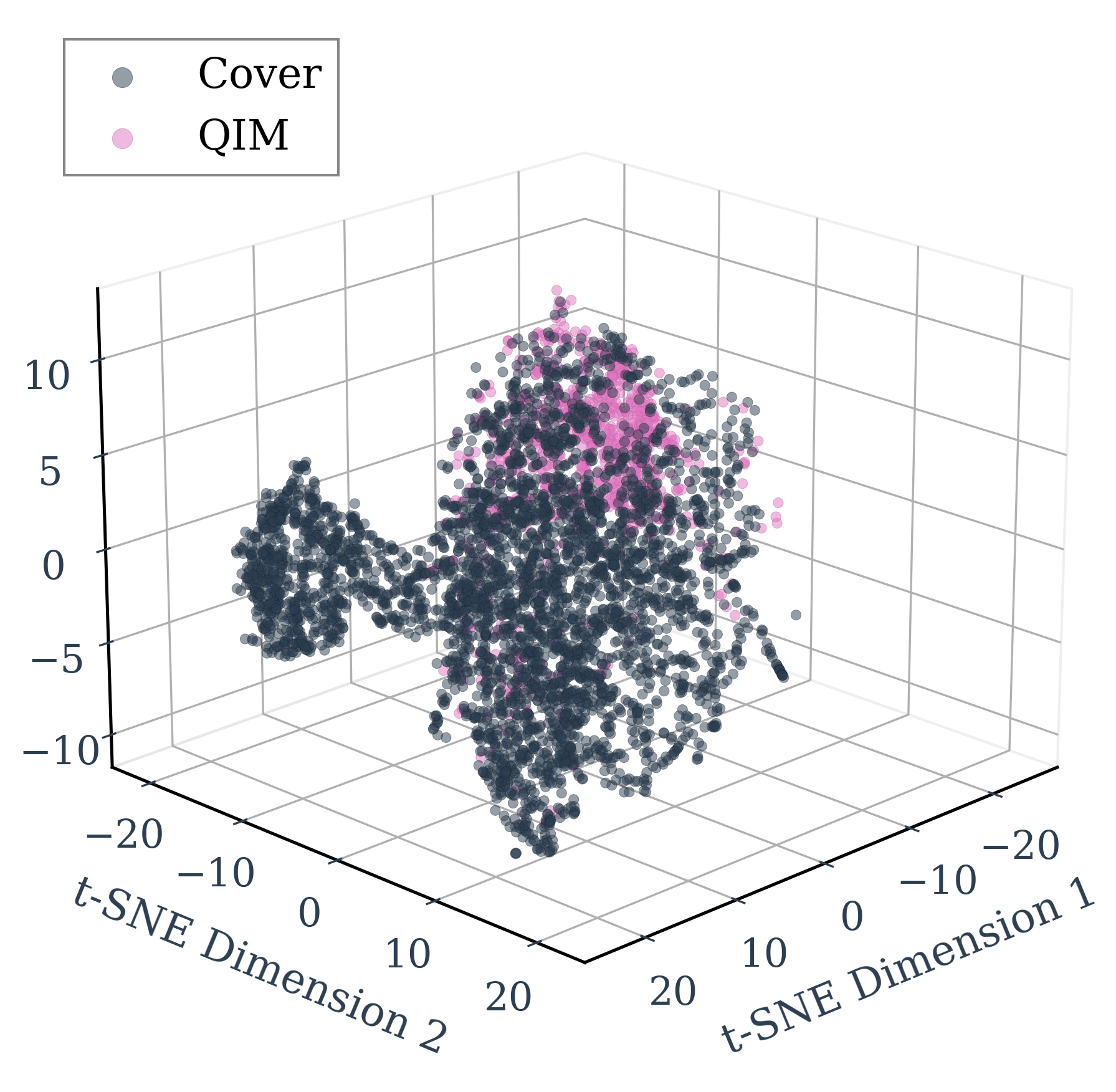}}%
    \caption{t-SNE visualization for DAEF-VS. Specialized architecture alone
    yields limited improvement for difficult domains.}
    \label{fig:tsne_daefvs_detailed}
\end{figure*}

\textbf{DAEF-VS.} Fig.~\ref{fig:tsne_daefvs_detailed} shows that despite its
specialized architecture, DAEF-VS fails to achieve consistent separation,
particularly for PMS.

\begin{figure*}[t]
    \centering
    \subfloat[AHCM]{%
      \includegraphics[width=0.24\textwidth]{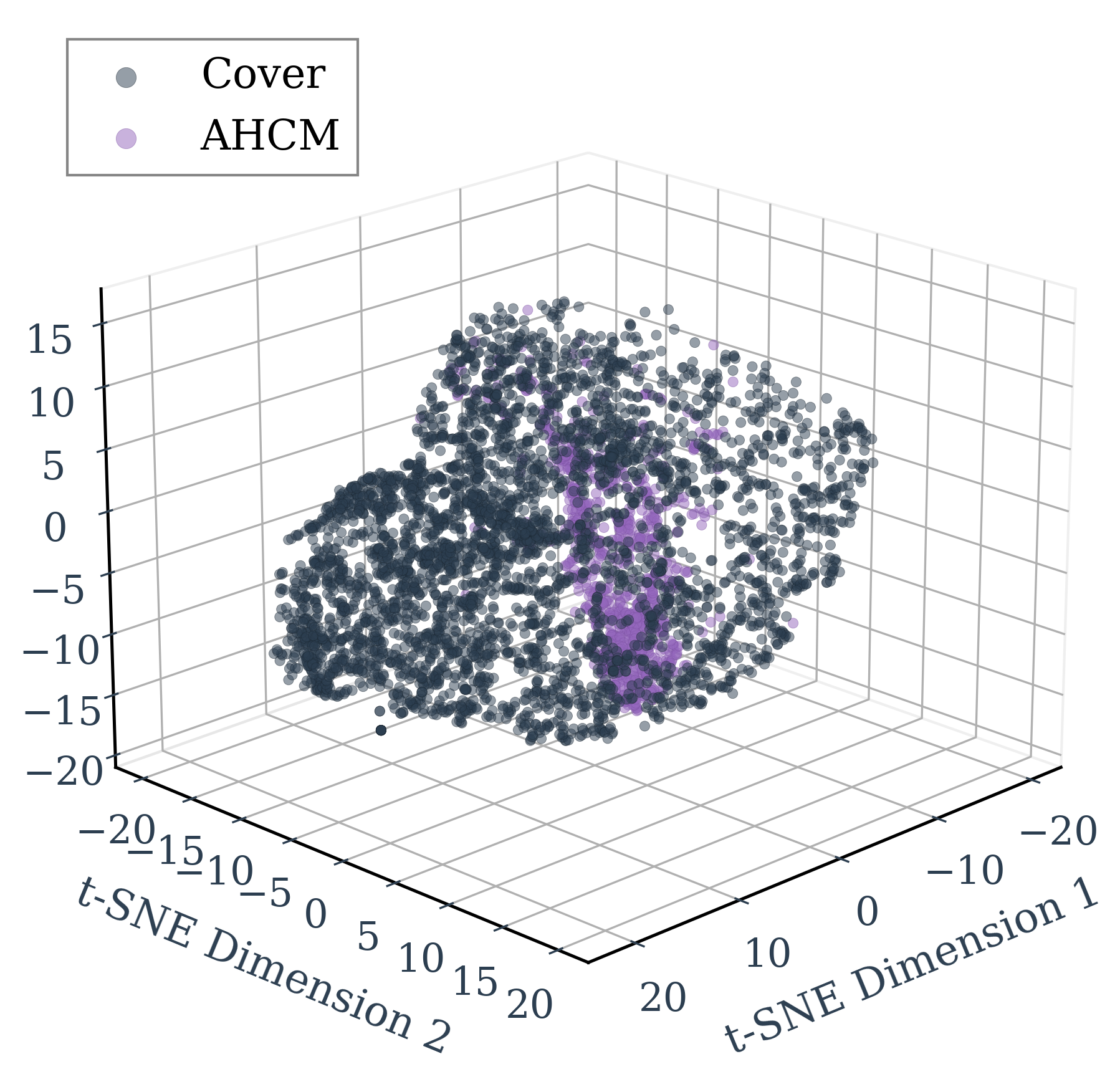}}%
    \hfill
    \subfloat[LSB]{%
      \includegraphics[width=0.24\textwidth]{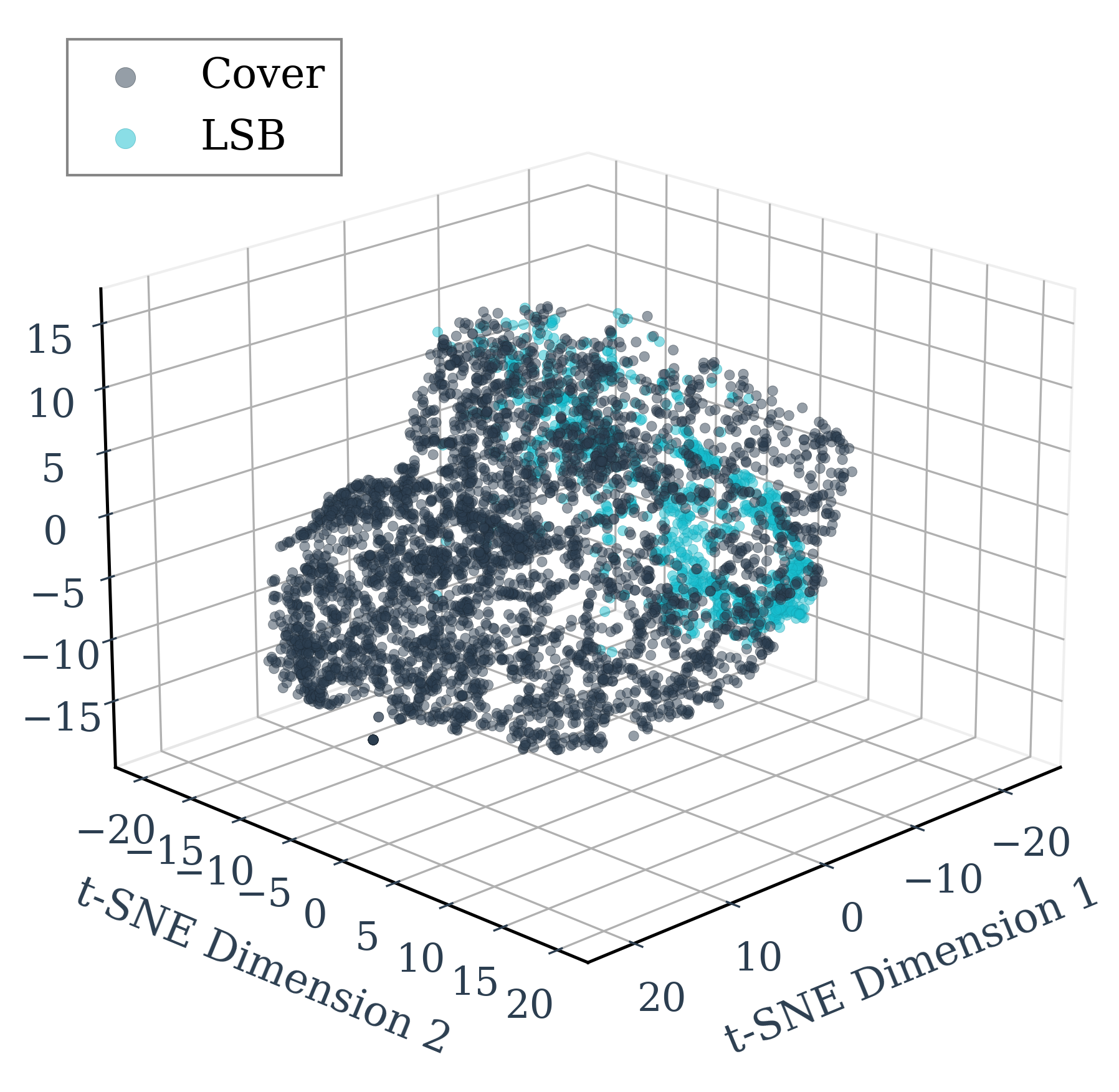}}%
    \hfill
    \subfloat[PMS]{%
      \includegraphics[width=0.24\textwidth]{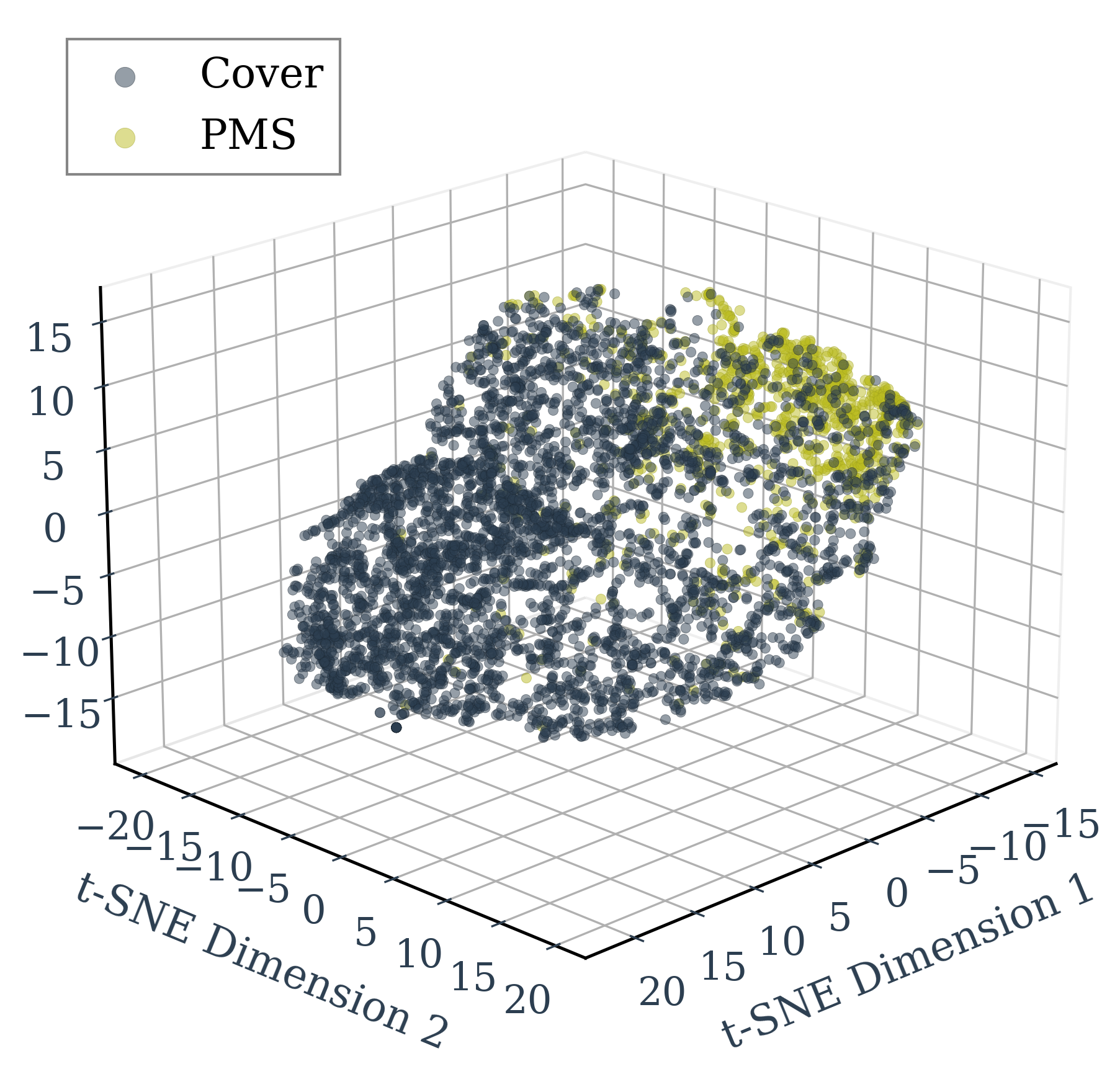}}%
    \hfill
    \subfloat[QIM]{%
      \includegraphics[width=0.24\textwidth]{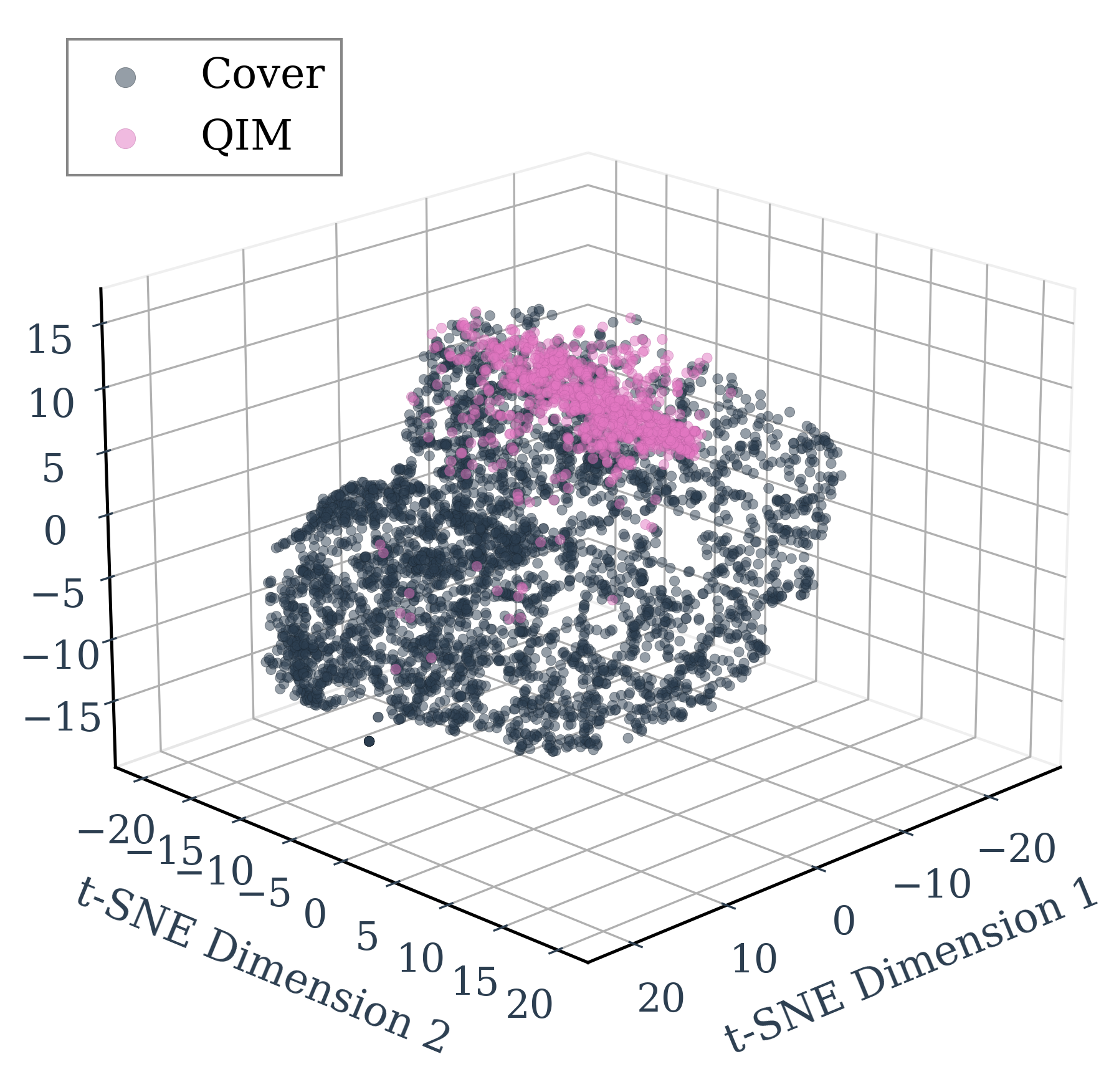}}%
    \caption{t-SNE visualization for DASM. Clear separation across all domains
    including PMS validates the effectiveness of domain-aware sharpness
    minimization.}
    \label{fig:tsne_dasm_detailed}
\end{figure*}

\textbf{DASM.} Fig.~\ref{fig:tsne_dasm_detailed} demonstrates superior feature
organization with clearly separated clusters across all domains. Even for PMS,
DASM maintains structured separation, directly correlating with our superior
detection accuracy.

The t-SNE visualizations reveal a clear progression: Adam produces entangled
features with poor discriminability; SAM improves separation for easier domains
but fails on PMS and QIM; DAEF-VS demonstrates that sophisticated architectures
alone cannot overcome optimization challenges; DASM achieves consistent separation
across all domains. The absence of mode collapse in PMS and QIM under DASM
validates our adaptive gap modulation in preventing easier domains from dominating
the optimization.

\subsection{Visual Analysis of Ablation Study}
\label{app:visual_ablation}

\begin{figure*}[t]
    \begin{minipage}[b]{0.35\textwidth}
        \centering
        \includegraphics[width=\linewidth]{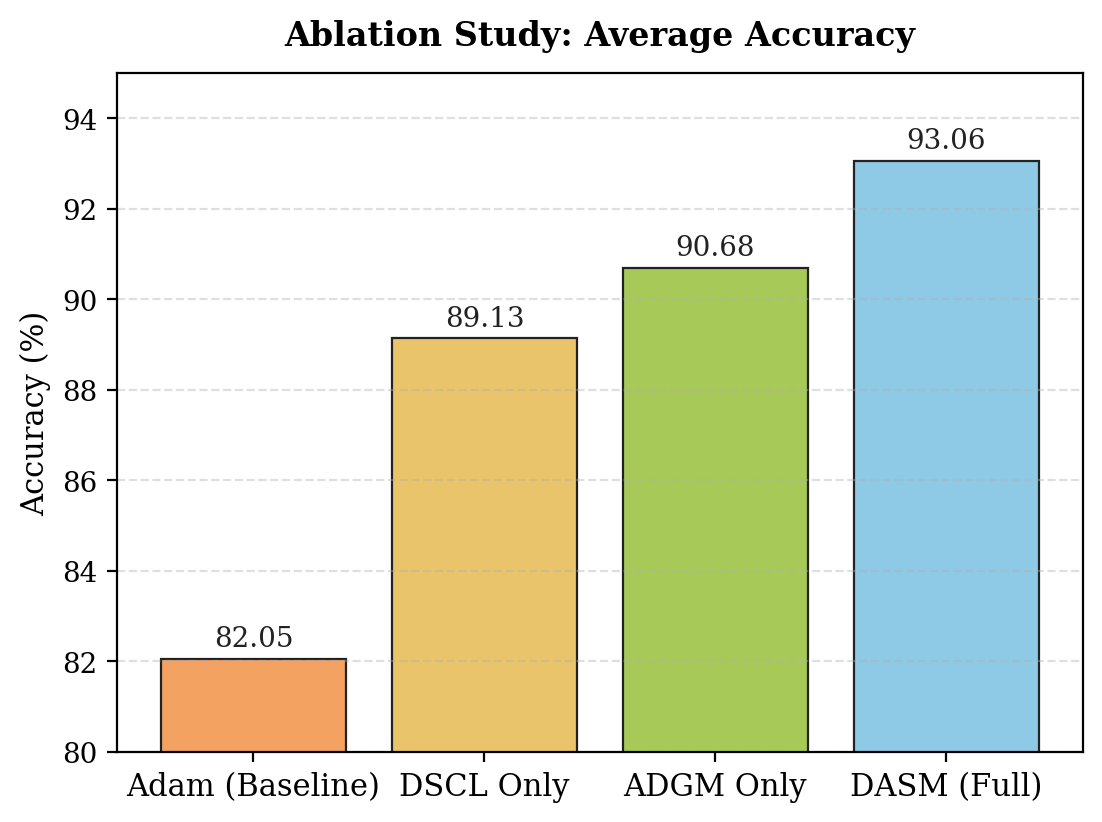}
        \caption{Ablation Study: Average Accuracy at ER=0.5.}
        \label{fig:ablation_bar_appendix}
    \end{minipage}
    \hfill
    \begin{minipage}[b]{0.60\textwidth}
        \centering
        \subfloat[Sensitivity to $\rho$]{\label{fig:sens_rho}%
          \includegraphics[width=0.48\linewidth]{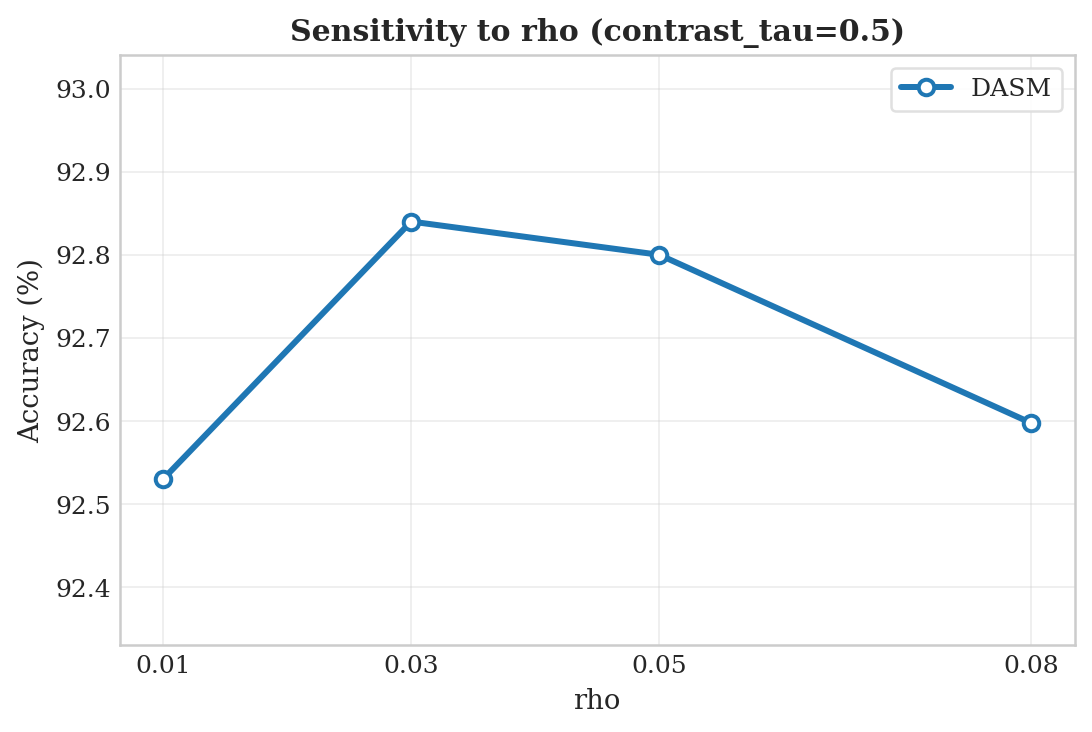}}%
        \hfill
        \subfloat[Sensitivity to $\tau$]{\label{fig:sens_tau}%
          \includegraphics[width=0.48\linewidth]{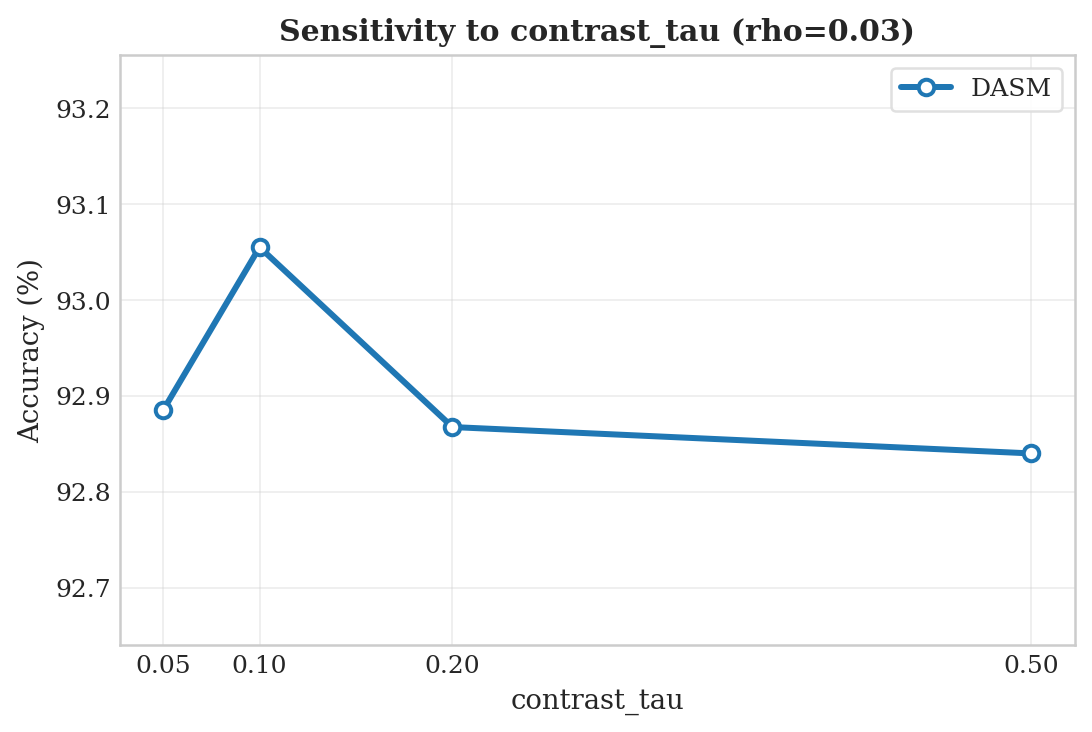}}%
        \caption{Hyperparameter sensitivity analysis of DASM.}
        \label{fig:sensitivity_line_appendix}    
    \end{minipage}
\end{figure*}

As illustrated in Fig.~\ref{fig:ablation_bar_appendix}, we provide a visual
performance comparison of the variants discussed in Section~\ref{subsec:ablation}.
The transition from the Adam Baseline (82.05\%) to the ``DSCL Only'' variant
(89.13\%) highlights the critical importance of preserving inter-domain feature
separability in multi-domain VoIP steganalysis. The ``ADGM Only'' variant further
improves the accuracy to 90.68\%, validating the efficacy of adaptive weighting in
guiding the optimizer toward a flat minimum that accounts for polarized domain
gaps. The complete DASM framework reaches the peak accuracy of 93.06\%,
demonstrating a clear synergy between $\mathcal{L}_{\text{DSCL}}$ and
$\mathcal{L}_{\text{ADGM}}$.

\subsection{Detailed Hyperparameter Sensitivity Trends}
\label{app:sensitivity_trends}

This section expands upon the sensitivity analysis in
Section~\ref{subsec:sensitivity}, characterizing the impact of $\rho$ and $\tau$
on detection performance.

\textbf{Analysis of Perturbation Radius $\rho$.}
Fig.~\ref{fig:sens_rho} shows the accuracy dynamics as $\rho$ varies within
$\{0.01, 0.03, 0.05, 0.08\}$ while fixing $\tau=0.5$. The results indicate that
$\rho=0.03$ is the optimal radius for sharpness-aware estimation in this task. A
minimal radius ($\rho=0.01$) fails to provide a sufficient regularization effect
to escape the sharp local minima associated with subtle steganographic artifacts.
Conversely, an excessively large radius ($\rho=0.08$) tends to over-smooth the
loss landscape, which may inadvertently obscure the microscopic statistical
deviations essential for distinguishing steganographic audio from cover audio. DASM
maintains high stability within the $[0.03, 0.05]$ range, suggesting a robust
neighborhood for locating flat basins.

\textbf{Analysis of Contrastive Temperature $\tau$.}
Fig.~\ref{fig:sens_tau} illustrates the sensitivity to $\tau$ with $\rho$ fixed
at 0.03. The peak performance at $\tau=0.1$ signifies the most effective scaling
for the similarity distribution. As $\tau$ increases toward 0.5, the contrastive
signal becomes diffused, which hinders the model's ability to pull apart domains
with extremely low discrepancies, such as PMS and Cover. On the other hand, an
overly low $\tau$ (0.05) sharpens the distribution to an extent that makes the
loss highly sensitive to hard-negative samples, potentially introducing
optimization noise. These trends confirm that $\tau=0.1$ provides the ideal
balance for enforcing domain separability without compromising optimization
stability.

\subsection{Sharpness Analysis}
\label{app:sharpness_details}

In this section, we provide the detailed quantitative measurements of the loss
landscape sharpness referenced in Section~\ref{sec:sharpness_main}.
Table~\ref{tab:zeroth_order_sharpness_appendix} lists the zeroth-order sharpness
values for individual domains as well as the aggregated metrics. The sharpness is
calculated as $\max_{\|\epsilon\|_2 \le \rho} \mathcal{L}(\theta+\epsilon) -
\mathcal{L}(\theta)$ with $\rho=0.05$ on the test set.

\begin{table}[t]
\caption{Zeroth-Order Sharpness Comparison ($\rho=0.05$). Lower Values Indicate
Flatter Minima. DASM Achieves the Lowest Sharpness and Variance.}
\label{tab:zeroth_order_sharpness_appendix}
\centering
\renewcommand{\arraystretch}{1.1}
\begin{tabular}{@{}l|cccc|c|c@{}}
\toprule
\multirow{2}{*}{Alg.} & \multicolumn{4}{c|}{Individual Domains} &
  \multirow{2}{*}{\shortstack{Mean\\[-2pt](Std)}} & \multirow{2}{*}{Total} \\
& QIM & PMS & LSB & AHCM & & \\
\midrule
Adam  & 2.326 & 2.272 & 3.142 & 1.596 & 2.33\,(0.55) & 0.847 \\
SAM   & 0.527 & 0.754 & 0.477 & 2.465 & 1.06\,(0.82) & 0.619 \\
DISAM & 2.611 & \textbf{0.314} & 2.948 & 1.791 & 1.92\,(1.02) & 1.895 \\
SAGM  & 9.675 & 9.114 & 9.186 & 5.238 & 8.30\,(1.78) & 8.248 \\
FSAM  & 3.271 & 0.750 & 3.409 & 2.509 & 2.49\,(1.06) & 2.300 \\
DGSAM & 3.670 & 0.643 & 3.839 & 3.289 & 2.86\,(1.30) & 2.831 \\
\textbf{DASM}
      & \textbf{0.229} & 0.371 & \textbf{0.147} & \textbf{0.262}
      & \textbf{0.25\,(0.08)} & \textbf{0.086} \\
\bottomrule
\end{tabular}
\end{table}

\textbf{Instability of Prior Methods.}
A critical observation from the table is the high variance exhibited by existing
domain generalization methods. For instance, while DGSAM achieves a reasonable
sharpness of 0.643 on PMS, it spikes significantly to 3.670 on QIM and 3.839 on
LSB, resulting in a high standard deviation of 1.296. Similarly, SAGM suffers from
extremely high sharpness across all domains with a mean of 8.303. This suggests
that without adaptive modulation, these optimizers may achieve ``pseudo-flatness''
by overfitting to specific domain directions while leaving the loss landscape sharp
and vulnerable in others.

\textbf{Isotropic Flatness of DASM.}
In contrast, DASM demonstrates true \textit{isotropic flatness}. It does not
merely minimize the average loss but regularizes the landscape uniformly. As shown
in the table, DASM achieves the lowest sharpness values consistently across every
single domain, such as 0.229 on QIM and 0.147 on LSB, with a negligible standard
deviation of 0.080. This confirms that the combination of contrastive learning and
adaptive gap modulation effectively eliminates sharp curvature in all principal
directions, providing a rigorous geometric guarantee for the model's robustness
against diverse steganographic distribution shifts.

\subsection{Nomenclature}
\label{app:notation}

Table~\ref{tab:notation} summarizes the mathematical notations used throughout
this paper.

\makeatletter
\setlength{\@dblfptop}{0pt}
\makeatother
\begin{table*}[!t]
\caption{Mathematical Notations and Hyperparameters.}
\label{tab:notation}
\centering
\small
\renewcommand{\arraystretch}{1.15}
\begin{tabular}{@{}l@{\hspace{6pt}}p{6.6cm}@{\hskip 2em}l@{\hspace{6pt}}p{6.6cm}@{}}
\toprule
\textbf{Symbol} & \textbf{Description} & \textbf{Symbol} & \textbf{Description} \\
\midrule
\multicolumn{4}{@{}l}{\textit{Problem Formulation}} \\
$\theta$              & Model parameters
  & $f_\theta(\cdot)$     & Neural network parameterized by $\theta$ \\
$\mathcal{D}$         & Training dataset: tuples $(x, y, d)$
  & $\mathbf{z}_i$        & L2-normalized feature of sample $x_i$ \\
$\mathcal{B}$         & Mini-batch sampled from $\mathcal{D}$
  & $\bar{\mathbf{z}}_k$  & Mean feature of domain $k$ in the batch \\
$B$                   & Batch size
  & $d_i$                 & Domain index for sample $i$ \\
$S$ / $K$             & Number of steganographic domains & & \\
\midrule
\multicolumn{4}{@{}l}{\textit{Loss Functions}} \\
$\mathcal{L}_{\text{total}}$ & Composite total loss
  & $\mathcal{L}_{\text{DSCL}}$  & Domain-supervised contrastive loss \\
$\mathcal{L}_{\text{CE}}$    & Cross-entropy classification loss
  & $\mathcal{L}_{\text{ADGM}}$  & Adaptive domain gap modulation loss \\
\midrule
\multicolumn{4}{@{}l}{\textit{Sharpness-Aware Optimization}} \\
$\rho$           & Perturbation radius
  & $\eta$           & Learning rate \\
$\hat{\epsilon}$ & Adversarial weight perturbation
  & $H$              & Hessian matrix of the loss landscape \\
\midrule
\multicolumn{4}{@{}l}{\textit{Contrastive Learning}} \\
$\tau$                & Temperature for similarity scaling
  & $N(i)$                & Negative samples from different domains \\
$P(i)$                & Positive samples sharing domain with anchor $i$
  & $S_i^+$, $S_i^-$     & Aggregated positive/negative similarities \\
\midrule
\multicolumn{4}{@{}l}{\textit{Domain Gap Modulation}} \\
$\mathbf{c}_k$              & Running centroid for domain $k$
  & $\tau_g$                    & Adaptive temperature for gap weighting \\
$\mathbf{c}_{\text{cover}}$ & Running centroid for cover domain
  & $w_k$                       & Adaptive weight for domain $k$ \\
$\mu$                       & EMA momentum coefficient
  & $\xi$                       & Numerical stability constant \\
$g_k$                       & Domain gap between domain $k$ and cover & & \\
\midrule
\multicolumn{4}{@{}l}{\textit{Evaluation Metrics}} \\
$d_A$ & Proxy A-Distance for domain discrepancy & & \\
\bottomrule
\end{tabular}
\end{table*}

\end{document}